\documentclass[10pt,superscriptaddress,nofootinbib]{revtex4-1}

\usepackage{setspace} 
\usepackage{amssymb}
\usepackage{amsfonts}
\usepackage{graphicx}
\usepackage{subfigure}
\usepackage{enumitem}
\usepackage{amsmath}
\usepackage{setspace}
\usepackage{fancybox}
\usepackage[bottom]{footmisc}
\usepackage[left=3.5cm,top=2.5cm,right=3cm,bottom=2cm]{geometry}
\usepackage{rotating}
\usepackage{color}
\usepackage{url}
\usepackage[english]{babel}
\usepackage{booktabs}
\usepackage{siunitx}
\usepackage{footnote}
\usepackage{footmisc}
\usepackage{multirow}
\usepackage{fancyhdr}
\makeatletter
\renewcommand{\@biblabel}[1]{\quad#1.}
\makeatother

\date{}
\begin{document}

%\doublespacing

\title{Contact patterns in a high school: a comparison between data collected using wearable sensors, contact diaries and friendship surveys}

\author{Rossana Mastrandrea}
\affiliation{Aix Marseille Universit\'e, Universit\'e de Toulon, CNRS, CPT, UMR 7332, 13288 Marseille, France}
\author{Julie Fournet}
\affiliation{Aix Marseille Universit\'e, Universit\'e de Toulon, CNRS, CPT, UMR 7332, 13288 Marseille, France}
\author{Alain Barrat}
\affiliation{Aix Marseille Universit\'e, Universit\'e de Toulon, CNRS, CPT, UMR 7332, 13288 Marseille, France}
\affiliation{Data Science Laboratory, ISI Foundation, Torino, Italy}

\begin{abstract} 

Given their importance in shaping social networks and determining how information or diseases propagate in a population, human interactions are the subject of many data collection efforts. To this aim, different methods are commonly used, from diaries and surveys to wearable sensors. These methods show advantages and limitations but are rarely compared in a given setting. As surveys targeting friendship relations might suffer less from memory biases than contact diaries, it is also interesting to explore how daily contact patterns compare with friendship relations and with online social links. Here we make progresses in these directions by leveraging data from a French high school: face-to-face contacts measured by two concurrent methods, sensors and diaries; self-reported friendship surveys; Facebook links. We compare the data sets and find that most short contacts are not reported in diaries while long contacts have larger reporting
probability, with a general tendency to overestimate durations. Measured contacts corresponding to reported friendship can have durations of any length but all long contacts correspond to reported friendships. Online links not associated to reported friendships correspond to short face-to-face contacts, highlighting the different nature of reported friendships and online links. Diaries and surveys suffer from a low sampling rate, showing the higher acceptability of sensor-based platform. Despite the biases, we found that the overall structure of the contact network, i.e., the mixing patterns between classes, is correctly captured by both self-reported contacts and friendships networks. Overall, diaries and surveys tend to yield a correct picture of the structural organization of the contact network, albeit with much less links, and give access to a sort of backbone of the contact network corresponding to the strongest links in terms of cumulative durations.
\end{abstract}

\maketitle

\section{Introduction}

Despite the wealth of communication and interaction means made available by our modern societies, direct face-to-face 
interactions between individuals remain an essential element of human behavior and of human societies. They contribute to 
shape human social networks and determine channels of information propagation, of opinion formation, as well as 
the potential transmission routes of 
infectious diseases, in particular of respiratory pathogens. Accurate descriptions of the corresponding contact patterns
represent therefore important tools in several respects: for the fundamental knowledge and understanding of human behavior and social networks,
as well as to inform models of epidemic spread and to design and evaluate control measures such as the targeting of specific groups 
of individuals with appropriate prevention strategies or interventions. 

Empirical data describing direct interactions between individuals are however by nature difficult to gather. Various techniques
have been developed to this aim, and many data sets collected and exploited, in particular in the epidemiological
context (see \cite{Read:2012} for a review): 
surveys and diaries~\cite{Bernard:2009,Edmunds:1997,Read:2008,Mossong:2008,Mikolajczyk:2008,Danon:2012,Danon:2013,Conlan:2011,Smieszek:2012,Smieszek:2014}, 
synthetic population models~\cite{Eubank:2004,Iozzi:2010,Fumanelli:2012} and, thanks to the increase 
in the availability and use of novel technologies, infrastructures based on various types of wearable sensors 
\cite{Hui:2005,ONeill:2006,Pentland:2008,Salathe:2010,Hashemian:2010,SocioPatterns,Cattuto:2010,Isella:2011,Hornbeck:2012,Fournet:2014,Sekara:2014,Obadia:2015}.

Methods based on surveys or diaries on the one hand, and sensing platforms based on wearable sensors on the other hand,
have each advantages and limitations. Well-studied questionnaires allow to gather informations not only on the existence of contacts but
also on additional characteristics, such as their context (home, work, travel), an estimate of their durations, the existence of repeated contacts
with the same individual, or even the distance from home at which the contacts take place~\cite{Danon:2013}. Questionnaires can also ask to
specify for each contact if it involved physical contact and distinguish periods of well-being and illness of the
respondent~\cite{VanKerckhove:2013}. Surveys however are costly and it is difficult to recruit
participants~\cite{Conlan:2011,Danon:2013}. Moreover, self-reporting procedures entail biases that are difficult to
estimate~\cite{Read:2008,Smieszek:2012,Smieszek:2014}, as participants might not recall all their contacts or might make incorrect estimates of their 
durations. For instance, according to \cite{Paulhus:2007}, factors such as ambiguity, emotions, and rapid responding together with the 
retrospective collection of survey data can induce an {\it extreme responding} bias in scale-rating diaries. 
Furthermore, time perception can be inaccurate in retrospective analysis, as in general people's 
recollections decay rapidly with time \cite{Bernard:1984}; in this respect, having a precise schedule of 
day activity can help to assign durations to meetings \cite{Collopy:1996}.

Wearable sensors on the other hand can be tuned to specifically detect close-range face-to-face 
proximity~\cite{Salathe:2010,SocioPatterns,Cattuto:2010}.
They afford an objective definition of contact, can detect even short encounters, and the decrease in the related costs
makes nowadays large-scale deployments feasible. They also give access to temporally resolved data sets, i.e., make possible a longitudinal study
of human contacts. The main limitation of automated sensing platforms based on sensors comes from the fact that they 
do not register contacts with individuals not participating
to the data collection (not wearing any sensor) and therefore provide data on the contacts among a closed population. 
Sampling issues can also arise if not all the members of the population of interest agree to wear the sensors~\cite{Cattuto:2010,Genois:2015}. 

Given these respective advantages and limitations of different methods, comparing data collected
by both types of methods in a given population is of great interest. To our knowledge, only one such study has been performed
to date, as it is rarely possible to collect data using both methods. Smieszek et al.~\cite{Smieszek:2014}
report on such a study in a high school context, showing for instance that many contacts 
registered by sensors are not reported in surveys, especially short ones, while long contacts are better reported.

Another issue of interest consists in the comparison between contact networks, 
corresponding to the actual behavior of individuals, and friendship relations or online social links. 
Many social studies are indeed performed through surveys of self-reported friendships, starting with \cite{Moreno:1953}, and the study of
online social networks has led to the whole field of computational social science \cite{Lazer:2009}. The question of 
how these various networks (contacts, reported friendship, online social links) overlap or complement each other
is quite largely open \cite{Sekara:2014,Barrat:2010}. For instance, social phenomena such as homophily \cite{McPherson:2001} are typically studied
through questionnaires, but recent studies have shown that behavioral contact networks 
can also provide interesting insights in this issue \cite{Fournet:2014,Stehle:2013}. 
Combining data from different sources to obtain a more complete picture of the contacts and interactions in a population
could therefore represent an interesting route to study various questions of relevance in social sciences or epidemiology. 

Here we present and analyze data sets corresponding to these various types of interactions and collected through concurrent 
data collection methods among more than $300$ high school students during one week in December 2013. 
Data regarding face-to-face contacts were measured
using both the SocioPatterns sensing platform \cite{SocioPatterns} based on wearable sensors and contact diaries
in which the students were asked to report the contacts they had. Surveys in which students were
asked to nominate their friends were also used, and students were finally asked to provide the network of their
contacts on Facebook. All data sets obtained using survey-like methods suffer from important sampling issues,
highlighting the interest of sensor-based methods in this respect.
We compare the contact patterns measured by sensors and reported by the students,
in a spirit similar to the work of~\cite{Smieszek:2012,Smieszek:2014}. We confirm the main findings of these previous
studies, for instance that longer contacts have major reporting probabilities,
and provide some additional insights: we find in particular that, despite the important discrepancies
between the data obtained by both methods, the mixing patterns between classes
are well identified even through contact diaries. We moreover compare the network of reported friendships
and the Facebook links with the contact patterns, finding that the longest contact durations correspond to reported friendships
but that reported friendship links can also correspond to short contacts. We finally perform a preliminary multiplex analysis
of the various possible links between individuals: contacts, reported friendships and Facebook links. We find that
reported friendships and Facebook links are not at all equivalent with respect to the durations of actual face-to-face contacts.

\section{Methods}

\subsection{Study design, data collection and description}

The data collection concerned high school students of specific classes called ``classes pr\'eparatoires'' in Lyc\'ee 
Thiers, Marseilles, France. 
These classes, specific to the French schooling system, gather students for studies that take place for two years after the end of the usual 
high school studies. The students study in a high school environment but are de facto mostly separated from the ``regular'' high school students:
their classes are located in a different part of the high school building and they typically take their lunches separately. 
At the end of these two years, students go through competitive exams yielding admission to various higher education colleges.

The classes have different specialization: ``MP'' classes focus more on mathematics and physics, ``PC'' classes on physics and
chemistry, ``PSI'' classes on engineering studies and ``BIO'' classes on biology.  We collected data among students of nine classes
corresponding to the second year of such studies: 3 classes of type ``MP'' (MP, MP*1, MP*2), 
two of type ``PC'' (PC and PC*), one of type ``PSI'' (PSI*) and 3 of type
``BIO'' (2BIO1, 2BIO2, 2BIO3).
All these students must prepare a small scientific project that they present at the final exam, and several students
could build a project based on their participation to the data collection, together with the use of the collected data in some small
scale analysis or numerical simulations. The active involvment of some students ensured a good participation of other students to the data collection, 
as also reported in other works \cite{Conlan:2011,Fournet:2014}. 

We collected data of different nature.
\begin{itemize}

\item we deployed the contact measurement platform developed by the SocioPatterns collaboration \cite{SocioPatterns},
which is based on sensors that are embedded in unobtrusive wearable badges and exchange 
ultra-low power radio packets in order to detect close proximity of individuals wearing them. 
As described in detail in \cite{SocioPatterns,Cattuto:2010,Isella:2011}, the power is tuned so that
the sensors can exchange packets only when within $1 -– 1.5$ meters of one another. Moreover, 
students were asked to wear the sensors on their chests using lanyards, ensuring that the devices of two 
individuals can only exchange radio packets when the persons are facing each other. The sensors are also tuned
so that the face-to-face proximity of two individuals wearing them can be assessed over an interval of $20$ 
seconds with a probability in excess of $99\%$. Contact data are thus collected with a temporal resolution of $20$ seconds:
two individuals are considered to be in contact during a $20s$ time window if their sensors exchanged at least one 
packet during that interval, and the contact event is considered over when the sensors do not exchange packets over a $20s$ interval.
The information on face-to-face proximity events detected by the wearable sensors is relayed to radio receivers installed throughout the high school: 
contacts occurring outside the school premises were not measured. 
Thanks to this infrastructure, we gathered contact data of the $327$ participating students (out of $379$ in the $9$ classes, i.e., 
a $86.3\%$ participation rate) during the week of Dec. 2-6, 2013.

\item At the end of one specific day, namely Dec. 5, 2013, we asked students to fill in paper contact diaries: they were
asked to give the list of other students they had had contact with (where contact was defined as close face-to-face proximity) during the day
in the high school, and
to give the approximate aggregated duration of the contacts with each nominated individual, to choose in one of four possible
categories: at most 5 minutes, between 5 and 15 minutes, between 15 minutes and 1 hour, more than one hour. $120$ students 
returned a filled in diary.

\item During the period of the deployment, we moreover asked the students to fill in a survey in which they were asked to give 
the names of their friends in the high school. We obtained such friendship surveys from $135$ students.

\item We asked students to use the Netvizz application (\verb+https://apps.facebook.com/netvizz/+) to create their local 
network of Facebook friendships (i.e., the use of the application by a student yields
the network of Facebook friendship relations between this student's Facebook friends). $17$ students gave us access
to their local network, from which we removed all users who were not concerned by the data collection.

\item Each participating student gave us moreover access to the following metadata: gender, class, class in the previous year, 
whether s/he was smoker, and if s/he was repeating the year.

\end{itemize}

We finally note that the present data collection was the third of this kind in this high school. The data coming from the previous two data collections
(performed in Dec. 2011 and Nov. 2012) were analyzed in \cite{Fournet:2014}. They concerned only the contacts recorded
by the SocioPatterns infrastructure (for three classes in 2011 and five classes in 2012), but no
contact diaries nor friendship surveys. 

\subsection{Ethics and privacy}
Before the study, students and teachers were informed on the details and aims of the study.  A signed informed consent was obtained for
each participant (no minors were involved as all students were at least $18$ at the time of the deployment). They received a wearable
sensor to wear during the school time. No personal information, besides the ones indicated in the previous section, were
collected. The {\it Commission Nationale de l'Informatique et des Libert\'es} (CNIL, http://www.cnil.fr), representing the French
national bodies responsible for ethics and privacy, and the high school authorities approved the study.

\subsection{Data analysis}

The collected data sets have different nature and different resolutions. The most detailed data comes from the SocioPatterns
infrastructure and consists in a temporal network of contacts between students, with a temporal resolution of $20$ seconds: the nodes of
this network represent students, and for each time window of $20$ seconds a link is drawn between pairs of students between whom
contacts are detected. This temporal network can be aggregated temporally over a given duration, for instance a day or the whole
duration of the data collection: weighted daily or global aggregated networks are then obtained. In each
aggregated network, a node represents
an individual and a weighted link between two nodes $i$ and $j$ represents the fact that the two corresponding individuals have been
in contact at least once during the aggregation time window. The weight $w_{ij}$ of a link between $i$ and $j$ is given by the total
time spent in contact by $i$ and $j$ during the aggregation window. The data can be further aggregated by grouping together
students of the same class: the resulting mixing patterns between classes are described by the so-called ``contact matrices''. We define contact
matrices of different types: 
\begin{itemize}
\item in the contact matrix of link densities, the element $X,Y$ is given by the density of links between classes $X$ and $Y$, i.e., 
the number of links $E_{XY}$ between individuals of class $X$ and individuals of class $Y$ normalized by the maximum possible
number of such links ($n_X n_Y$ if $X \ne Y$, $n_X(n_X-1)/2$ if $X=Y$, where $n_X$ is the number of students in class $X$).

\item  in the contact matrix of contact durations, the element $X,Y$ is defined as $W_{XY}= \sum_{i \in X, j \in Y} w_{ij}$ (and 
$W_{XX}= \sum_{i,j \in X} w_{ij}/2$): it gives the total time spent in contact between students of class $X$ and students of class $Y$.
These elements can also be normalized, e.g., $W_{XY}/n_X$ gives the average time spent by a student of class $X$ with students of class $Y$.
\end{itemize}
As for the aggregated networks, these matrices can be defined on any temporal aggregation time-window.

The contact diaries, on the other hand, do not provide temporally resolved data. As each participating student reports contacts 
with other students, giving a certain aggregated duration for each, the resulting data set is a weighted directed network:
in this network, each node is a student and a link is drawn from $i$ to $j$ with weight $w_{ij}^{diary}$ if student $i$ 
reported contacts of total duration $w_{ij}^{diary}$ with student $j$. Note that links are not necessarily reciprocated, i.e., 
student $j$ might not report a contact with $i$ even if $i$ reported such a contact and, even when they are reciprocated,
the reported durations might not coincide, i.e., $w_{ij}^{diary}$ is not necessarily equal to $w_{ji}^{diary}$. We will first perform
a systematic study of these discrepancies.
Then, in order to compare the contact diary data with the sensor data as in \cite{Smieszek:2014}, we will
consider a symmetrized version of the network, in which a link exists between $i$ and $j$ if at least one of the two students 
reported a contact, and the weight of the link is taken as the maximum of $w_{ij}^{diary}$ and  $w_{ji}^{diary}$. From such symmetrized network,
we can also aggregate the data by class and obtain contact matrices.

The friendship survey yields, as the contact diaries, a directed network between students: indeed, a student $i$ might nominate $j$ as a friend
without being nominated by $j$. As we did not ask students to quantify the intensity of their friendships, we obtain a directed unweighted
network of reported friendship relations. We can also symmetrize the network in order to compare it with the aggregated contact network, and
obtain a link density contact matrix between classes.

\begin{figure}[!ht]
\includegraphics[width=0.5\textwidth, trim =30 150 60 50,clip=true]{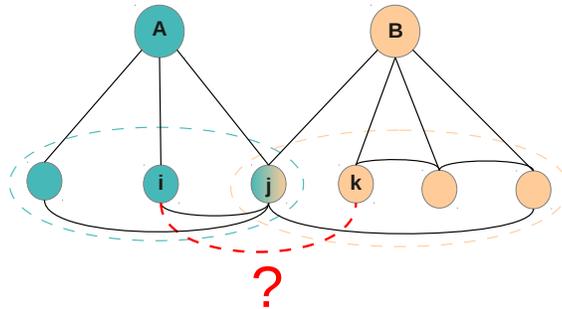}
\caption{{\bf Facebook local networks.} the local Facebook friendship networks provided by students $A$ and $B$ are shown in 
black. In particular, we know that $i$ and $j$ are friends on Facebook but not $j$ and $k$, as $i$ and $j$ are both friends of  
$A$ and $j$ and $k$ are both friends of $B$. On the other hand, we do not know if $i$ and $k$ are friends or not:
the red dashed line represents the lack of knowledge about the potential existence of this relationship.}
\label{FBprob}
\end{figure}

Finally, the data set gathered from the local Facebook friendship networks has a slightly more complex character. Indeed, for each individual 
who gave us access to his/her local network, we obtain the Facebook friendship links between his/her Facebook friends. However,
as only few students provided us with such data, the presence or absence of a Facebook friendship link between many pairs of students
remained unknown. To understand this point, let us take the example of two students $A$ and $B$, with sets of Facebook friends
$F_A$ and $F_B$. Let us now take two students $i$ and $j$ in the union $F_A \cup F_B$. If both $i$ and $j$ are in $F_A$, or both are in $F_B$, 
we know if they are Facebook friends or not. If however $i \in F_A\backslash F_B$ and $j \in F_B\backslash F_A$, i.e., 
$i$ is friend with $A$ but not with $B$ and $j$ is friend with $B$ but not with $A$, we do not have access to the existence or absence of a friendship link
between $i$ and $j$ (see Fig. \ref{FBprob}). As a result, the data set can not be represented as
a network but consists in a list of pairs of students for which
we know if they are  Facebook friends or not (list of ``known-pairs'') and a list of pairs for which the presence or absence of such a link is unknown.

\section{Results}
\subsection{Contact patterns from sensor data}

During the data collection, $67,613$ contact events were registered between the $327$ students carrying wearable sensors, 
with a cumulative duration of $3,770,160$s ($\sim$ 1047 hours). As the statistical characteristics of these contacts are
very similar to the ones obtained in the previous data collections in the same setting \cite{Fournet:2014}, we only provide
here a summary of these results and refer the interested reader to the Supporting Information (SI) for details.

As commonly observed in such data \cite{Barrat:2013,Barrat:2014}, the duration of contacts was highly variable:
the average duration of a contact was $56$ seconds, with however $75\%$ of 
the contacts having duration smaller than $1$ minute and on the other hand $2\%$ lasting more than $5$ minutes.
The distribution of these durations, shown in the SI, is very broad, spanning
several orders of magnitude, with a coefficient of variation (CV) equal to $2.7$. 
Moreover, the time intervals between successive
contacts are also broadly distributed, spanning several orders of magnitude: most intercontact durations are short,
but very long durations are also observed, and no characteristic timescale 
emerges~\cite{Hui:2005,Karagiannis:2007,Scherrer:2008,Cattuto:2010,Salathe:2010,Barrat:2014}.
This bursty behavior is a well known feature of human dynamics 
and has been observed in a variety of systems driven by human actions~\cite{Barabasi:2010}.

The contact network aggregated over the whole data collection has $327$ nodes and $5818$ weighted edges. As expected in 
such networks, the average shortest path length (SPL) is small ($2.16$) and the clustering coefficient is large
($\approx 0.5$, against  $\approx 0.11$ in a random network with the same number of nodes and edges). 
As also found in other similar data sets \cite{Isella:2011,Fournet:2014}, 
the distribution of degrees (the degree of a node is the number of other nodes it is connected to) is 
narrow ($CV \approx 0.38$). The average degree, i.e., the average number of students each student had contact with, is $\approx 35$.
On the other hand, the distribution of links weights, i.e., of the total time spent in contact by pairs of students, is broad:
the average cumulated duration is of $648$ seconds ($\sim 11$ minutes), but most weights are much
smaller ($59\%$ of the links have a weight smaller than $2$ minutes) and large values are also observed ($12\%$ of the weights
are larger than $15$ minutes and $4\%$ are larger than $1$ hour). Overall, the distribution spans several orders of magnitude (CV = $3.87$)
and no characteristic interaction timescale can be naturally defined.  

The population of interest is structured into $9$ classes. We compute the contact matrices defined above to describe the mixing patterns
between these classes (see SI). As also observed in \cite{Fournet:2014}, most contacts occur within classes, and students
of different classes have very few contacts: 
$62,342$ contacts ($92.2\%$ of the total), representing a cumulated duration of $3,505,380s$ ($93\%$ of the total contact time 
of students), were recorded between pairs of students belonging to the same class. An additional substructure of three groups
of three classes each emerges moreover: (i) classes MP, MP*1, MP*2, (ii) classes PC, PC*, PSI*, (iii) classes 2BIO1, 2BIO2, 2BIO3.
More contacts are observed between students of two classes in the same group than between students of two different groups. 
This substructure corresponds to a grouping of classes according to the studied field, but could also be reinforced
by the fact that the classrooms of each group are physically close in the high school.

We finally mention that the number of contacts fluctuates strongly 
throughout the day, along a robust daily pattern driven by the occurrence of class breaks and lunches. As in \cite{Fournet:2014},
the daily contact matrices are very similar to each other (with cosine similarities between matrices ranging from $92.7\%$ to $98\%$).
Overall, although the contacts occurring in different days are not all the same, the temporal daily fluctuations and the
mixing patterns between classes are robust across different days.

\subsection{Contact diaries}

As mentioned above, the contact diaries yield a directed weighted network of reported contacts between students. 
Only $120$ students filled in a contact diary, yielding a network of reported contacts of $120$ nodes
and $502$ directed, weighted edges among them. Note that we ignore the contacts reported with students who did not
fill a diary. Moreover, in the following we consider only students for whom the sensors also registered contact
data during the day concerned by the contact diary.
This corresponds to $109$ students and $416$ directed weighted links.

$158$ contacts were reported by only one student (non-reciprocated links), while the reciprocated links correspond to $129$ pairs of students
reporting both a contact with each other. Moreover, reciprocated links were sometimes reported with different 
durations by the two students involved.
Table \ref{table:diary} gives the corresponding statistics. In $81$ cases out of $129$, both students involved reported
the same duration category. In $35$ cases, the reports of the two involved students differed by only one category. These results 
are similar to the ones of \cite{Smieszek:2012}. Following \cite{Smieszek:2012}, we moreover compute the probability $P$ to
report a contact of a certain duration, under the hypothesis that such probability depends only on the duration. If $N_c$ is the real number
of contacts, and $N_{both}$ is the number of pairs of students reporting both the contact, then $N_{both} = N_c P^2$, while the number of contacts reported by only
one student is $N_{one} = 2 N_c P (1-P)$; as a result, the estimate of $P$ is given by $N_{both}/(N_{both} + N_{one}/2)$.
We obtain that the overall reporting probability is $P \approx 62\%$. Assuming that the correct duration of a reported
contact is the highest reported value, we obtain that the probability to report a contact 
is $40\%$ for contacts of less than $5$ min, $54\%$ for contacts between $6$ and $15$ min, $61\%$
for contacts between $16$ and $60$ min, and $72\%$ for contacts with aggregate duration longer than one hour.
These numbers are in-between the values reported in \cite{Smieszek:2012} and \cite{Smieszek:2014}, respectively.
They share with these previous studies the fact that the reporting probability increases strongly with the contact duration.

\begin{table}[!ht]
\medskip
\begin{tabular}{l|cccc|c}
\toprule
\multicolumn{6}{c}{\bf Reported duration: higher value} \\ [1ex] \cmidrule{1-6}
    \multirow{2}{*}{\bf Reported duration:}&{\bf 1-5 min } & {\bf 6-15 min} &  {\bf 16-60 min} & {\bf more than 60 min} & {\bf Row Tot}\\ [1ex]
 {\bf lower value}& & & && \\
\midrule
\hline
 \multirow{2}{*}{\bf Not reported}    &  {\bf 38} $(24\%)$    & {\bf 31} $(20\%)$ & {\bf 33} $(21 \%)$ & {\bf 56} $(35\%)$ & {\bf 158} ($100\%)$   \\
                                      &  $ (75\%)$         & $ (63\%) $           & $ (56\%) $         &  $ (44 \%)$       & $ (55\%)$\\ \cmidrule{1-6}
\multirow{2}{*}{\bf 1-5 min}          &   {\bf 13} $(42\%)$&  {\bf 10} $( 32\%)$  & {\bf 5} $(16\%)$   & {\bf 3} $(10\%)$  & {\bf 31} $(100\%)$  \\
                                      &     $(25\%)$       & $(21\%)$             & $(9\%)$            & $(2\%)$           & $(11\%)$\\ \cmidrule{1-6}
\multirow{2}{*}{\bf 6-15 min}         &                    &  {\bf 8} $(32\%)$    & {\bf 12} $(48\%)$  & {\bf 5} $(20\%)$  & {\bf 25} $(100\%)$   \\
                                      &                    & $(16\%)$             & $(20\%)$           & $(4\%)$           & $(9\%)$\\ \cmidrule{1-6}
\multirow{2}{*}{\bf 16-60 min}        &                    &                      &{\bf 9} $(41\%)$    & {\bf 13} $(59\%)$ & {\bf 22} $(100\%)$   \\
                                      &                    &                      & $(15\%)$           & $(10\%)$          & $(8\%)$\\ \cmidrule{1-6}
\multirow{2}{*}{\bf more than 60 min} &                    &                      &                    & {\bf 51} $(100\%)$& {\bf 51} $(100\%)$ \\
                                      &                    &                      &                    & $ (40\%)$         & $(17\%)$\\ \cmidrule{1-6}
\hline
\multirow{2}{*}{\bf Column Tot}       & {\bf 51} $(18\%)$  &  {\bf 49} $(17\%)$   & {\bf 59} $(20.5\%)$&{\bf 128} $(44.5\%)$& {\bf 287} $(100\%)$  \\
                                      &     $(100\%)$      & $(100\%)$            & $(100\%)$          & $(100\%)$          & $(100\%)$\\
\bottomrule
\end{tabular}
\caption{{\bf Cross-tabulation of pairs of contact reports from the contact diaries}. 
Each pair of participants with at least one contact reported gives a single observation. For instance, there were $12$ pairs of students $(i,j)$
such that $i$ reported contacts with $j$ with total duration between $6$ and $15$ min while $j$ reported a duration between $16$ min and $1$ h.
Each percentage within a cell represents the percentage with respect to the row (right of the cell entry) and column (below the cell entry) totals.
}\label{table:diary}
\end{table}

\subsection{Comparing contact diaries and sensor data}

In this section, we compare the data collected by the wearable sensors with the contacts reported by the students using contact diaries.
We therefore consider on the one hand the weighted network of the contacts registered by the sensors on Dec. $5^{th}$, and on the other hand
the symmetrized version of the network obtained from the contact diaries, in which the highest value of the aggregated contact duration
reported by two students is retained. 

Table \ref{table:contact_vs_diaries} reports some properties of these networks. As not all students filled in contact diaries, we moreover
report the networks' properties when restricted to the nodes present in both ($109$ nodes). The density and average
degree of the contact network obtained by the sensors are almost twice as large as the ones obtained using contact diaries, but the degree
distributions have similar shapes (see SI).
The cliques are also larger in the contact network, 
while the average shortest path length is smaller (the distributions of shortest path lengths is shown in the SI): nodes seem farther apart
in the contact diary network than in the sensor data network. The average clustering on the other hand is similar.

\begin{figure}[!ht]
\includegraphics[width=\textwidth]{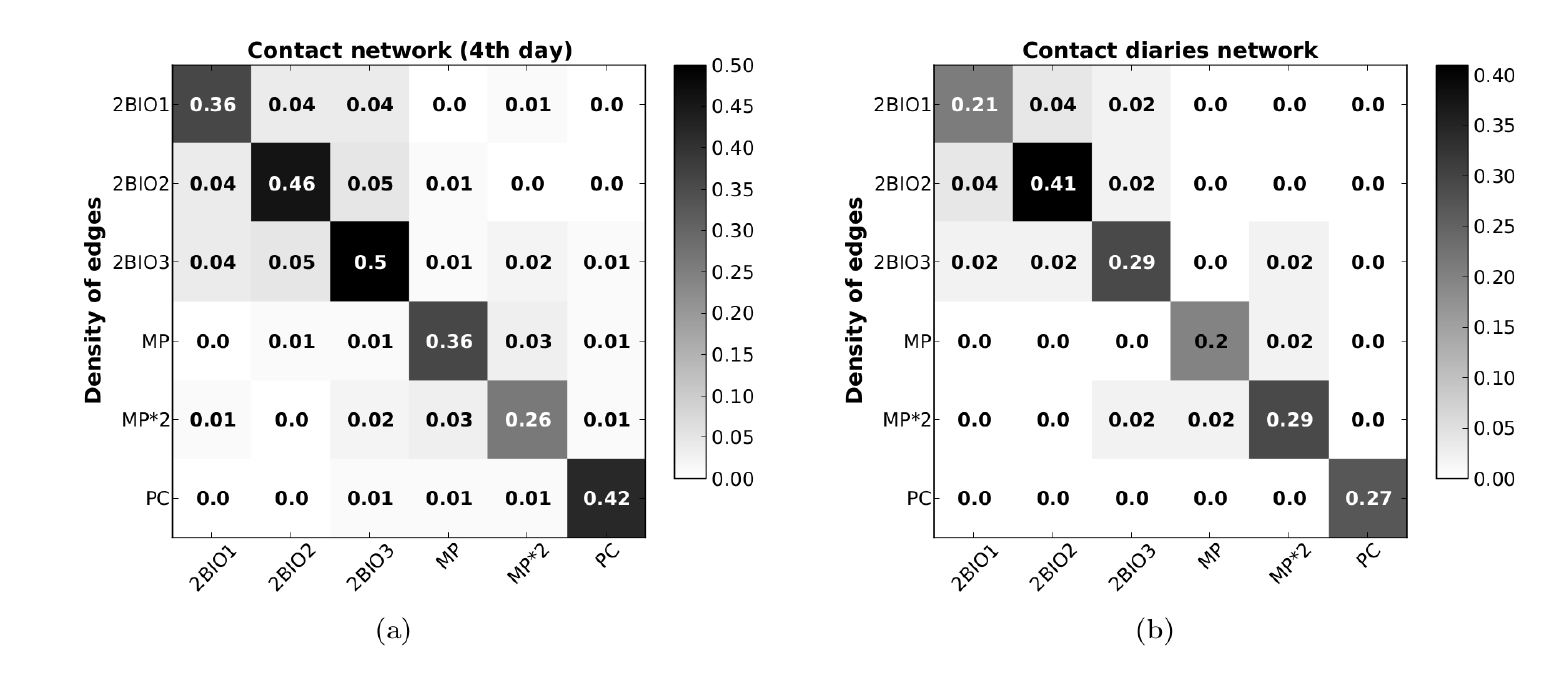}
\caption{{\bf Contact matrices of link densities.} We compare here the contact matrices of link
densities built from (a) the network of contacts obtained using the sensor data collected on Dec. $5^{th}$ 
and (b)  the network of contacts as reported in the contact diaries. We discarded here the 
data corresponding to the MP*1, PC* and PSI* classes as too few students from these classes filled in a contact diary ($1$ for
 MP*1, $0$ for PC* and PSI*). The similarity between these two matrices is of $97\%$.
\label{mat_sur}}
\end{figure}

Interestingly, the strongly structured character of the contact network, as highlighted by the dominance of the diagonal elements of the
contact matrices and the existence of groups of classes, is well preserved in the contact diary network, as shown in Fig. \ref{mat_sur}. A high
similarity is obtained between the link density contact
matrices computed on both networks: despite the low sampling of the contact diaries, a 
sensible information on the mixing patterns of different classes is obtained.

\begin{savenotes}
\begin{table}[!ht]
\medskip
\begin{tabular}{c|cc|c|cc}
\hline
\toprule
    &{\bf Sensors} & {\bf Contact diaries} & & {\bf Sensors} & {\bf Contact diaries}\\
\hline
\midrule
 {\bf Nodes} &      $295$  &  $120$   & & $109$ & $109$ \\
{\bf Links} &       $2162$ &  $348$   & & $488$ & $287$\\
{\bf Density} &    $0.05 $ &  $0.05$  & & $0.08$ & $0.05$\\
{\bf Avg. Degree} &$15 (8)$ & $6 (2)$ & & $9 (4)$ & $5 (2)$  \\
{\bf Avg. Clustering} & $0.38 (0.18)$ & $0.45 (0.25)$ && $0.45 (0.21)$ & $0.44 (0.26)$\\
{\bf Avg. SPL} &  $2.81 (0.8)$ &  $5.36 (2.73)$ && $2.94^* (1.03)$ & $5.36 (2.66)$ \\
{\bf Maximal Clique} & $9$ & $5$ && $8$ & $5$\\
\bottomrule
\end{tabular}
\caption{{\bf Comparison of properties for contact and  networks.} All network properties for the memory-survey network are computed on 
its symmetrized version. In this summary table we assume that if a contact is reported by at least one of the two nodes, it exists. 
The right side of the table is performed after matching the two networks. Matching is done by removing the nodes who did not 
participate to the survey and the ones who did not have contacts recorded by sensors on the 4th day of the study. A $^*$ close 
to the SPL average means that after the match some isolated nodes appeared. In this case, we computed the average on the 
connected pairs only. Standard deviations are given in parentheses.}\label{table:contact_vs_diaries}
\end{table}
\end{savenotes}

It is clear from the different numbers of links in the two networks that there are discrepancies between 
sensor-based data and contact diaries, as also found in \cite{Smieszek:2014}. Overall,
$70.4 \%$ of the links obtained from contact diaries correspond to contacts 
registered by the sensors, while only $41.4 \%$ of the contacts registered by the sensors find a match in the contact diary.
We now investigate these discrepancies in more details.

\begin{figure}[!ht]
\includegraphics[width=\textwidth]{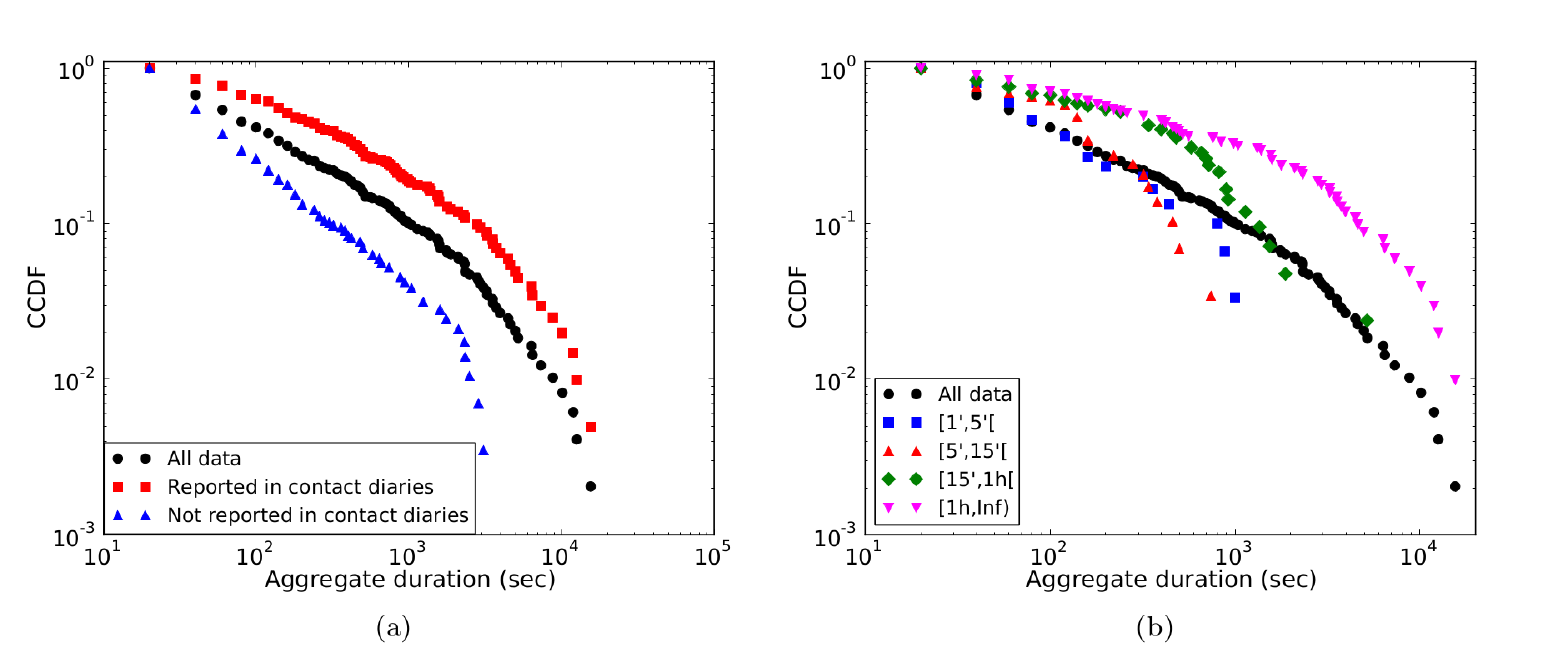}
\caption{{\bf Sensors vs. contact diaries: distributions of cumulative durations registered by the sensors.}  
(a) Cumulative distributions of the aggregate durations of contacts registered by the sensors for 
(i) all $488$ links between the $109$ nodes belonging to both networks; (ii) the $202$ links
that were also reported in the diaries; (iii) the $286$ links that were not reported in the diaries.
(b) Cumulative distribution of aggregate durations of contacts registered by the sensors 
for the different categories of links reported in the diaries.
\label{durations_diaries}
}
\end{figure}

Figure \ref{durations_diaries}(a) and Table \ref{table:durations_diaries} first compare the distributions of the cumulative durations of the links
registered by the sensors, distinguishing between the links which were reported in the contact diaries and those which were not. For reference, the figure
also reports the distribution of durations for all the links registered by the sensors.
Both distributions are broad, spanning several orders of magnitude. However, the distribution of durations for the links finding a match
in the contact diaries is much broader, with much larger average duration and standard deviation~\footnote{Wilcoxon tests 
for each pair of distributions reject the null hypothesis of equality of the distributions}. 
In particular, links not reported tend to correspond to smaller durations, and all the links with a duration 
above a certain threshold (close to $1$ hour) were reported in the diaries. This result indicates that individuals 
tend to remember preferentially the longest interactions they had during the day, and that the information concerning very long contacts,
as measured by sensors, is mostly recovered when using diaries.

\begin{table}[!ht]
\medskip
\begin{tabular}{c|c|m{.3\textwidth}|m{.3\textwidth}}
\toprule
    &{\bf All data} & {\bf Links present in the contact diaries} &  {\bf Links absent from the contact diaries} \\
\hline
\midrule
 {\bf Mean (s)} &   $480$  &  $934$   & $156$   \\
{\bf Std. dev.}    &   $1470$ &  $2152$  & $413$ \\
{\bf $\#$links}    &   $488$ & $202$    & $286$ \\
\bottomrule
\end{tabular}
\caption{Average and standard deviation of the distributions of aggregate durations for different sets of links (as in Fig. \ref{durations_diaries}).
}
\label{table:durations_diaries}
\end{table}

We moreover investigate in Figure \ref{durations_diaries}(b) the diversity of the cumulative durations registered by the sensors
for the links reported in the diaries in each category. Strikingly, all distributions are rather broad and, given a reported category, both much shorter
and much longer durations can be registered by the sensors. In particular, the distributions corresponding to the two first categories 
(less than $5$ min and between $6$ and $15$ min) are similar.
However, the distributions become consistently broader
for categories corresponding to larger durations, and durations (as registered by the sensors) above a certain threshold
are reported only in the highest duration category of the diaries.

Table \ref{table:sensors_vs_diaries} gives more details through a cross-tabulation of the aggregate durations of the contacts 
as registered by the sensors or reported in the diaries. If we consider the duration registered by the sensors
as accurate, we obtain that $32\%$ of the contacts in the first category ($1-5$ min) were reported, against $71\%$ and $69\%$
for the next categories, and $100\%$ for contacts lasting more than $1$ hour.

\begin{table}[!ht]
\medskip
\begin{tabular}{l|ccccc|c}
\toprule
\multicolumn{6}{c}{\bf SENSORS} \\ [1ex] \cmidrule{1-7}
    \multirow{2}{*}{\bf SURVEY}&{\bf Not detected}&{\bf 1-5 min } & {\bf 6-15 min} &  {\bf 16-60 min} & {\bf $\mathbf{>}$ 60 min} & {\bf Row Tot}\\ [1ex]
 & & & && \\
\hline
\midrule
 \multirow{2}{*}{\bf Not reported} & {\bf unknown} (n/a) &  {\bf 258} $(90\%)$ & {\bf 16} $(5.5\%) $  & {\bf 12} $(4.5 \%)$ & {\bf 0} $(0\%)$ & {\bf 286 } ($100\%)$   \\
 &                                                      (n/a)&$ (68\%)$    & $ (29\%) $    & $ (31\%) $            &  $ (0 \%)$          & $ (53\%)$\\ \cmidrule{1-7}
\multirow{2}{*}{\bf 1-5 min}  & {\bf 36} $(44\%)$&  {\bf 38} $(46.5\%)$ &  {\bf 7} $( 8.5\%)$  & {\bf 1} $(1\%)$    & {\bf 0} $(0\%)$  & {\bf 82} $(100\%)$  \\
&                                                   $(42\%)$    &  $(10\%)$    & $(12.5\%)$                & $(3\%)$   & $(0\%)$               & $(11\%)$\\ \cmidrule{1-7}
\multirow{2}{*}{\bf 6-15 min} & {\bf 5} $(28\%)$  &  {\bf 9} $(50\%)$      &{\bf 4} $(22\%)$  & {\bf 0} $(0\%)  $ & {\bf 0}$(0\%)$ & {\bf 18 } $(100\%)$  \\
&                                                     $(6\%)$              & $(2\%)$              & $(7\%)$       & $(0\%)$  & $(0\%)$     & $(5\%)$\\ \cmidrule{1-7}
\multirow{2}{*}{\bf 16-60 min}& {\bf 17} $(29\%)$&  {\bf 24} $(40.5\%)$     & {\bf 12} $(20\%)$ & {\bf 5} $(8.5\%)$ & {\bf 1} $(2\%)$ & {\bf 59} $(100\%)$   \\
&                                                    $(20\%)$       &   $(6\%)$     & $(21.5\%)$             & $(13\%)$   & $(7\%)$         & $(10\%)$ \\ \cmidrule{1-7}
\multirow{2}{*}{\bf $\mathbf{>}$ 60 min}  &{\bf 27} $(21\%)$  &{\bf 51} $(40\%)$ &{\bf 17} $(13\%)$ & {\bf 20} $(16\%)$ & {\bf 13} $(10\%)$ & {\bf 128} $(100\%)$ \\
&                                                      $(32\%)$ &   $(13.5\%)$  & $(30\%)$  &   $(53\%)$  & $ (93\%)$ & $(22\%)$\\ \cmidrule{1-7}
\hline
\multirow{2}{*}{\bf Column Tot}  & {\bf 85} $(15\%)$ & {\bf 380} $(66\%)$ &  {\bf 56} $(10\%)$ &  {\bf 38} $(7\%)$ & {\bf 14} $(2\%)$ & {\bf 573} $(100\%)$  \\
&                                                 $(100\%)$  &    $(100\%)$          & $(100\%)$              & $(100\%)$              & $(100\%)$                & $(100\%)$\\
\bottomrule
\end{tabular}
\caption{{\bf Sensors vs. contact diaries: cross-tabulation of the number of links in each duration category.} 
The percentages within a cell are computed with respect to the row (right of the cell entry) and column (below the cell entry) totals.}
\label{table:sensors_vs_diaries}
\end{table}

Among the $202$ links found in both networks, discrepancies emerge moreover between the durations reported by the students
and registered by the wearable sensors. In particular, $60$ ($29.7\%$) links correspond to the same duration category in both cases, while
$133$ ($65.8\%$) are overestimated in the diaries with respect to the sensor data, and only 
$9$ ($4.5\%$) are assigned a shorter duration in the diaries than in the sensor data \footnote{Overall, the 
Kendall's $\tau$ computed for the list of links ranked according to the durations either registered or reported
yields a rank-correlation of $\approx 45\%$.}.
This outcome is in agreement with results from social studies about self-reported diaries biases stating that
individuals tend to perceive the time spent in some activities (talk, work, play) differently from the reality \cite{Collopy:1996} 
and frequently to overestimate it \cite{Hyett:1979}.

\vskip .5cm

We finally investigate the discrepancies between sensor data and diaries at the individual level. 
In contrast with the data reported in \cite{Smieszek:2014}, we observe a significant correlation ($0.4$) between 
the degree of the nodes in the two networks,
despite important fluctuations are present (see SI). More in detail, and 
when considering in addition the directed character of the network built from the diaries,
we do not find any significant correlation between the out-degree of a 
node $i$ in the diaries' network (number of contacts reported by the corresponding student)
and its degree in the sensor data network ($k_i^{sensors}$). However, we find a significant
positive correlation (equal to $0.4$) between $k_i^{sensors}$ and the in-degree of 
$i$ in the diaries' network, which represents the number of other students who
have reported a contact with him/her. Similar correlation values are obtained when considering
only contacts larger than a given threshold (see SI).

Interestingly, this indicates that, in terms of actual contacts, the overall picture of each individual
{\it as reported by all the others} is more accurate than the records of the individual him/herself.
We refine this observation by investigating in more detail the relationship between out-degree in the contact diary network and the
degree in the sensor data network. We start from the idea, supported by the evidence provided above, that among contacts of different
durations individuals tend to record more easily their longest contacts
in a contact diary. We then compute for each student the 
coefficient of variation ($CV$) of the longest durations of his/her encounters with other students,
which gives the extent of the variability of these durations: $CV_i \le 1$ means that the 
student $i$ has contacts of similar durations with other students, while $CV_i > 1$
corresponds to a large variability, i.e., that $i$ divides his/her contact time in a 
heterogeneous way among the other students s/he has met during the day.
We obtain $42$ students with $CV \le 1$ and $67$ with $CV > 1$, and a significant correlation (close to $0.35$) between
the out-degree in the contact diary network and $k_i^{sensors}$ in the first group, but no significant correlation
in the second group \footnote{The correlation for the in-degree is present and around $0.4$ in both groups.}. 
In other words, for students who have encounters of similar maximum durations with other individuals ($CV \le 1$),
the contact diary data reported by these students correlate with the data registered by the sensors. For students
whose maximum contact durations are heterogeneous ($CV > 1$) on the other hand, no correlation 
in the diary-reported and sensor-measured
degrees is observed. This is perfectly in line with our initial hypothesis: 
For $CV \le 1$ the number of links remembered is then correlated with the real number of links in the contact network while, if $CV > 1$,
there might be an arbitrary large number of links with only
short contacts that are not remembered in the diaries: in the latter case, the number of links that are reported
is not correlated with the total number of links registered by the sensors.

\subsection{Multiplex network of students' relationships}

The face-to-face proximity measured by different methods as described in the previous sections represent only one type of relationship
between individuals. Other social ties exist, such as friendships and online relationships, and contribute to form a multiplex network in
which the nodes represent students and links of different nature coexist between them. These links might a priori be related (one
has more contacts with a friend, or becomes friend with someone after meeting him/her often, etc...)  but might also differ substantially:
one can be very good friend with someone and meet him/her only rarely because of specific constraints (such as different schedules
of classes).
It is thus of interest to compare the different layers of this multiplex network and to investigate what information on actual
contacts can be gathered from data describing friendship relations. In particular, friendship survey data might be more reliable than contact
diaries: it might be easier to remember the names of one's friends than the contacts occurred during a day and moreover, friendships
evolve on slower timescales so that friendship surveys can more easily be gathered on several days without memory biases.

\begin{figure}[ht]
\includegraphics[width=\textwidth]{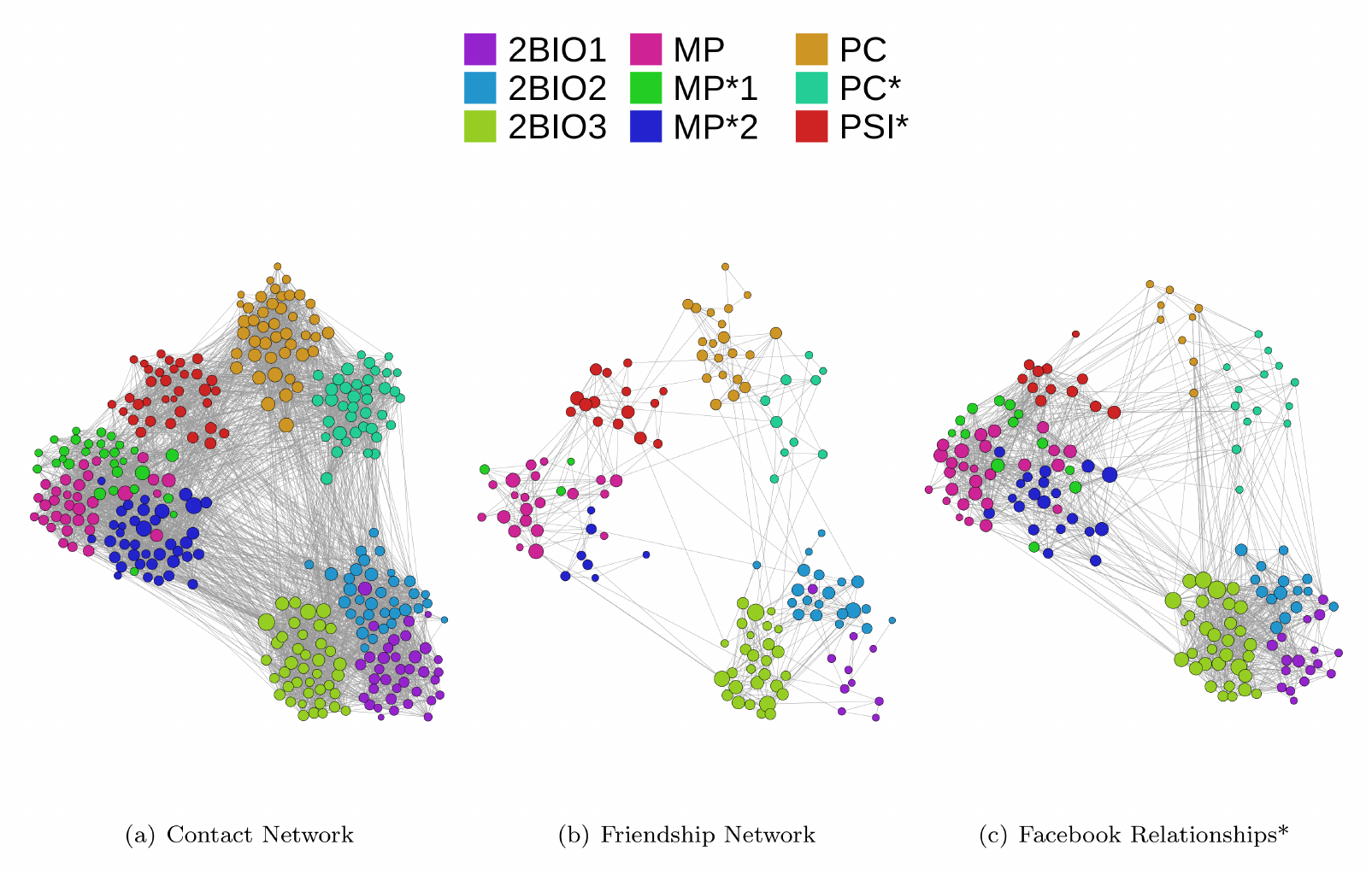}
\caption{{\bf Contact and friendship networks.} The color of each node represents its class and size represents its 
degree in the corresponding network (here we consider a symmetrized version of the network of reported friendships)
.$^*$ Strictly speaking, the Facebook data do not provide a network as we do not have information about the presence or absence
of a link between many pairs of nodes (see Figure \ref{FBprob}). Figure created using the Gephi software
http://www.gephi.org.}
\label{snap}
\end{figure}

As described in the Methods section, we have collected data concerning the friendships and online Facebook links of
a number of students, and we compare in this section the properties of the resulting data sets. 
Figure \ref{snap} displays the network of contacts registered by the sensors during the week of 
data collection, as well as the network of reported friendships
and the Facebook links \footnote{As explained in the Methods section, in the case of 
Facebook other links than the ones represented might exist:
we are only representing the "known-pairs", so that Fig. \ref{snap} might be an underestimation of the real number 
of existing Facebook links. For this reason, standard network metrics cannot be computed in this case.}, 
using the same position for the nodes in the three representations. The friendship survey and Facebook data involve respectively
$41\%$ and $48\%$ of the students who participated to the data collection, so that the resulting networks have also clearly much less links
than the contact network. Overall, the networks appear substantially different, although the grouping of nodes in classes seems relevant in all three cases.
We perform a more detailed comparison in the next paragraphs.

\paragraph{Contact network versus friendship-survey network} 

\begin{figure}[!ht]
\includegraphics[width=\textwidth]{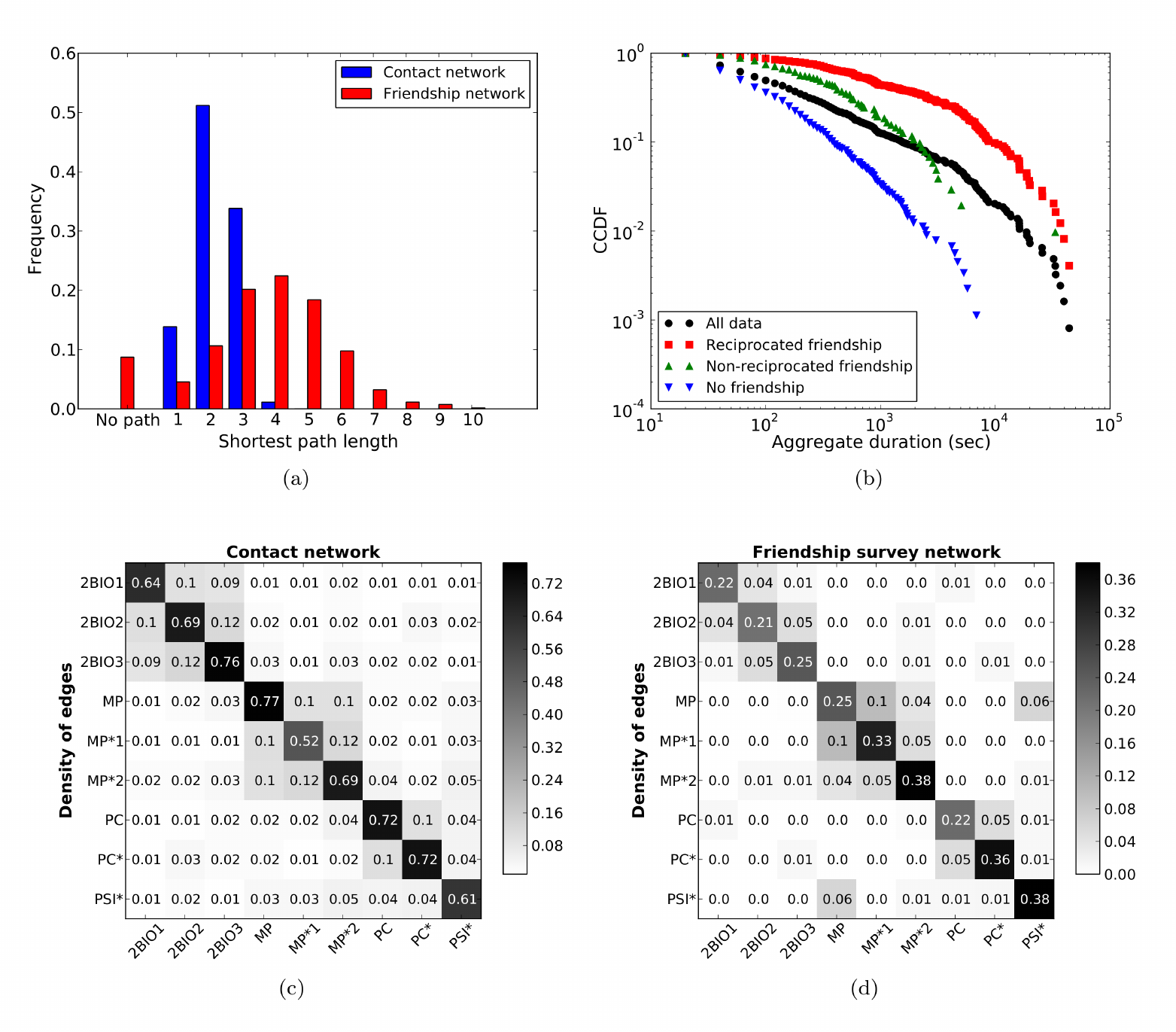}
\caption{ 
{\bf Comparison of the networks of contacts and friendships.}
(a) Shortest path length distributions for both networks;
(b) Distributions of aggregate durations, as measured by the sensors, for different kinds of links in the contact network: (i) all links,
(ii) links $i-j$ for which only one of $i$ or $j$ reported a friendship with the other, (iii) links for which both students reported
the friendship, and (iv) links for which no friendship was reported;
(c) and (d): Contact matrices of link densities. We compare here the contact matrices of link
densities built from (c) the global aggregated network of contacts obtained using the sensor data 
and (d) the symmetrized network of reported friendships. The similarity between these two matrices is $\approx 95\%$.
 \label{SimCont}}
\end{figure}

As for the contact diaries, the network built from the friendship surveys is directed: a student $A$ might report another student $B$ as a friend
while $B$ does not mention $A$. In the present data set, the network comprises 
$689$ directed links of which $137$ are not reciprocated (and $276$ pairs of students declare a friendship towards each other). 
In the following, we will for the sake of simplicity 
mostly consider a symmetrized version of the network in which a link is drawn between two students if at least one of them has reported a friendship with the other.
The resulting network has $135$ nodes and $413$ links. 
Table \ref{Stat} reports a comparison between the main features of the networks
of reported friendships and of contacts. Here we consider the network of contacts registered by the sensors, aggregated over the whole data collection,
and we report the properties of the whole networks and of the networks once restricted to the same set of nodes
(as many students who wore a sensor did not fill in the friendship survey). Similarly to what we found in the contact diary data, 
we observe a much denser network for the sensor data than for the friendship network (with narrow degree distributions,
see SI). This is quite expected as one naturally encounters 
many persons whom one would not list as friends in a survey. The contact network has 
as well larger cliques and smaller shortest path lengths, as shown in Fig. \ref{SimCont}(a): 
most pairs of students are only at distance $2$ in the contact network,
and the maximal distance is $4$, while pairs of students can be separated by as much as $10$ hops in the network of friendships. 

\begin{table}[!ht]
\medskip
\begin{tabular}{c|m{.1\textwidth}m{.15\textwidth}|c|m{.1\textwidth}m{.15\textwidth}}
\toprule
    &{\bf Contact network} & {\bf Reported Friendships} & & {\bf Contact network} & {\bf Reported Friendships}\\
\hline
\midrule
 {\bf Nodes} &      $327$  &  $135$   &  & $134$ & $134$ \\
{\bf Links} &       $5818$ &  $413$   &  & $1235$ & $406$\\
{\bf Density} &    $0.11 $ &  $0.05$ &  & $0.14$ & $0.05$\\
{\bf Avg. Degree} &$36 (13)$ & $6 (3)$ && $18 (8)$ & $6 (3)$  \\
{\bf Avg. Clustering} & $0.5 (0.14)$ & $0.53 (0.29)$ && $0.55 (0.18)$ & $0.54 (0.29)$\\
{\bf Avg. SPL} & $2.16 (0.6)$ & $4.06^* (1.6)$ && $2.22 (0.69)$ & $4.02^* (1.6)$ \\
{\bf Largest clique size} & $23$ & $8$ && $14$ & $8$\\
\bottomrule
\end{tabular}
\caption{ {\bf Summary statistics for the global contact network and the network of reported friendships.} 
All network properties for the network of friendships are computed on its symmetrized version. 
Values on the right part of the table are obtained after retaining only the students present in both networks.
 $^*$The network of friendships is not connected so the average is computed taking into account only the pairs 
of nodes belonging to the same connected component. Numbers in parentheses represent the standard deviations. }\label{Stat}
\end{table}

Despite the very different densities of the two networks, Figure \ref{SimCont} shows that the matrices of link densities
giving the mixing patterns of the different classes have a clear structural similarity, as was the case for the contact matrices
emerging from the contact diaries. Although the values of each cell differ between the two matrices, the structure of classes
and of blocks of classes of the contact patterns occurring between students in the high school is well mirrored in the network of declared friendships,
with a very high similarity between the two matrices.

More in detail, $86\%$ of the reported friendships find a corresponding link in the network of contacts, while only
$24\%$ of the links of the contact network correspond to a friendship link. These numbers change if we restrict the contact data to
stronger links, i.e., to contacts of larger aggregate duration: if we consider only links
with an aggregate duration of more than $1$ min (resp. $3$ min) we find that $82\%$ (resp. $68\%$) of the declared friendship links have 
a corresponding link in the contact network, while $39\%$ (resp. $50\%$) of the contact network links correspond to friendships. 

We investigate moreover in Fig. \ref{SimCont}(b) and Table \ref{table:durations_friendships} the diversity of the cumulative durations registered by the
sensors for the reported friendships: we compare the cumulative distributions of the aggregate durations of contacts between pairs of students, distinguishing 
between students who both reported a friendship with each other, pairs of students with only one directed link of friendship reported, and pairs of students
who did not report any friendship with each other. For reference, we show on the same graph the distribution of the aggregate durations 
of all contacts registered between the $134$ students common to both networks. All distributions are broad: even pairs of students
who have both reported a friendship link might have spent little time in contact. However, the aggregate duration of contacts of declared
friends have a larger average and a broader distribution, especially if the friendship was reported by both (Wilcoxon tests reject the null 
hypothesis of equal distributions at $5\%$ significance level). In particular, all links in the contact
network with aggregate duration larger than a certain threshold (close to 2 hours and a half) correspond to a declared friendship.
Thus, even if a reported friendship link can correspond to effective contacts of very different durations, the global network of 
friendships includes the most important contacts in terms of durations. We also note that both in- and out-degrees in the friendship
survey network are positively correlated with the degree in the contact network: we obtain correlations of resp. $0.4$ and $0.3$,
rising to $0.51$ and $0.45$ if only links with aggregate contact durations above a threshold of $1$ min are taken into account in the contact network.
These results tend to indicate a correspondence between long contacts and reported friendships.

\begin{table}[!ht]
\medskip
\begin{tabular}{c|m{.15\textwidth}|m{.2\textwidth}|m{.2\textwidth}|m{.2\textwidth}}
\toprule
    &{\bf All contact network links} & {\bf Reciprocated friendship} & {\bf Non-reciprocated friendship } & {\bf No reported friendship}\\
\hline
\midrule
 {\bf Mean (s)} &      $926$  &  $3584$  & $942$ & $189$ \\
{\bf Std. dev.} &       $3438$ &  $6714$     & $3349$ & $517$\\
{\bf Number of links} & $1235$ & $245$ & $103$ & $887$ \\
\bottomrule
\end{tabular}
\caption{ Mean and standard deviation of distribution of aggregate durations for different sets of links 
(as in Fig. \ref{SimCont}(b)}\label{table:durations_friendships}
\end{table}

\begin{figure}[!ht]
\includegraphics[width=0.5\textwidth]{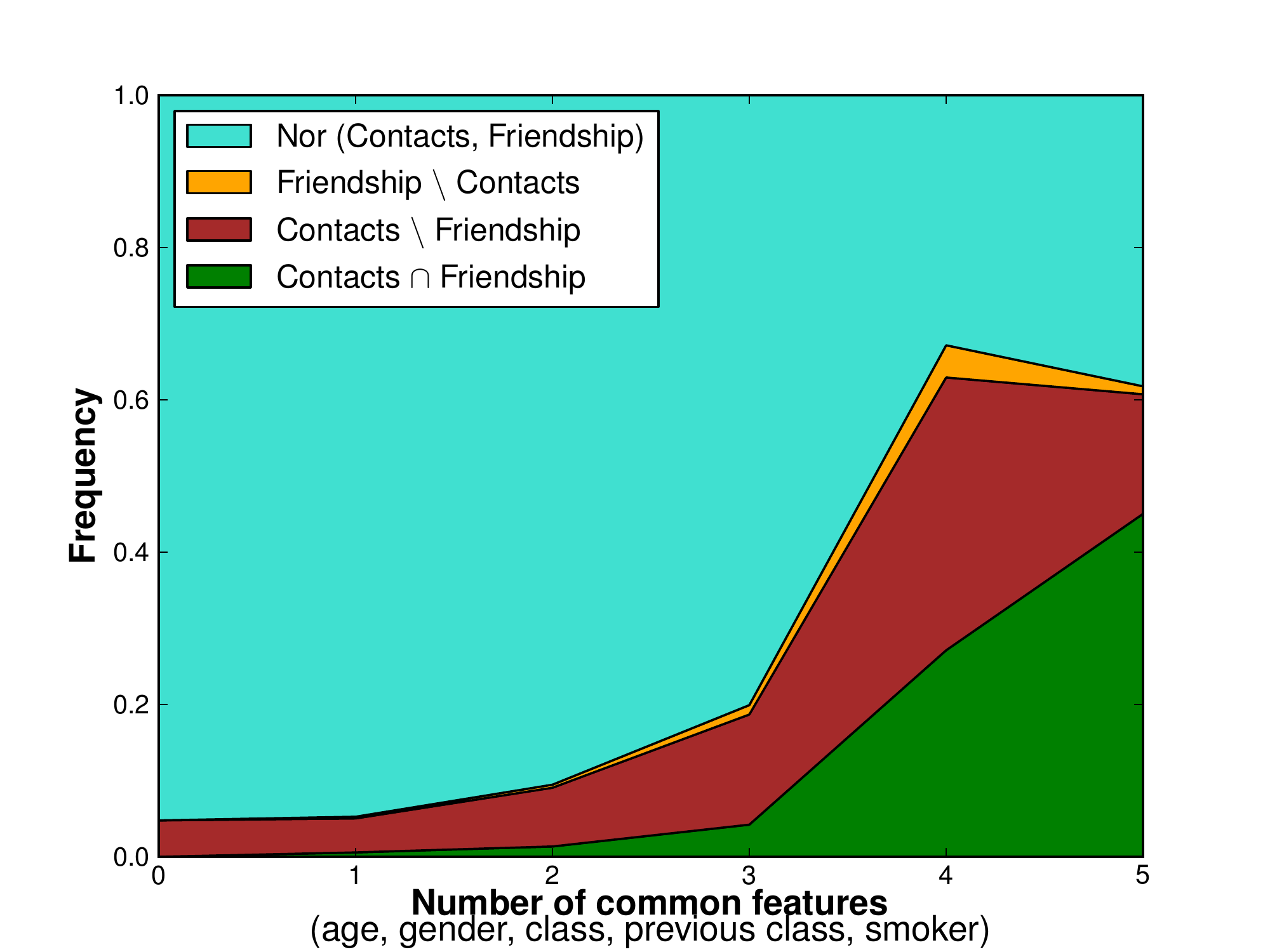}
\caption{{\bf Fraction of friendship and contact links as a function of the number of features shared by two students.} 
} \label{friendship_features}
\end{figure}

We finally investigate in Fig. \ref{friendship_features} how the metadata can provide more intuition on the existence of both contacts
and friendship links. Each pair of students can indeed share between $0$ and $5$ characteristics among age, gender, class, class in the previous year,
and smoking behavior. If two students share less than $4$ features, the largely most probable situation is that they did not have any contact and are not friends
either. On the other hand, if they share $4$ or $5$ features, at least one link is found between them with higher probability.
In particular, friendship relations are observed almost only between students sharing $4$ or $5$ features, especially if they did not have any contact.
Contacts among non-declared friends on the other hand can also be found for pairs of students sharing few features, highlighting the more random character
of such links.

\paragraph{Face-to-face contacts and Facebook links}

We now perform a similar analysis as in the previous subsection, but focusing on a comparison between contact and Facebook data.
As explained in the Methods section, only $17$ students provided us with the network of their Facebook friends, so that we could
not build the entire network of Facebook relationships between the students but rather work with a list of pairs of students
(``known-pairs'') for which we know if they have a friendship relation on Facebook or not. The corresponding data set
includes $4515$ known-pairs involving $156$ students, with $1437$ Facebook links (and $3078$ known pairs of students with no Facebook link).
Moreover, these $156$ students have $1118$ links in the aggregated contact network. $52\%$ of the Facebook links find a corresponding link in the contact 
network, and $67\%$ of the  $1118$ links of the contact network are between Facebook friends.

\begin{figure}[!ht]
\includegraphics[width=\textwidth]{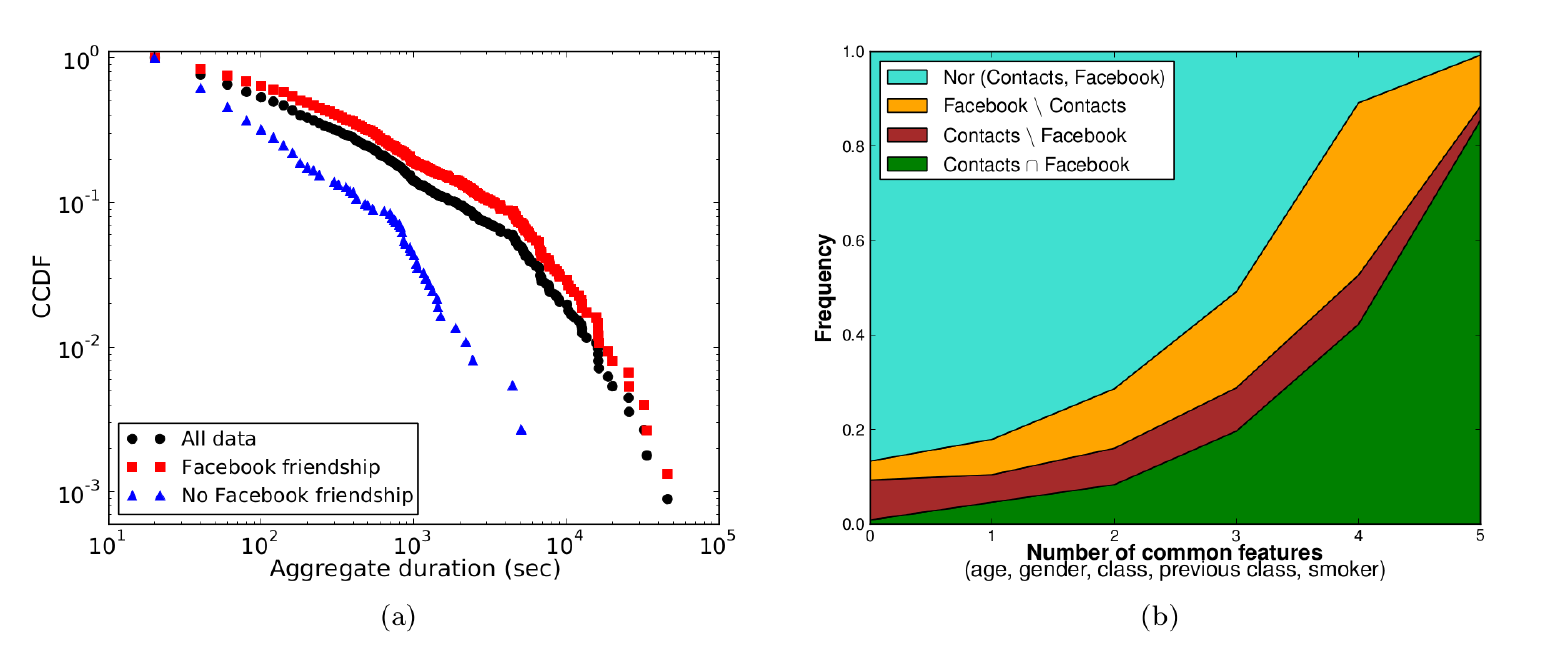}
\caption{ {\bf Contact vs. Facebook links.} (a) Distribution of aggregate durations for the different sets of links.
(b) Fractions of pairs of students belonging to specific groups (no link, link in both the contact network and Facebook, link
in only one of the two) as a function of the number of common features.\label{FC}}
\end{figure}

More information about the aggregate durations of links  present in the contact network
and corresponding, or not, to a Facebook link is provided in Fig. \ref{FC}(a) and Table
\ref{table:durations_fb}: both distributions are broad, as in the case of reported friendships (Fig. \ref{SimCont}), but 
the links between students who are not friends on Facebook have a clearly narrower duration distribution, and links
with aggregate duration larger than a certain threshold correspond all to contacts between Facebook friends. 
The aggregate durations of contacts between students who are friends on Facebook display however
a less broad distribution (which has also a smaller average) than the ones of students with a reciprocated reported
friendship (compare with Fig. \ref{SimCont}(b)).

\begin{table}[!ht]
\medskip
\begin{tabular}{c|c|c|c}
\toprule
    &{\bf All data} & {\bf Facebook friendship} &  {\bf No Facebook friendship}\\
\hline
\midrule
 {\bf Mean (s)} &      $917$ &  $1275$ & $186$   \\
{\bf Std. dev.} &        $3032$  &  $3634$ & $464$ \\
{\bf Number of links} & $1118$ & $750$   & $368$ \\

\bottomrule
\end{tabular}
\caption{Mean and standard deviation of the distributions of aggregate durations for different sets of links in the contact network. }
\label{table:durations_fb}
\end{table}

This last point indicates a different nature of survey-reported and online friendships, which is further investigated by comparing Fig. \ref{FC}(b) 
with Fig. \ref{friendship_features}: the number of shared features of a pair of students  has a still strong but clearly smaller influence on the fraction of such
pairs having a link in the contact network or on Facebook, and a substantial fraction of pairs of students with none or only one feature in common
have a Facebook link. Facebook links which do not correspond to contacts are also observed between pairs of students with any number of shared features.
Overall, the relation between Facebook links and shared features is thus more similar to the one of contact links than the one of reported
friendships.

\paragraph{Contacts and friendship networks as a multiplex}

Friendship relations, online friendship and face-to-face contacts can be combined to provide a more complete picture
of the relationships between students in the high school. Each pair of students can be characterized by one, two, three or none of these three possible
links, forming a multiplex network with three layers\footnote{Strictly speaking, we have a multiplex set of nodes and links
and not a network, as for the Facebook layer we do not have information about many pairs of nodes.}. 
We here perform a simple analysis of this multiplex, in which we consider only students
who are part of all three corresponding data sets: they represent $82$ nodes, 
$992$ links in the contact network (aggregated over the whole study), $326$ reported friendship links and $1026$ Facebook links.

\begin{figure}[!ht]
\includegraphics[width=\textwidth]{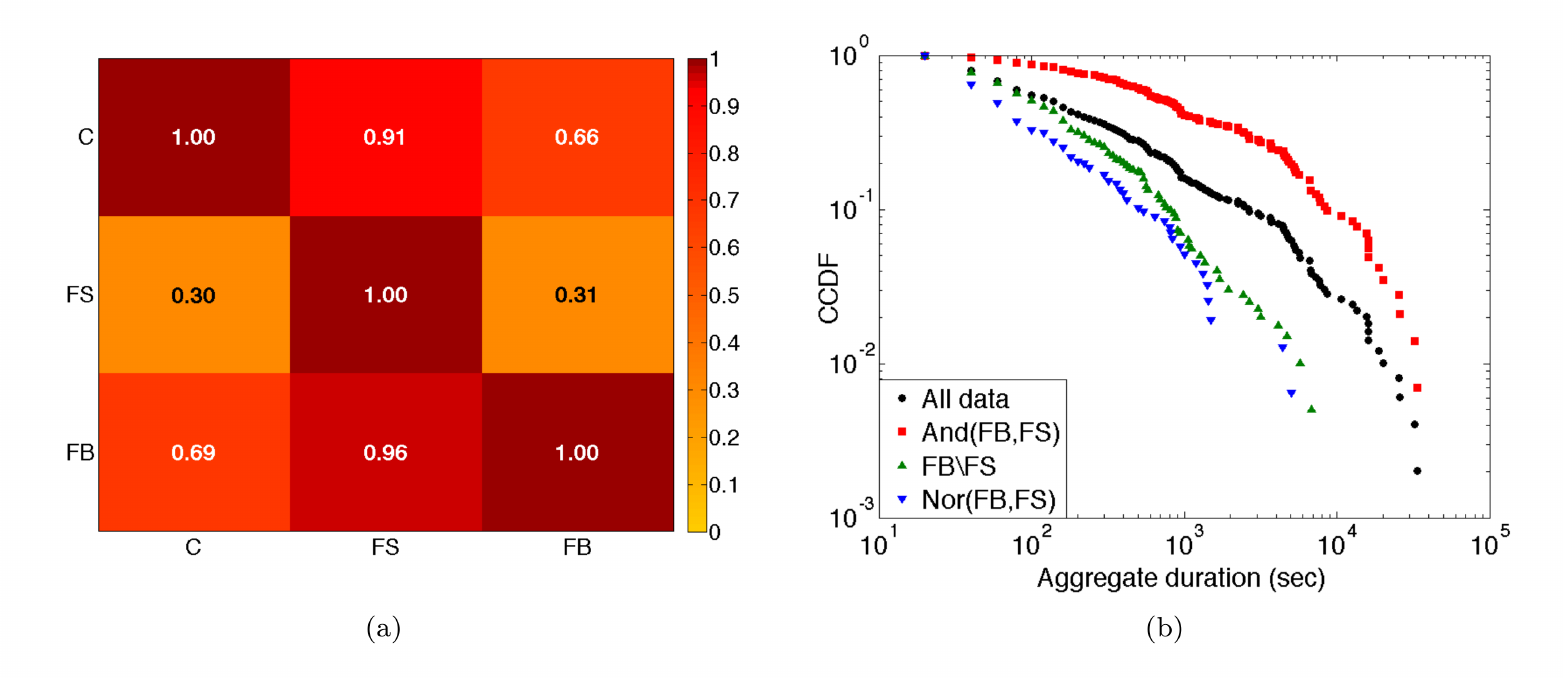}
\caption{ {\bf Multiplex Analysis.}
(a) Conditional probability to find a link in one layer (row index) given its existence in another one (column index); ``C'' stands for 
contact network, ``FS'' for friendship survey, ``FB'' for Facebook; 
(b) Distribution of aggregate durations in the contact network for different sets of links. 
 \label{multi}}
\end{figure}

Figure \ref{multi}(a) reports in a matrix form the conditional probability to observe a link between two students in a layer of the multiplex, 
given that the two students are linked in another layer \cite{Battiston:2013}.
The probabilities that a link is present in Facebook or that
a contact was registered, given that a friendship has been reported, are very high. The probabilities that a
friendship is reported if a contact has been registered or given that a Facebook link exists are much lower.
 Overall, contacts and Facebook
links have similar properties in comparison with the reported friendships.

The strong difference between reported and Facebook links, already noted above, is emphasized by the results of 
Fig.  \ref{multi}(b) and Table \ref{table:durations_FB_friendships} showing the cumulative distribution of aggregate
contact durations, as measured by the wearable sensors, for different sets of nodes pairs.
Although all distributions are broad, important differences are present: the duration distribution is much broader
for the pairs of students who are both reported friends and friends on Facebook; on the other hand, the distribution of the aggregate contact
durations for pairs of students who are only friends on Facebook is much narrower than the distribution of all contact durations, and similar
to the one obtained for the pairs of students who are neither friends on Facebook nor reported friends. Links of a duration above a certain threshold
are observed only between reported friends. With respect to the length of registered contacts, having a link on Facebook 
is therefore not at all equivalent to being reported friends: if reported friendship is not also present,
such a link tends to correspond to rather short face-to-face contacts. This emphasizes the need of an analysis taking into account the three layers
and not only the online friendship one (note that many pairs of students have a link on Facebook but did
not report a friendship relation, while the opposite is true only for $11$ pairs of students).

\begin{table}[!ht]
\medskip
\begin{tabular}{c|m{.14\textwidth}|m{.17\textwidth}|m{.14\textwidth}|m{.17\textwidth}}
\toprule
    &{\bf All data} & {\bf Facebook link and reported friendship} &  {\bf Facebook link only} & {\bf No friendship}\\
\hline
\midrule
 {\bf Mean (s)} &      $1155$  &  $3304$   & $342$ & $184$  \\
{\bf Std. dev.} &       $3549$ &  $6001$     & $820$  & $515$\\
{\bf Number of links} & $992$ & $285$  & $395$ & $301$ \\
\bottomrule
\end{tabular}
\caption{ Mean and standard deviation of distribution of aggregate durations for different sets of links. 
Only $11$ pairs of students have a reported friendship but no Facebook links, so that we do not give the corresponding statistics.
}\label{table:durations_FB_friendships}
\end{table}

\section{Discussion}
In this article, we have presented an analysis of several data sets concerning interactions between students in a high school, 
collected both through a decentralized sensing platform based on wearable sensors and through diaries and surveys. The fact that these data sets
were collected at the same time and in the same population allowed us to compare them and to quantify the overlap and complementarity
of data sets of different nature. 

Data collected thanks to the wearable sensors developed by the SocioPatterns collaboration yielded a temporally resolved network of
contacts between the $327$ participating students. This network had features similar to data collected in the same context in previous
years \cite{Fournet:2014} and in other contexts \cite{Barrat:2013,Barrat:2014}, with heterogeneous contact durations and inter-contact times, repeated
daily activity patterns, and an aggregated contact network structure shaped by the division of students in classes.
This confirms the robustness
of the properties of contact patterns measured in different years in a given context as found in \cite{Fournet:2014}.

Although the use of sensors to measure contact patterns has become more widely available and affordable in the last years, such deployments
are not always feasible. Other methods, in particular based on contact diaries, have been and are still widely used. We have thus, in the
same spirit as \cite{Smieszek:2014}, compared the network of contacts reported in diaries by the students with the contact network
measured by the sensors. We mostly confirmed the results of \cite{Smieszek:2014} and obtained some further insights:
\begin{itemize}
\item  many students who accepted to participate to the data collection by wearing sensors did not fill the diary, probably due to the extra burden
at the end of a school's day; 

\item similarly to the data of \cite{Smieszek:2012}, not all contacts between two individuals were reported by both; when both
students reported the contact, the approximate reported duration was most often the same;

\item as found in \cite{Smieszek:2014}, most short contacts detected by sensors were not reported in diaries,
while the reporting probability was high for contacts with long enough aggregate durations; 

\item the distribution of aggregate durations measured by the sensors were broad for contacts both reported and not reported in the diaries;
the distribution was however broader for contacts reported, and all contacts of a long enough duration (as measured by sensors) were reported in the diaries;

\item the contact durations reported by students tended to overestimate the durations measured by sensors; 

\item despite the lower sampling in the diary data, the overall structure of the sensor contact network was recovered in the diary data,
with a similar contact matrix. The density of the diary-based contact network was however much smaller due to the many short contacts not reported;
as a result, paths between individuals appear longer in the diary-based contact network;

\item we observed a correlation between the degrees of a node (number of distinct persons encountered) in both networks; when considering the
diary-based contact network as directed, we observe that the degree measured by sensors is not correlated with the out-degree of a node (i.e., the number
of contacts reported by the individual) but is correlated with its in-degree (i.e., the numbers of other persons reporting a contact with him/her).

\end{itemize}
Our results are in line with existing literature about the limits of self-reported surveys, which can stem from
response style and response set biases \cite{Lanyon:1997}:
The former include general distortions which do not depend on the specific content of the survey, such as
{\it acquiescence} (the tendency to endorse all statements) and {\it extreme} or {\it central responding} (tendency to indicate the
extreme or the central values in scale-rating surveys); the latter include biases strongly related to the survey topic such as the so-called
{\it social desirability} (tendency to answer according to a community \lq\lq desired\rq\rq{} response). 
These biases can in particular lead to an
overestimation of the durations and numbers of contacts \cite{Paulhus:2007}. Overestimation of the contact durations could also be related to the small
number of people met (node degree), as suggested by \cite{Hyett:1979} who found underestimation only for very large degrees (here the
distribution of degrees of the contact networks have mean $\sim 9$ and standard deviation $\sim 4.3$).

Overall, diaries reporting contacts give therefore a different
picture of the contact network measured by sensors in two respects: on the one hand, population
sampling issues are much more severe; on the other hand, most short contacts are lost. However, a kind of 'backbone' of the contact network
is reported, in which one can deduce the overall structure of the network and the most important contacts (in terms of duration) are present.
As the present data and the ones of \cite{Smieszek:2014} concern similar contexts (high schools), it would be of great
interest to confirm the robustness of these findings in other contexts. In cases such that collecting diary-based is easier than using wearable sensors, 
such knowledge on the differences between these two kinds of data could help obtain a more complete picture of the actual contact patterns.

Friendship relations represent another kind of data for which surveys are commonly used; the resulting data sets are a priori less prone to
biases due to imperfect memory, and might also be easier to collect in a given population than contact diaries. Surveys might indeed
be given and collected on a less constrained time frame than contact diaries, as friendships evolve on longer time scales than contacts.
Similarly, online social networks might in some cases be easier to collect automatically. How much the networks of reported friendships
or of online links correspond to or complement actual face-to-face contact networks is however not well known \cite{Sekara:2014,Barrat:2010}. 
We have therefore asked the participating
students to fill in surveys to report their friendships in the high school, and to give us access to the networks Facebook links of their Facebook
friends. As in the case of contact diaries, the data suffered from low sampling rates, showing the lesser acceptability of these requests (most probably
because of the added burden for students with a large number of courses to follow) with respect to the use of wearable sensors. 
The network of reported friendship was much less dense than the contact network, as could be expected: most declared friends had at least one face-to-face
contact during the data collection, while many contacts also occurred between pairs of individuals who did not report a friendship link. 
Despite this difference in the networks, the contact matrices of link densities between classes, as measured by the reported friendships, had a
structure similar to the one deduced from the actual contacts. We also found differences in the properties of the contact network links between declared friends
and between pairs of non-friends: links with an aggregate contact duration of any length are found between pairs of friends,
but the distributions of aggregate durations is much narrower for individuals who are not friends. In fact, all contact network links
of large enough duration occur between friends who both reported each other in the survey. In terms of network structure, the network of reported friendships 
gives information of quality similar to a contact diary with respect to the sensor-measured contact network (but lacks any
estimation of the contact durations).

Interestingly, Facebook links yield a different picture than reported friendship ones. First, sampling issues prevent us from 
building a complete network of Facebook relationships. The probability that a contact is observed between two individuals, knowing that
they are linked on Facebook, is also much smaller than the same probability conditioned on a reported friendship. 
In addition, the distribution of aggregate contact durations between individuals who are linked on Facebook but did not 
report a friendship to each other is much narrower than the one obtained for individuals
who reported a friendship: overall, Facebook links seem to have a more 'casual' character than friendship links, in agreement
with the intuition that Facebook links are easier to establish than real friendships.

In conclusion, the collection of data with different methods has shown that, at least in the context of a high school in which students
have a rather heavy work load (given they have competitive exams at the end of the year), the use of wearable sensors to gather data 
on contacts between individuals was more easily accepted than the burden of filling in contact diaries or surveys, and yielded
therefore a better population sampling. Moreover, most short contacts detected by sensors were not reported in diaries and did not correspond
to friendship relations, leading to much more dilute interaction networks based on diaries or surveys than on measured contacts. As a consequence,
the distance between individuals in the former networks are overestimated, which could have strong consequences when using such data
in data-driven models of dynamical processes (e.g., epidemic spreading). However, as all contacts of long enough durations were reported in the diaries
and surveys, and as the diary-based and survey-based networks correctly revealed the structural organization of the population in classes,
one could argue that the reported links carry enough information to feed data-driven models \cite{Genois:2015}. Future work will address this issue by performing
numerical simulations of spreading processes on the various networks and comparing the outcomes. Another interesting avenue
could be to understand how to combine data obtained with different methods to obtain a more complete picture of the interactions
of a population: on the one hand, diaries could be used if not all individuals of a population wore sensors, to compensate for
the resulting sampling bias; on the other hand, they could yield insights on how a given
population (for which contacts are measured with sensors) enters in contact with the external world. Indeed, even when a high-resolution contact
network has been measured among a given population and is used in a detailed data-driven simulation of epidemic spread for the evaluation
of containment measures \cite{Gemmetto:2014,Gauvin:2015}, the interactions of the population of interest with the outside
community is modeled in a simplistic way: using data from diaries and taking into account underreporting could
help refine the model and test the robustness of the results.

% Do NOT remove this, even if you are not including acknowledgments.

\section*{Acknowledgments}

We are grateful to the SocioPatterns collaboration \cite{SocioPatterns} for providing access to the SocioPatterns 
sensing platform that was used in collecting the contact data and to Ciro Cattuto for numerous
discussions. We thank the students of Lyc\'ee Thiers in Marseilles, France, who accepted to participate to the data collection.

This work was supported by the A*MIDEX project (n° ANR-
11-IDEX-0001-02) funded by the ''Investissements d’Avenir'' French Government program, managed by
the French National Research Agency (ANR), to A.B. and R.M. A.B. is also partially supported by the French ANR project
HarMS-flu (ANR-12-MONU-0018) and by the EU FET project Multiplex 317532.

% Either type in your references using

% 

\clearpage
\newpage

\section{Supplementary Material}

\setcounter{figure}{0}
\renewcommand{\thefigure}{S\arabic{figure}}

\subsection{Properties of the contact network measured by wearable sensors}

\begin{figure}[!ht]
\hspace{-10mm}
{\includegraphics[width=0.5\textwidth]{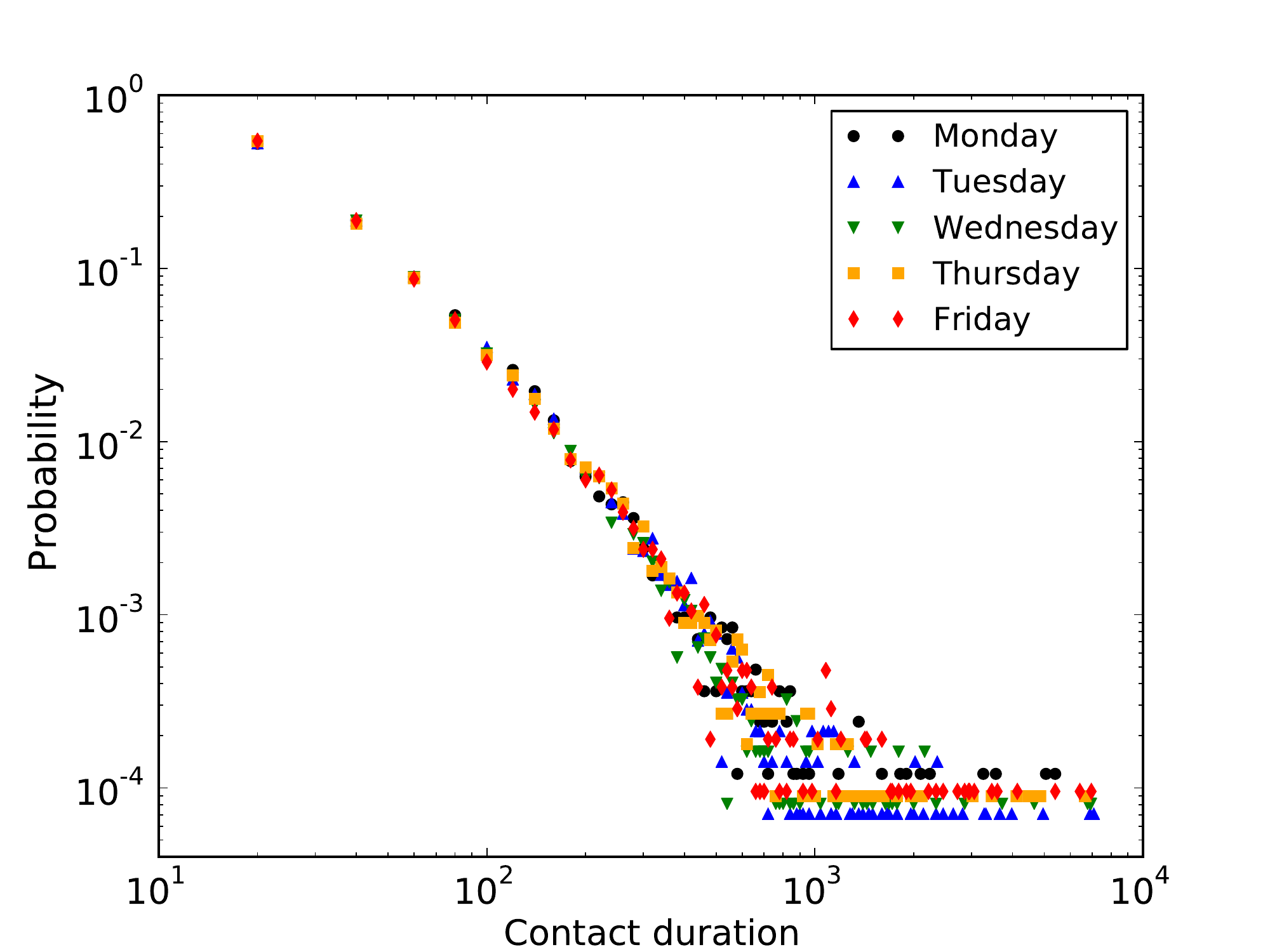}}
{\includegraphics[width=0.5\textwidth]{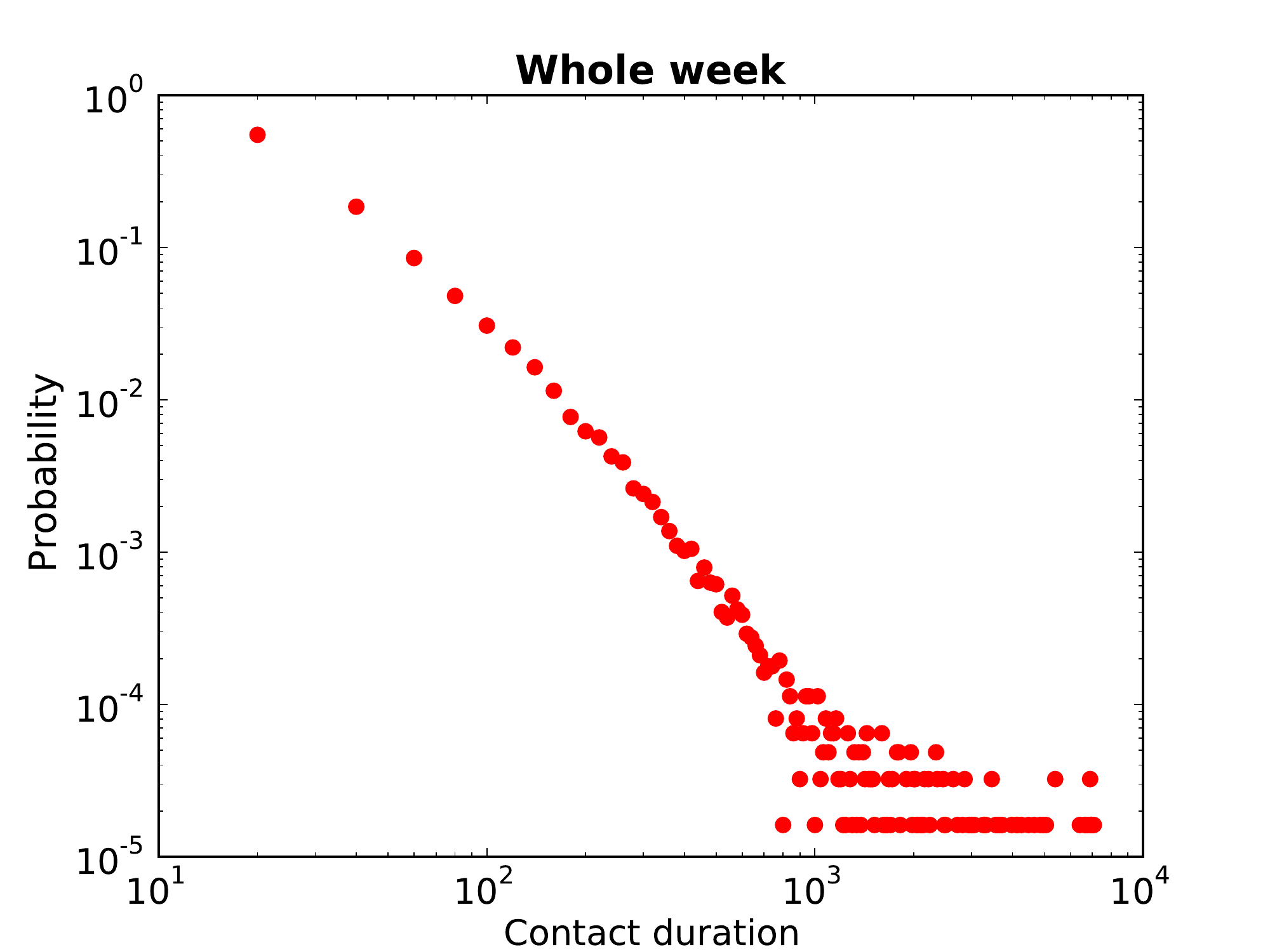}}
\caption{{\bf Distribution of contact durations in the temporal network of contacts measured by the SocioPatterns sensing platform.}
The left panel shows the distributions of contact durations for each day of the week, the right panel for the whole data set.}
\end{figure}

\begin{figure}[!ht]
\hspace{-10mm}
{\includegraphics[width=0.5\textwidth]{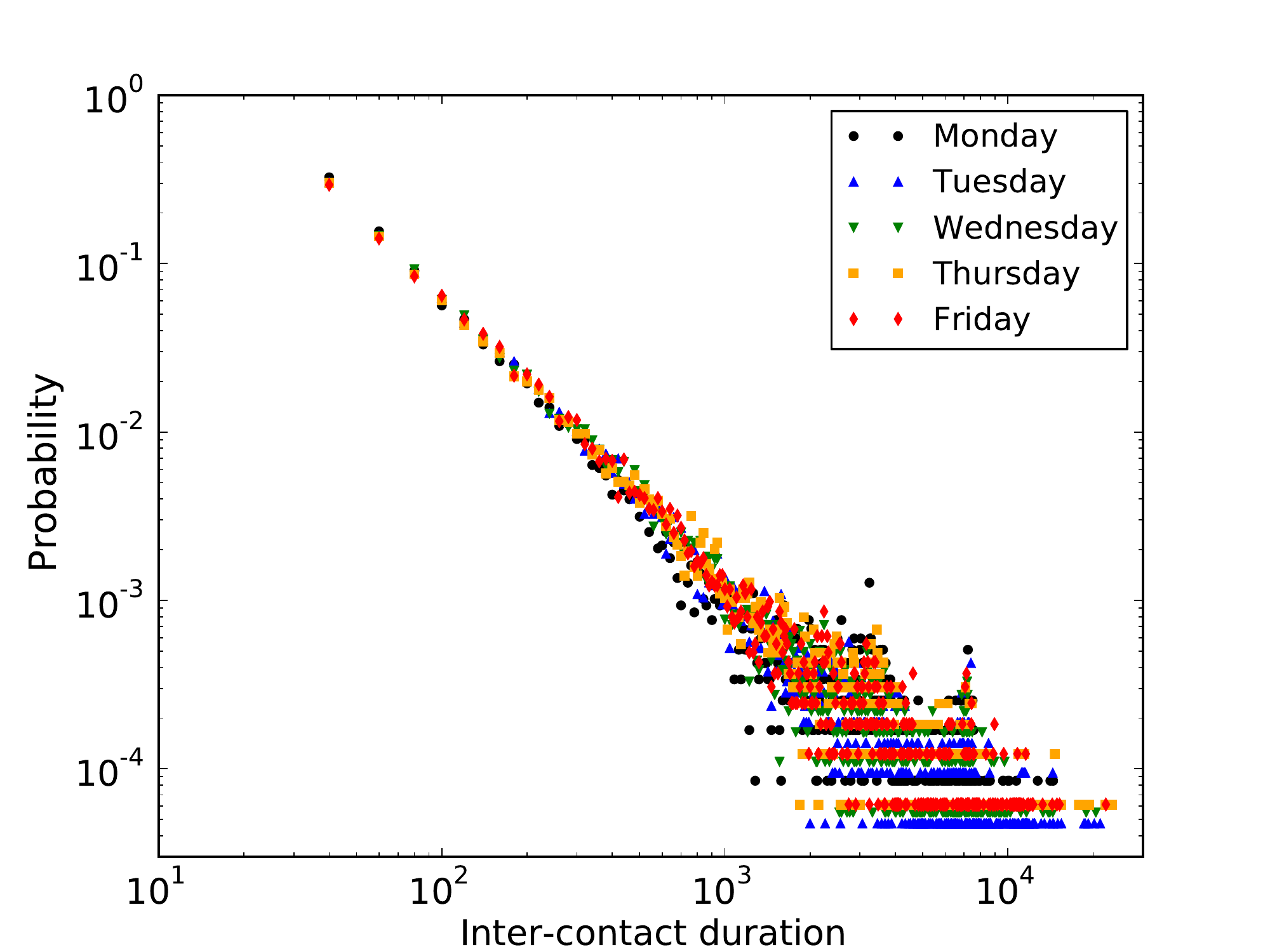}}
{\includegraphics[width=0.5\textwidth]{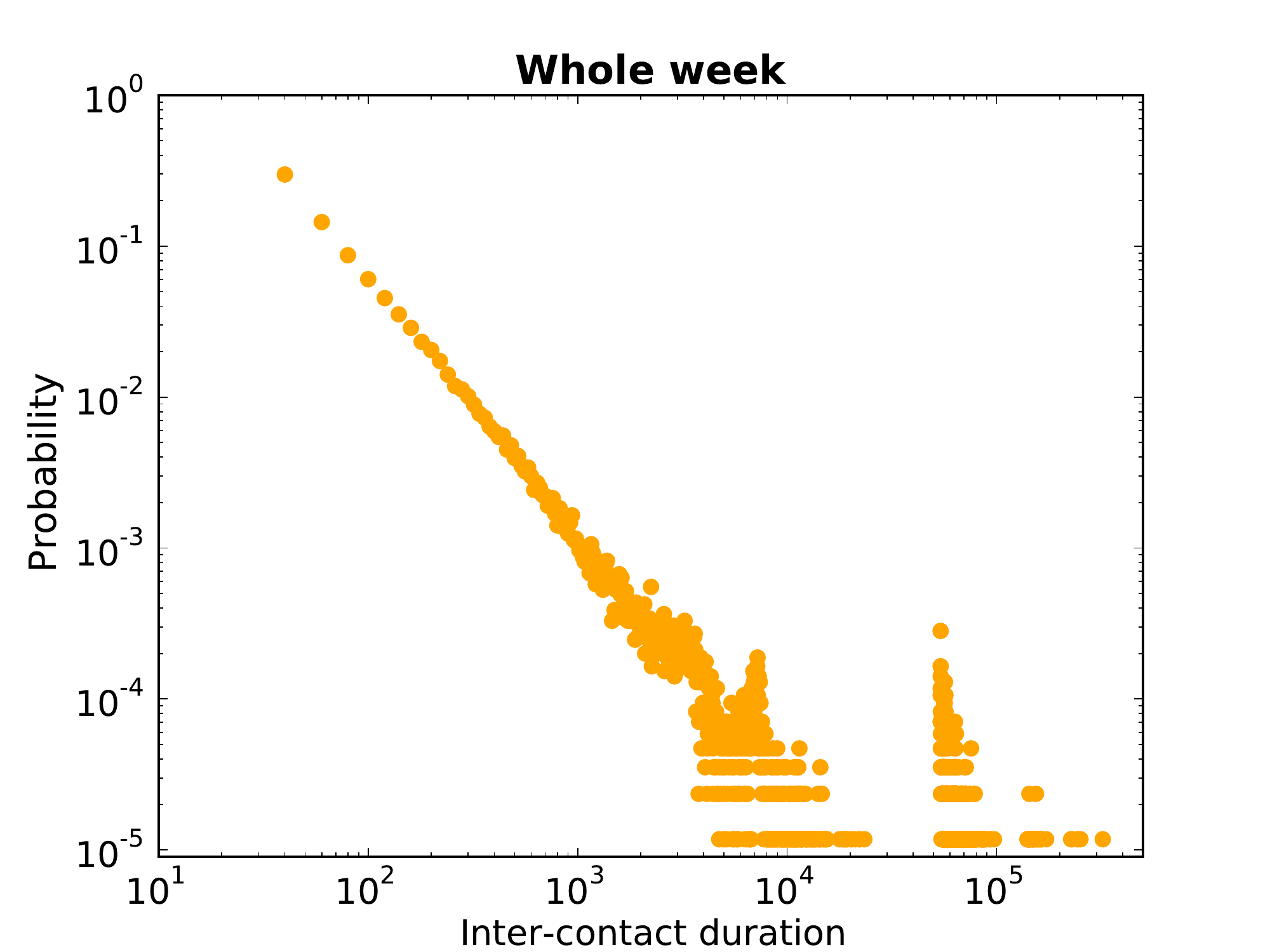}}
\caption{{\bf Distribution of inter-contact durations.}
The left panel shows the distributions of time elapsed between successive contacts of a given pair of individuals in each day.
The right panel displays the distribution measured on the whole data set. The peaks on the right correspond to inter-event durations
including one or two nights.}
\end{figure}

\begin{figure}[!ht]
\hspace{-10mm}
\subfigure[]%Distribution of weights for each day]
{\includegraphics[width=0.5\textwidth]{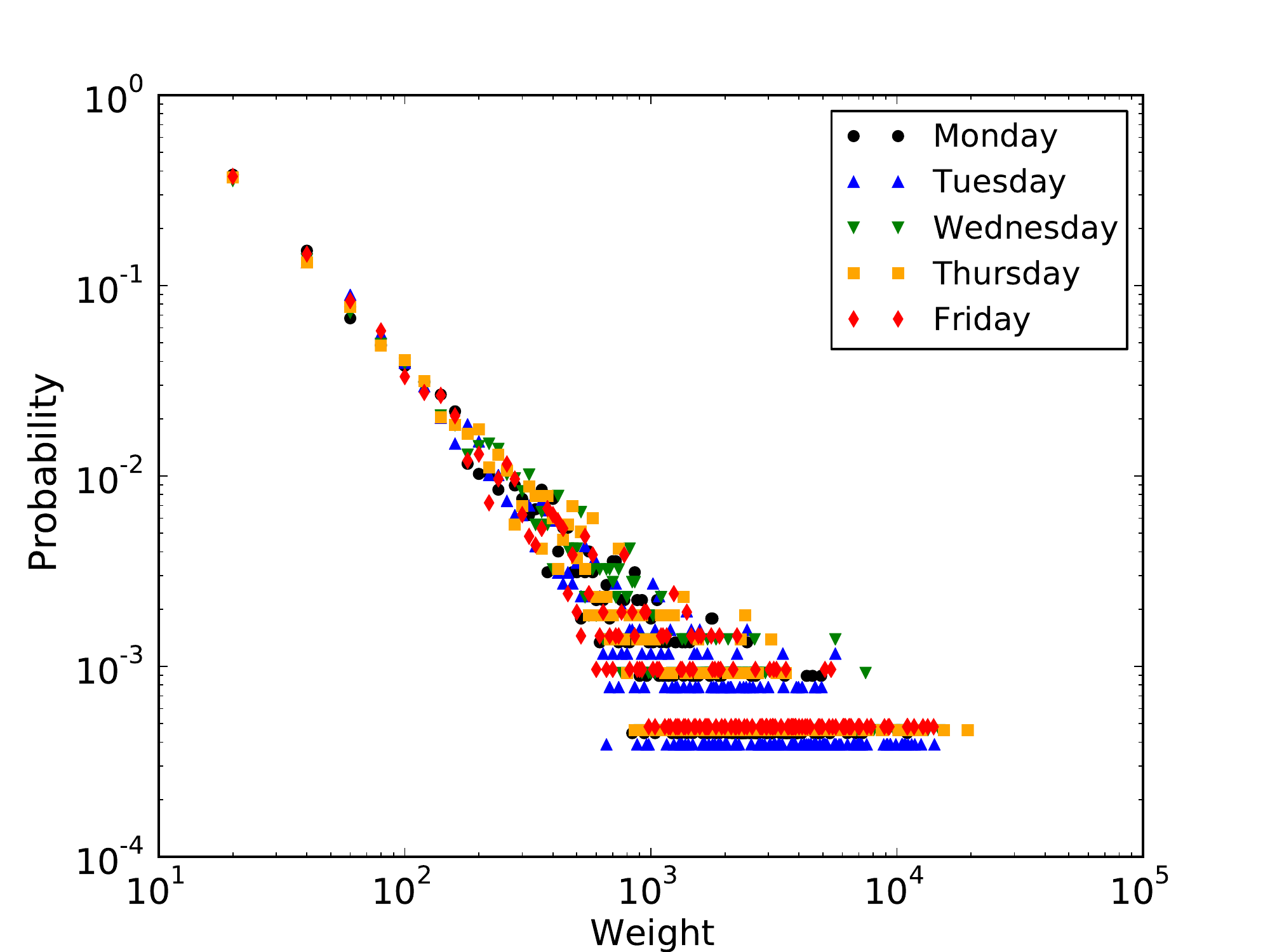}}
\hspace{-5mm}
\subfigure[]%Distribution of weights in the aggregated network]
{\includegraphics[width=0.5\textwidth]{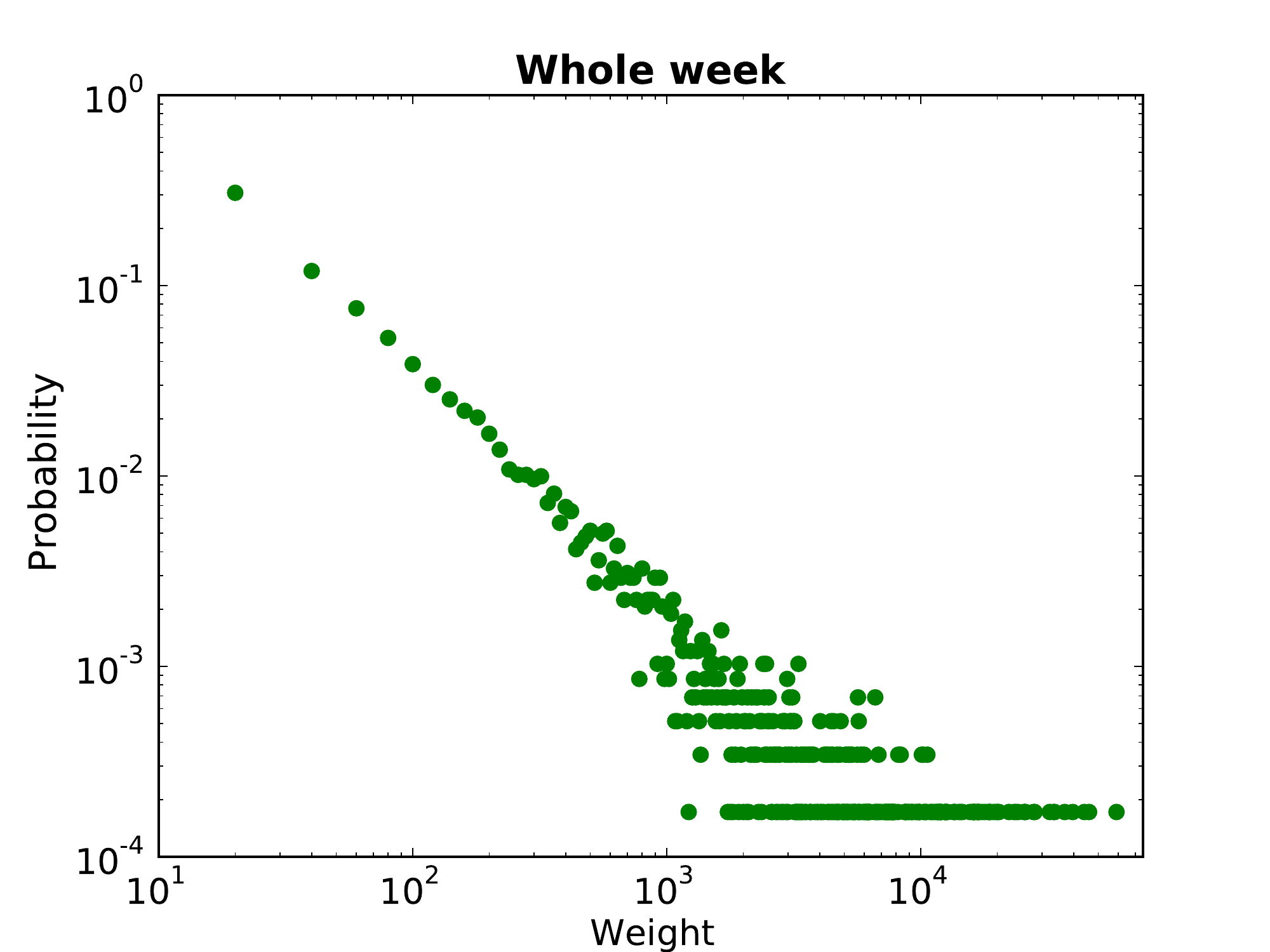}}
\caption{{\bf Distribution of edge weights,} i.e., of aggregated contact durations between individuals,
(a) in the daily aggregated contact networks, (b) in the contact network aggregated over the whole week.}
\end{figure}

\begin{figure}[!ht]
\hspace{-10mm}
\subfigure[]%Distribution of degrees for each day]
{\includegraphics[width=0.5\textwidth]{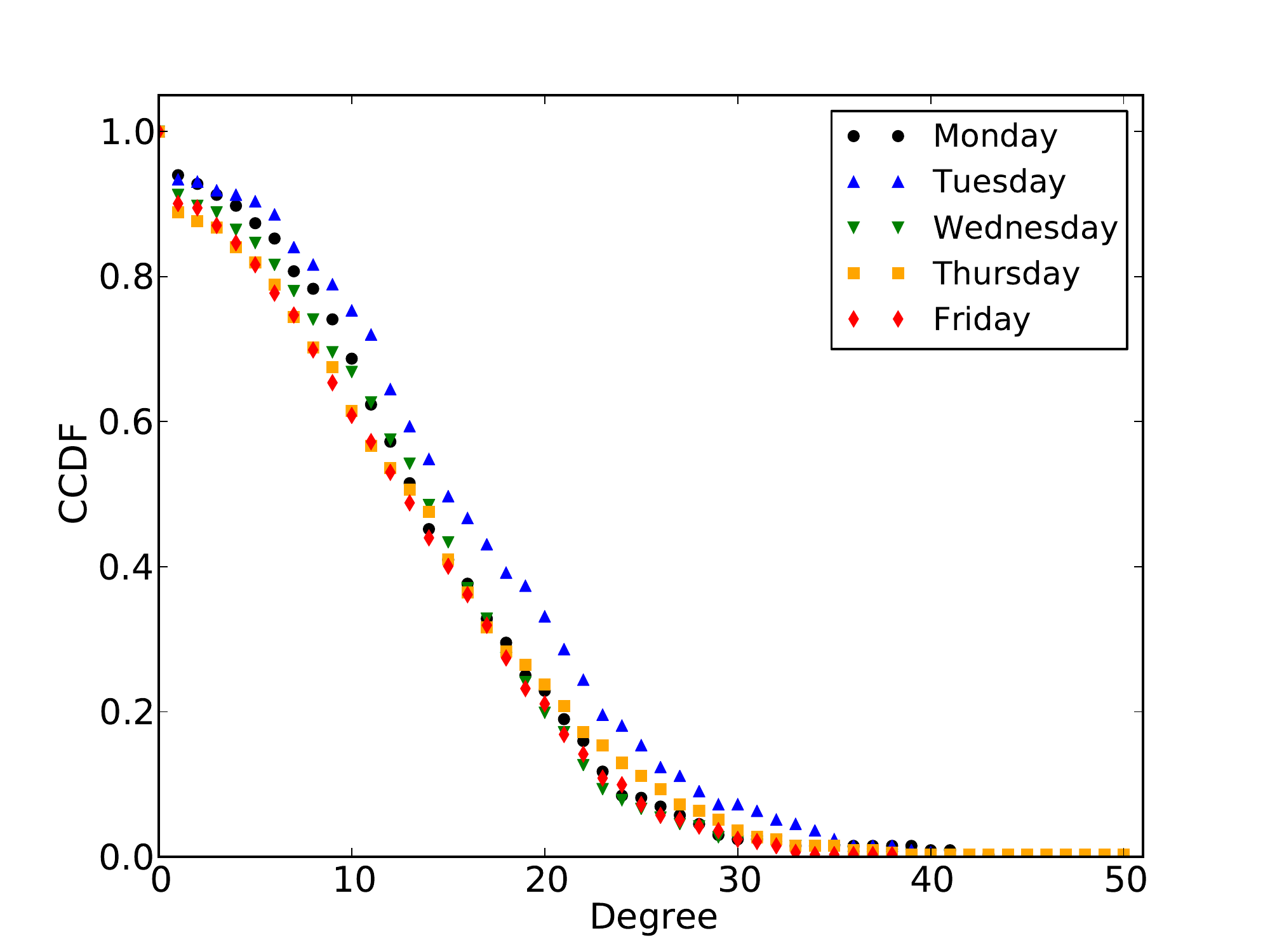}}
%\hspace{-5mm}
%\subfigure[]%Distribution of degrees for each day]
%{\includegraphics[width=0.5\textwidth]{2013_anticumulative_degree_distribution_week1all2.pdf}}
\hspace{-9mm}
\subfigure[]%Distribution of degrees in the aggregated network]
{\includegraphics[width=0.5\textwidth]{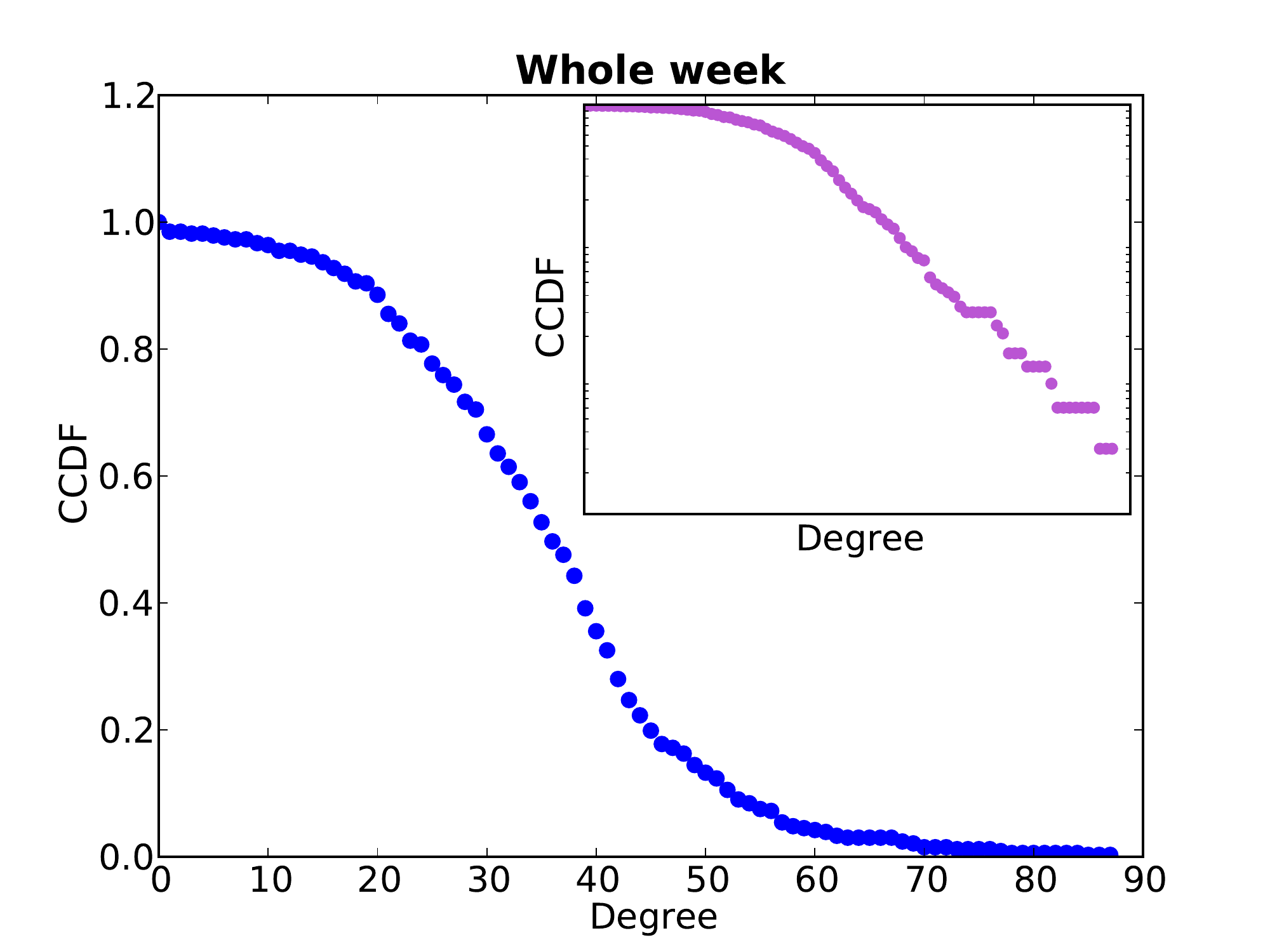}}
\caption{{\bf Complementary cumulative distribution function of degrees,} i.e., of distinct individuals with whom a person has had contacts, 
(a) in the daily aggregated networks 
%(b) rescaled by the average degree of each daily network, 
(b) in the contact network aggregated over the whole week. The inset shows the same distribution in lin-log scale.}
\end{figure}

\begin{figure}[!ht]
\hspace{-10mm}
{\includegraphics[width=0.5\textwidth]{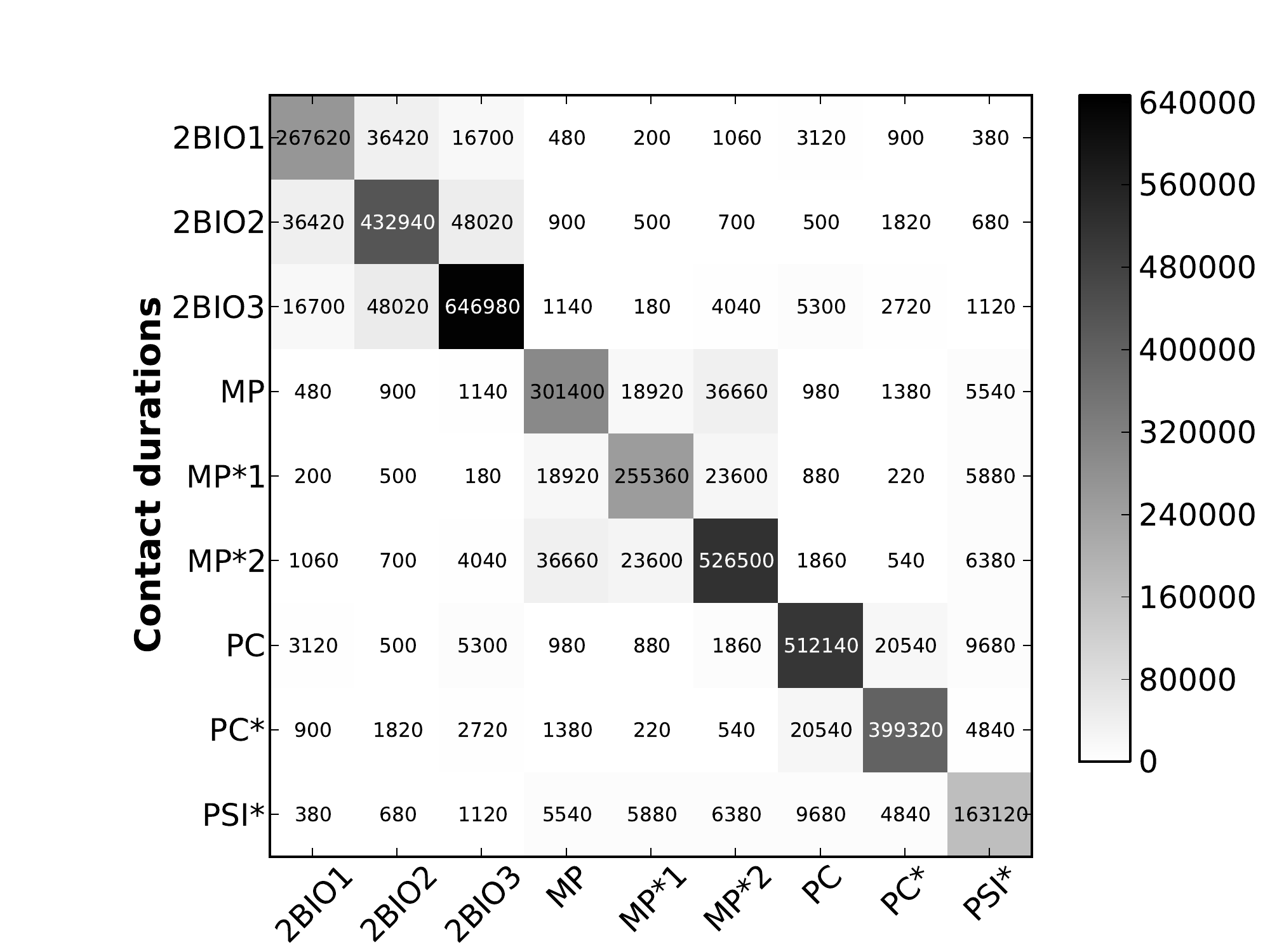}}{\includegraphics[width=0.5\textwidth]{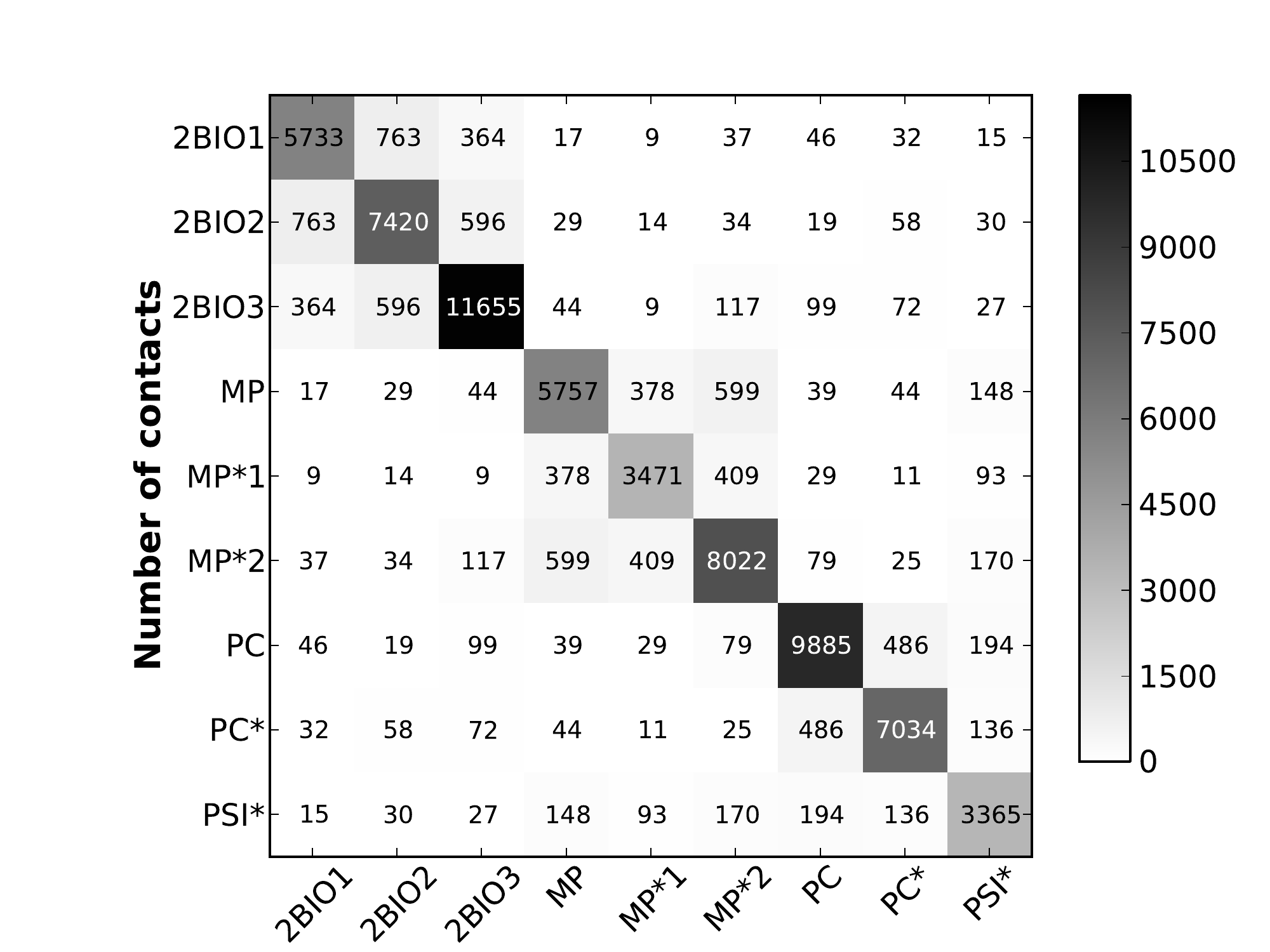}}

\hspace{-10mm}{\includegraphics[width=0.5\textwidth]{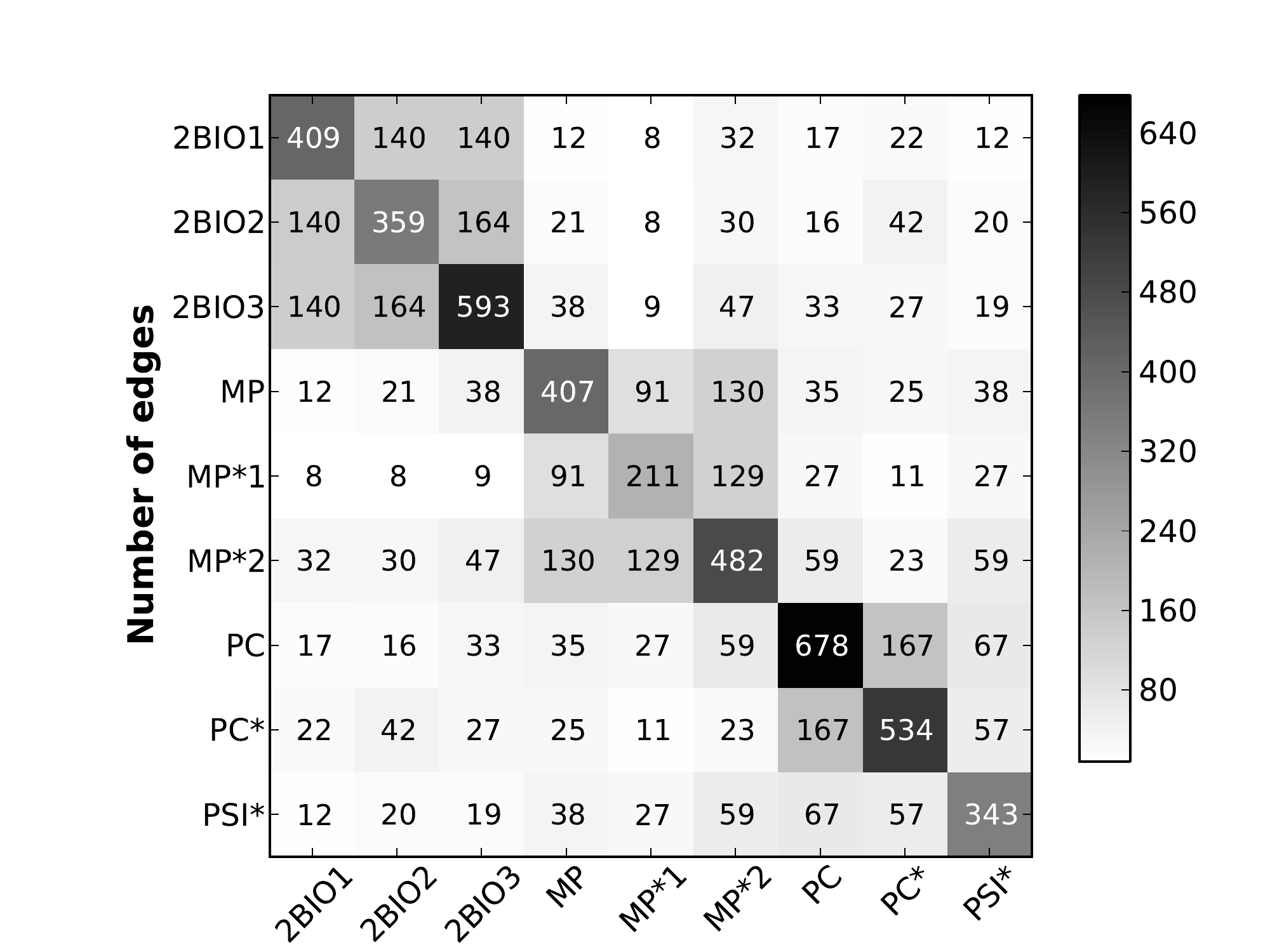}}{\includegraphics[width=0.5\textwidth]{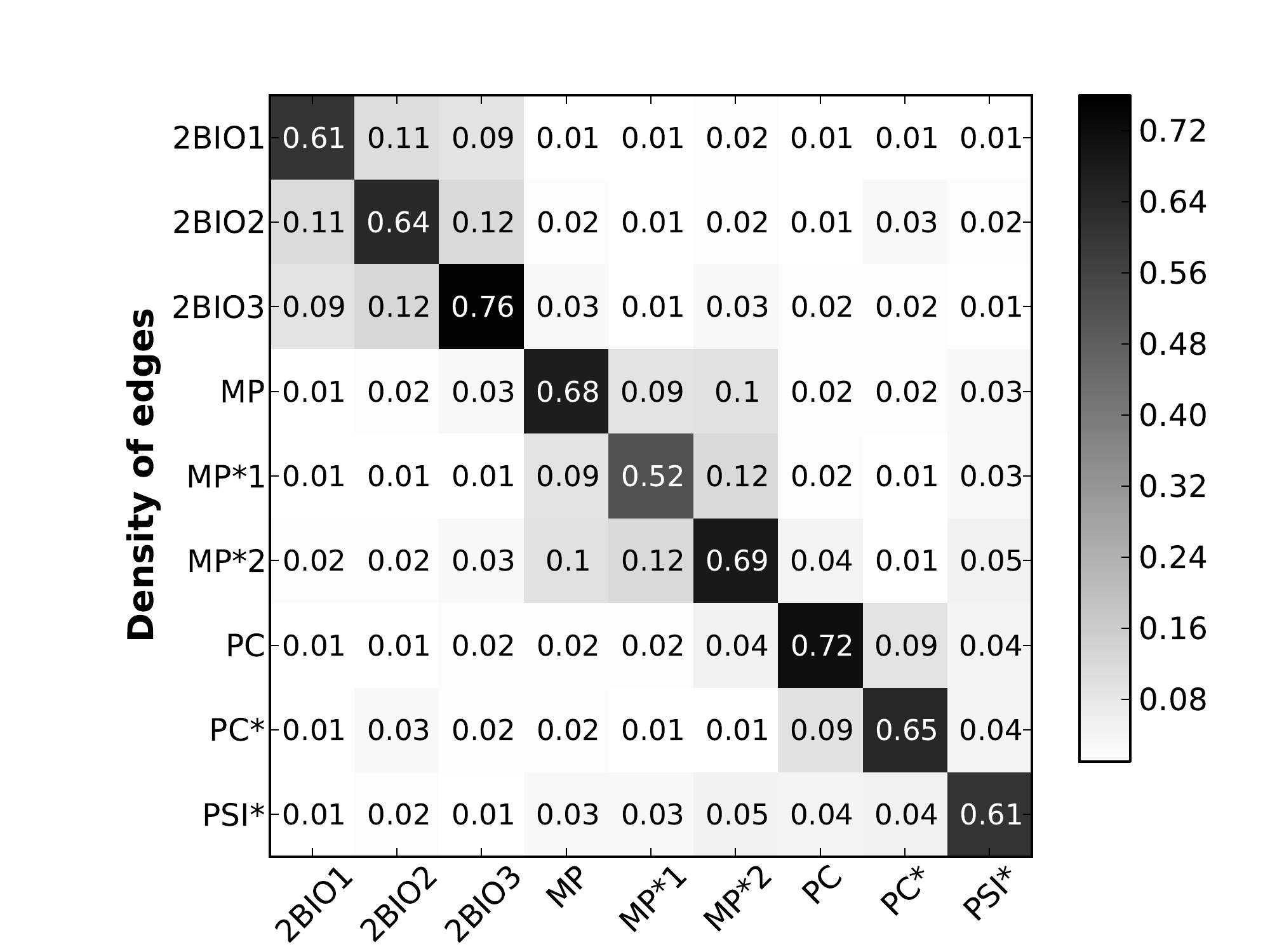}}
\caption{{\bf Contacts matrices} 
giving the cumulated durations in seconds and the numbers of contacts between classes (first row) and the numbers and densities of links between classes 
in the contact network (second row), for data aggregated over the whole week of data collection.}
\end{figure}

\begin{figure}[!ht]
%\hspace{-10mm}
%\subfigure[]
\includegraphics[width=0.5\textwidth]{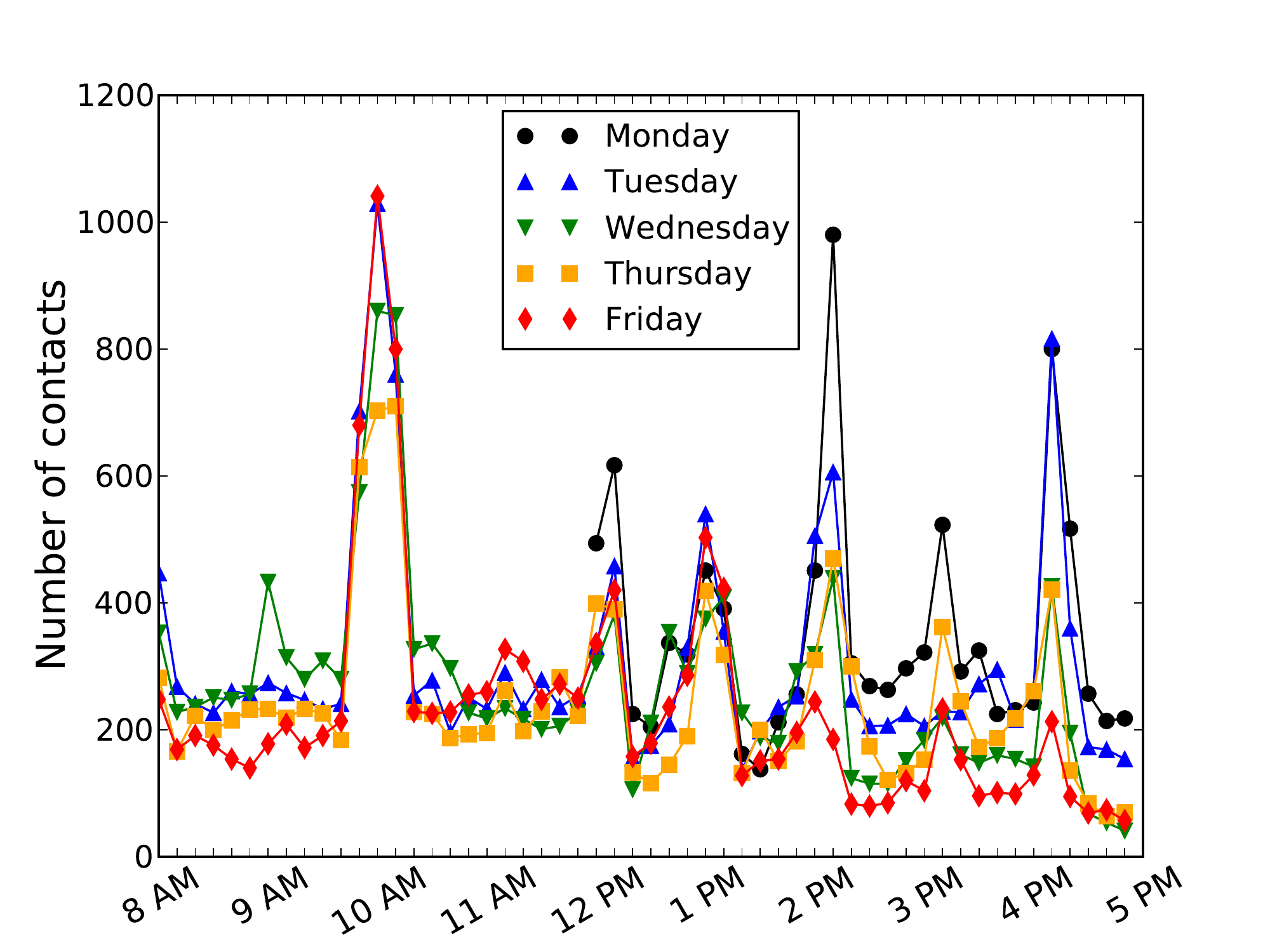}
%\hspace{-5mm}
%\subfigure[]
%{\includegraphics[width=0.5\textwidth]{2013_number_contacts_per_time_unit_week1.pdf}}
\caption{{\bf Activity timelines.} Number of contacts collected by the sensors per 10-minutes periods for each day.}
%(b) Number of contacts per one-hour time-windows.
\end{figure}

\begin{figure}[!ht]
\hspace{-10mm}
{\includegraphics[width=0.35\textwidth]{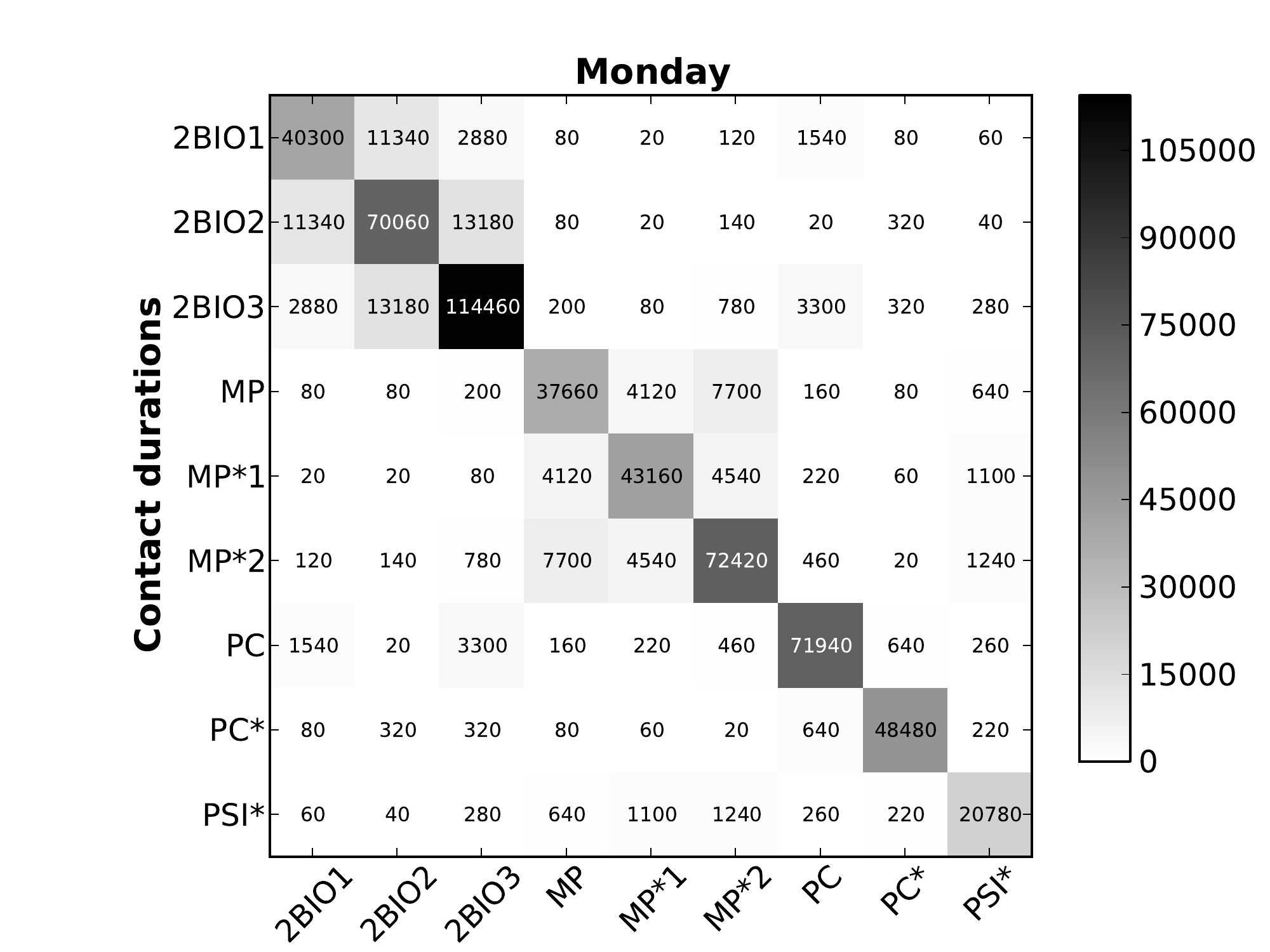}}{\includegraphics[width=0.35\textwidth]{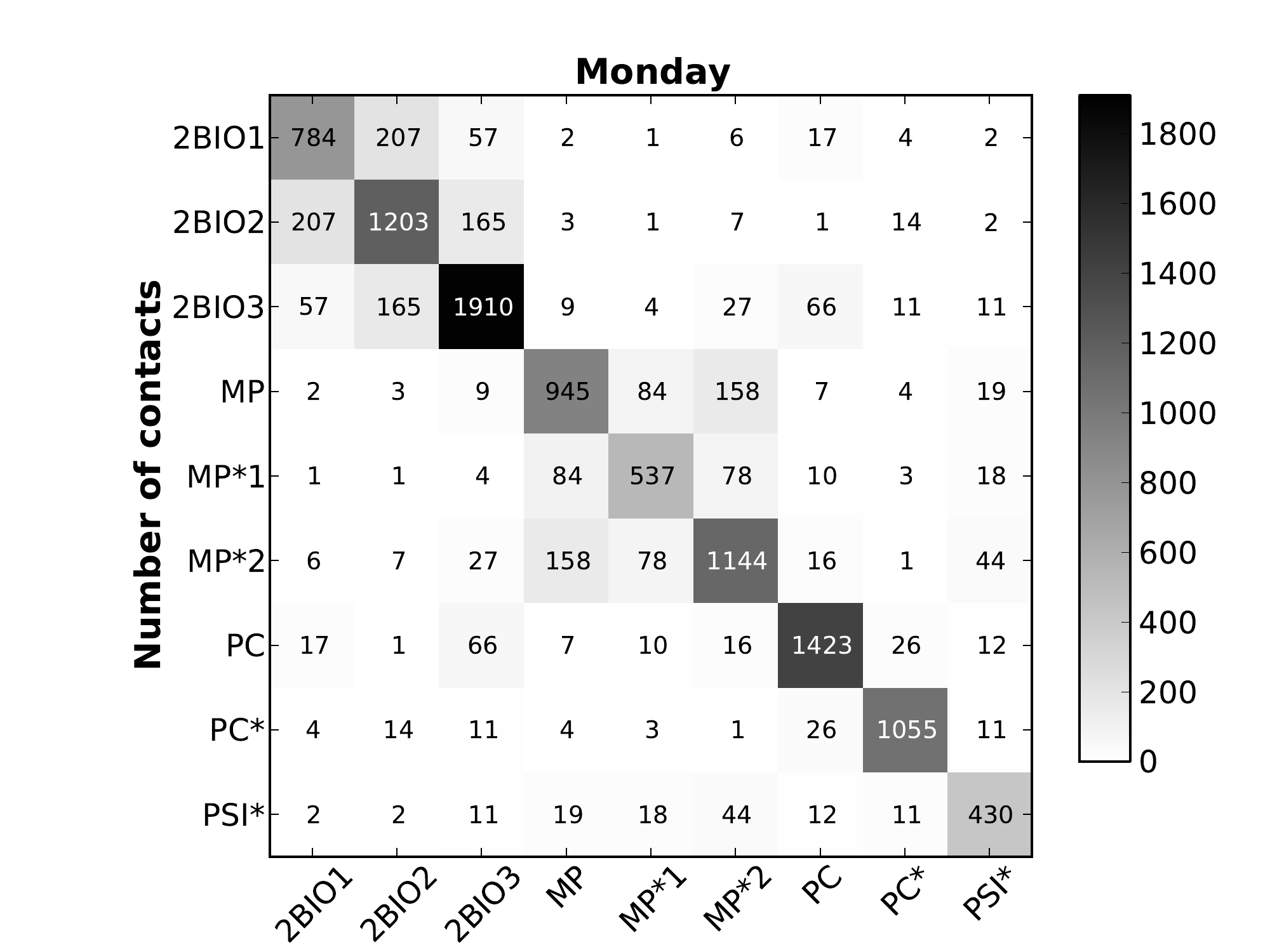}}
{\includegraphics[width=0.35\textwidth]{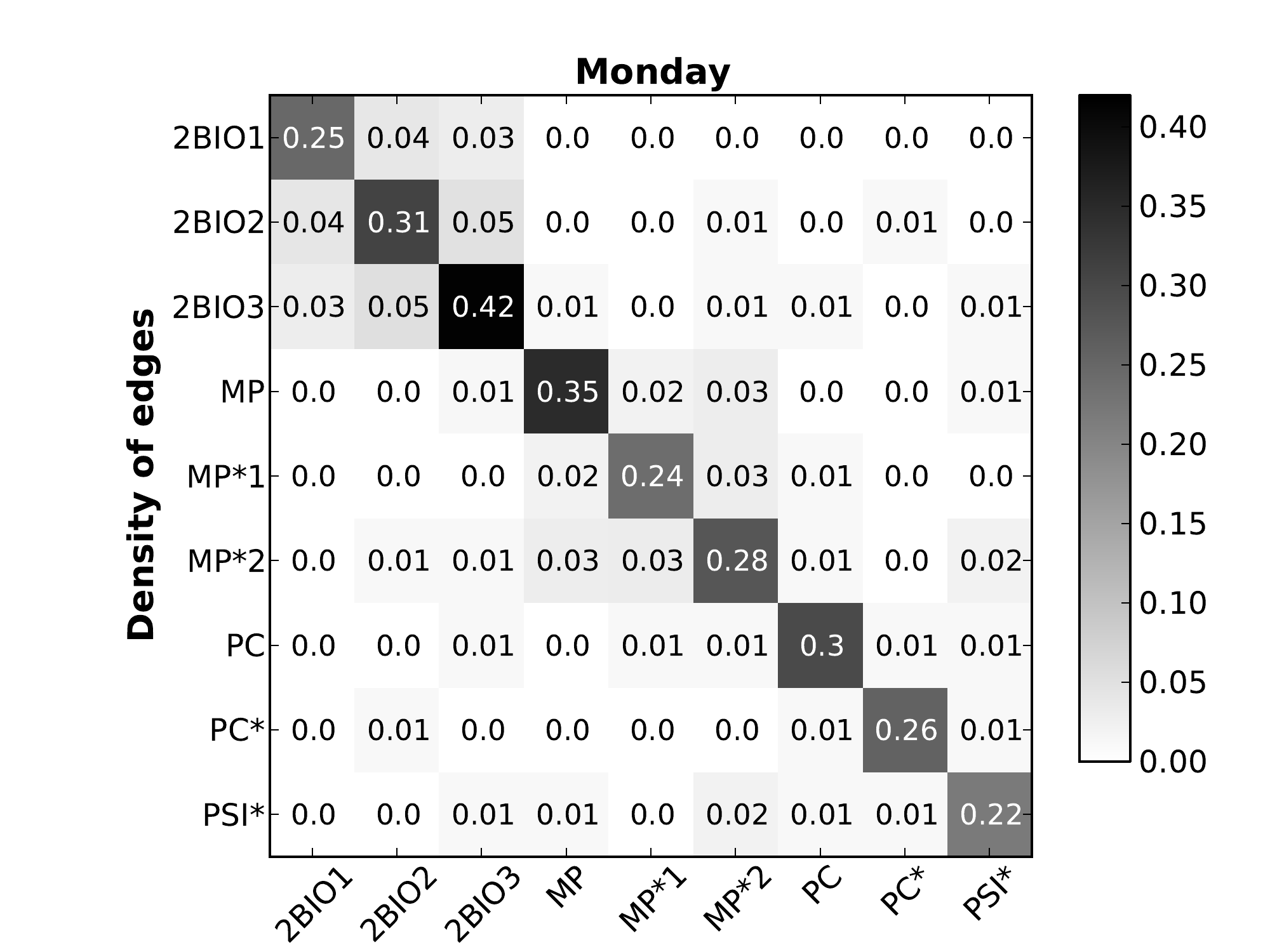}}

\hspace{-10mm}
{\includegraphics[width=0.35\textwidth]{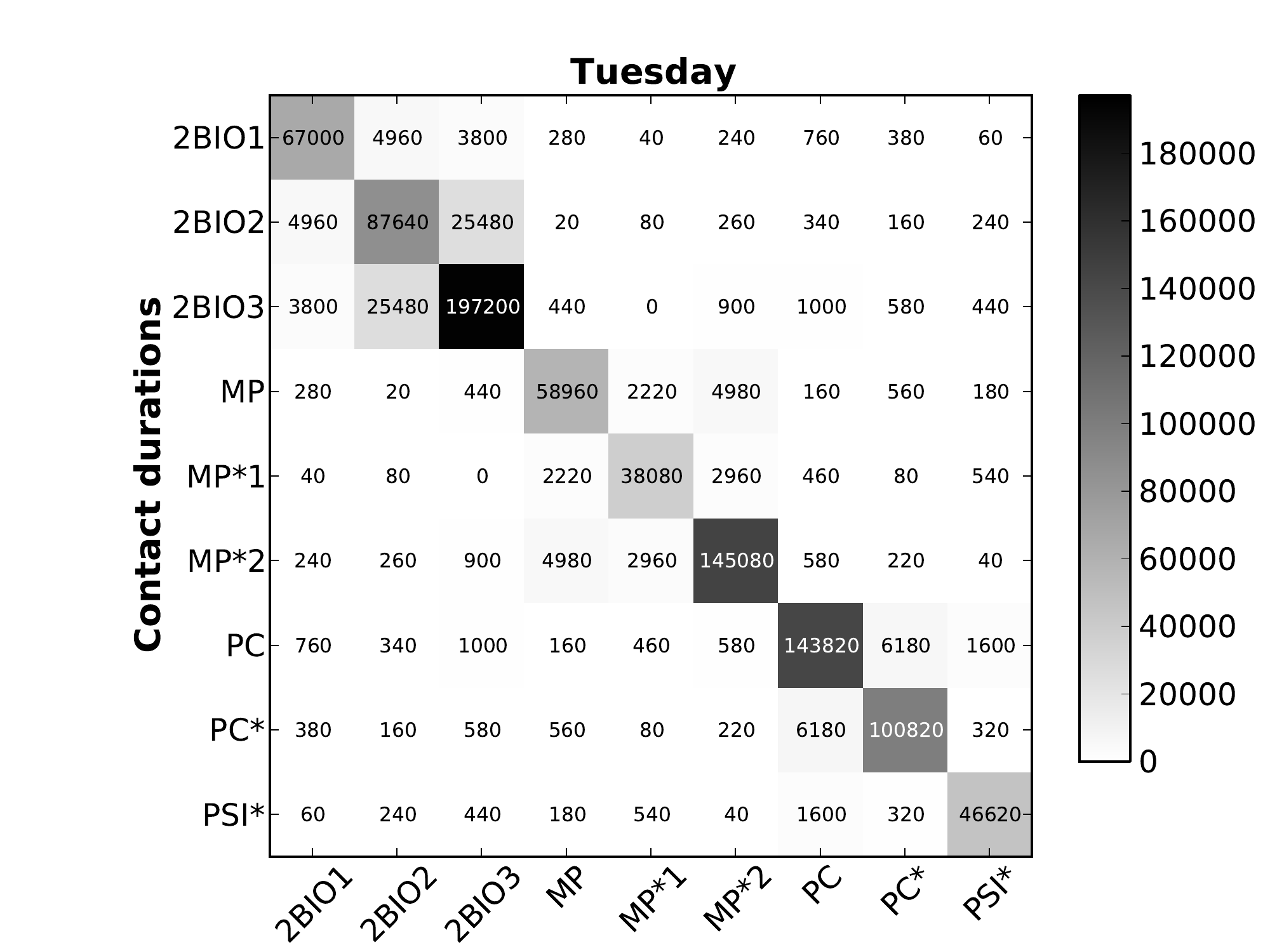}}{\includegraphics[width=0.35\textwidth]{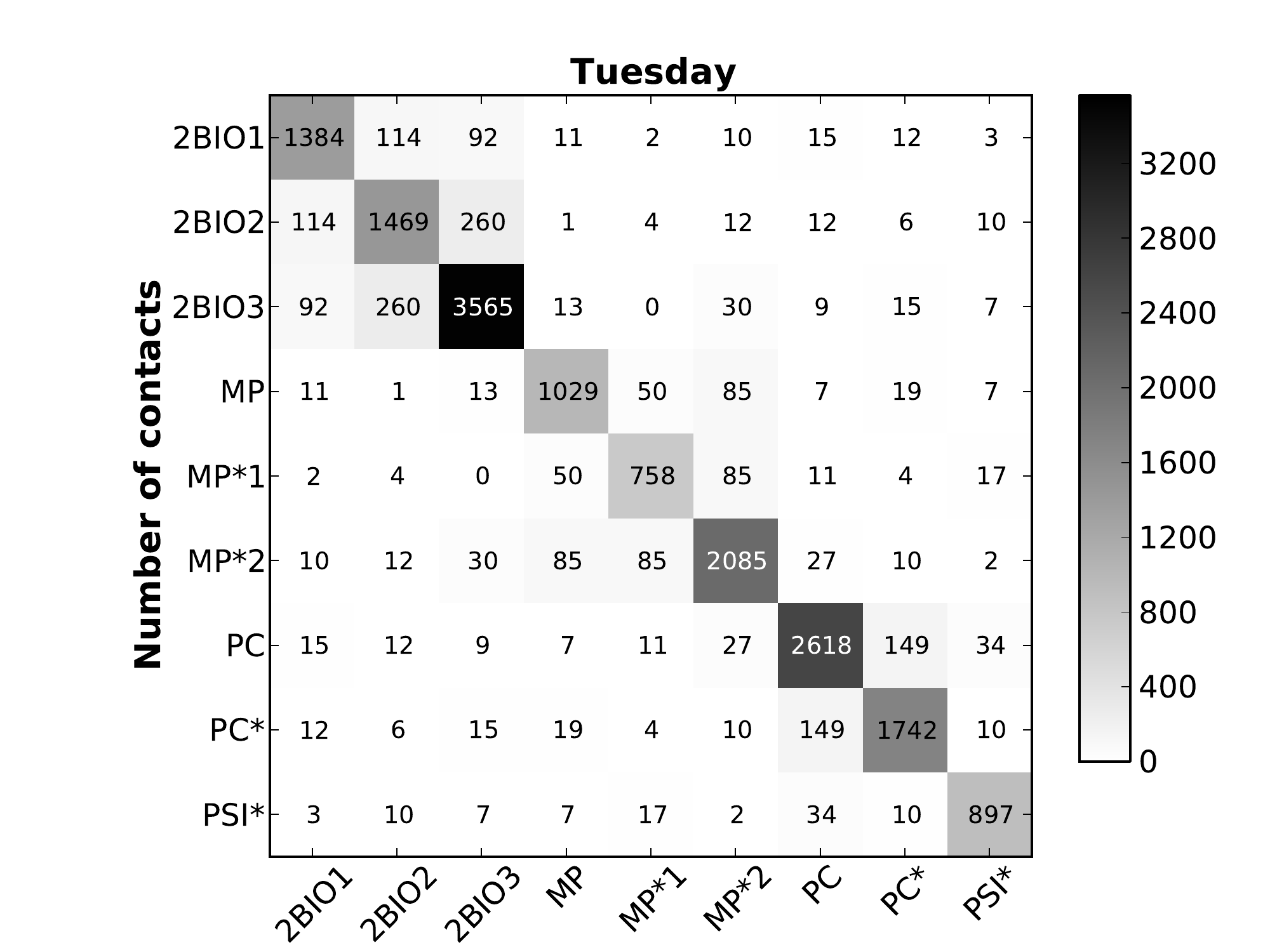}}
{\includegraphics[width=0.35\textwidth]{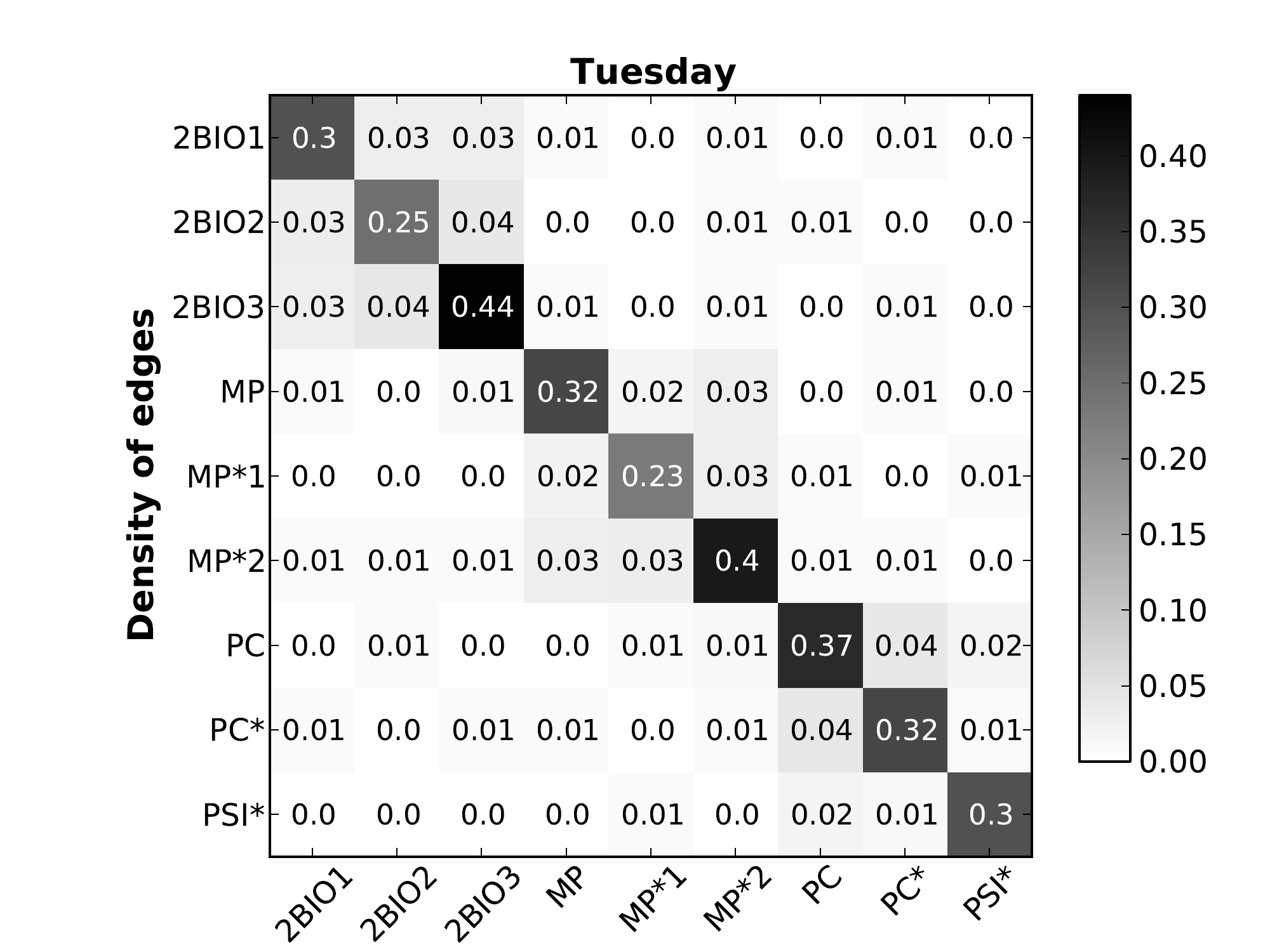}}

\hspace{-10mm}
{\includegraphics[width=0.35\textwidth]{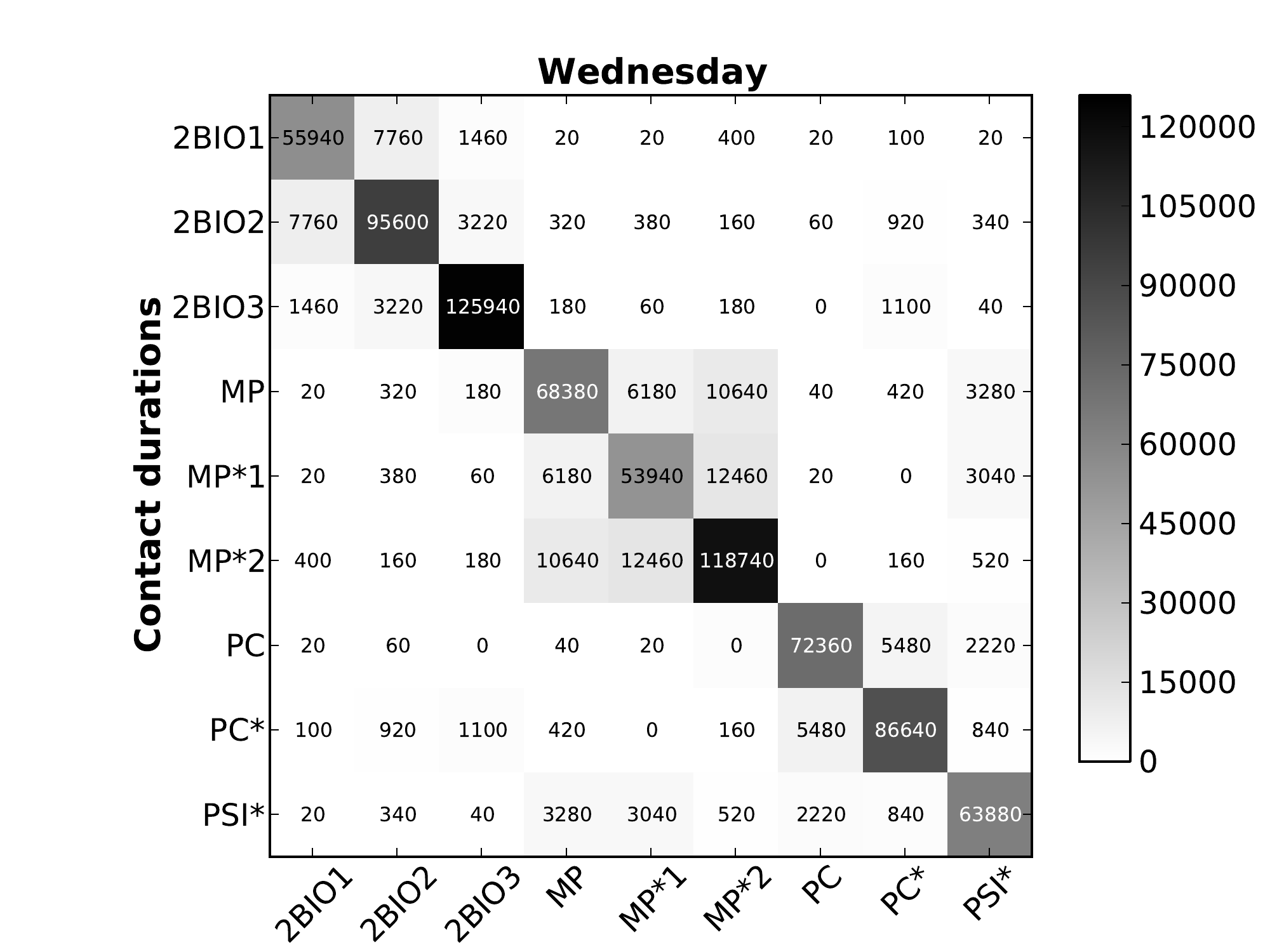}}{\includegraphics[width=0.35\textwidth]{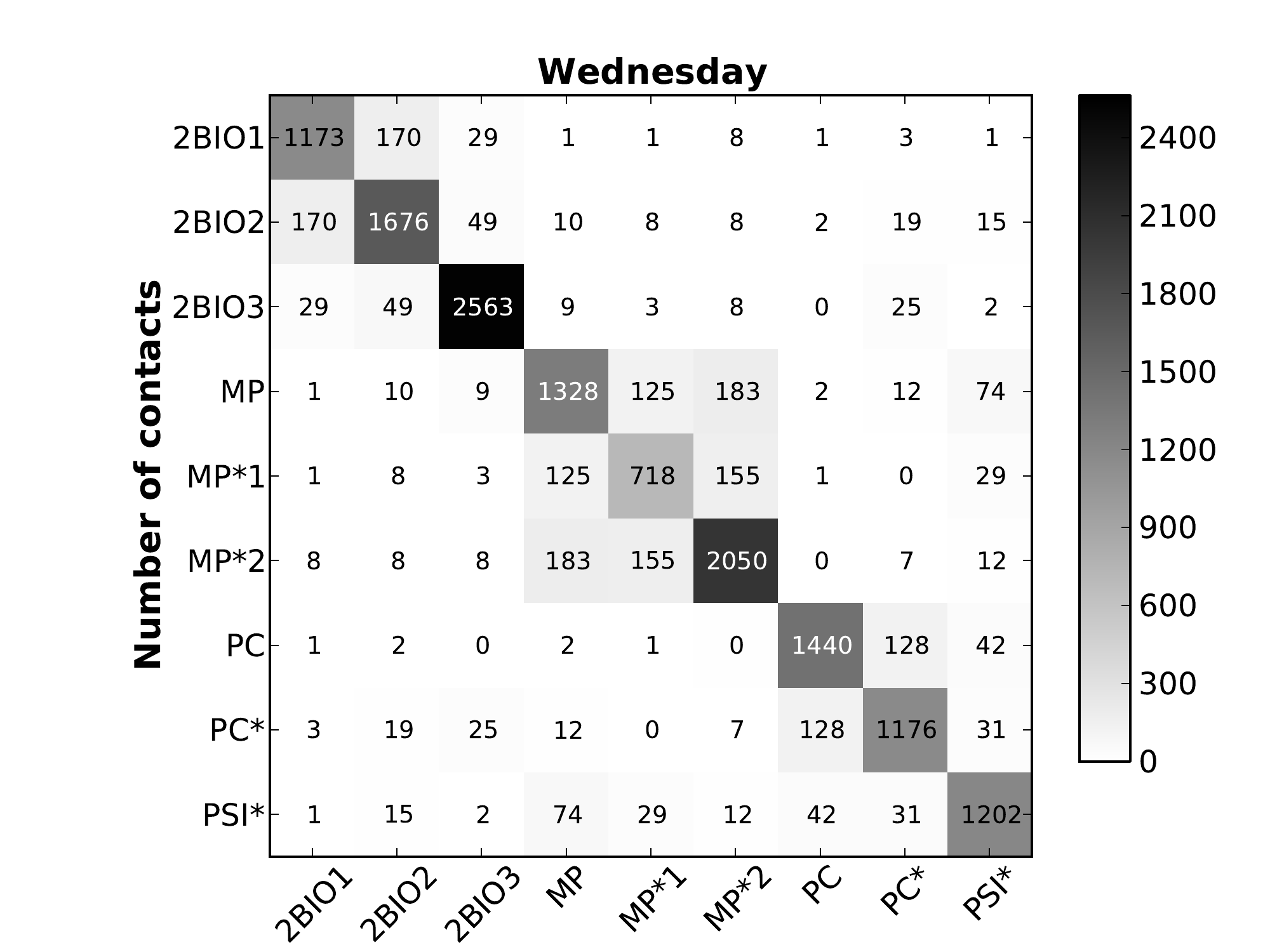}}
{\includegraphics[width=0.35\textwidth]{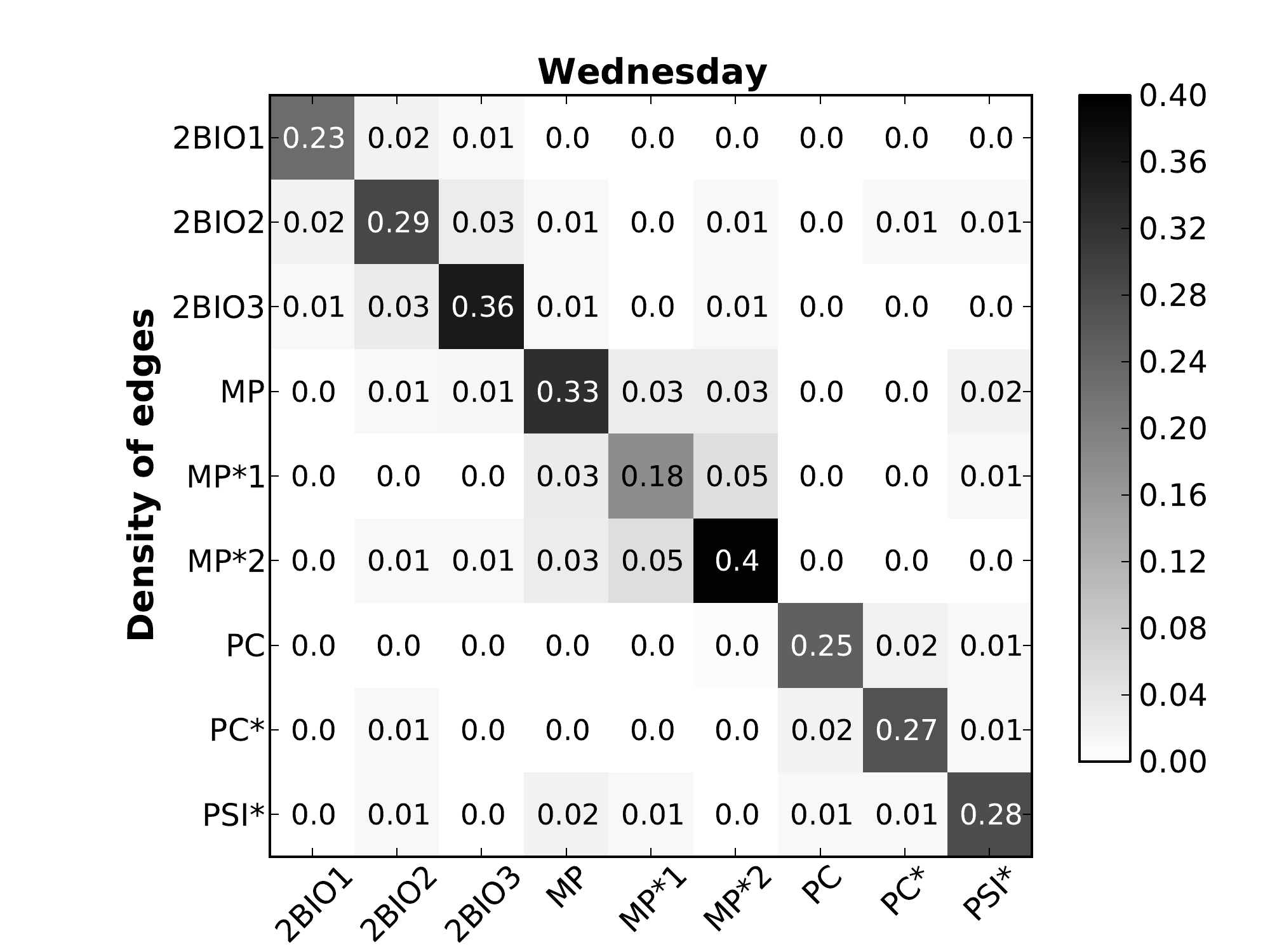}}

\hspace{-10mm}
{\includegraphics[width=0.35\textwidth]{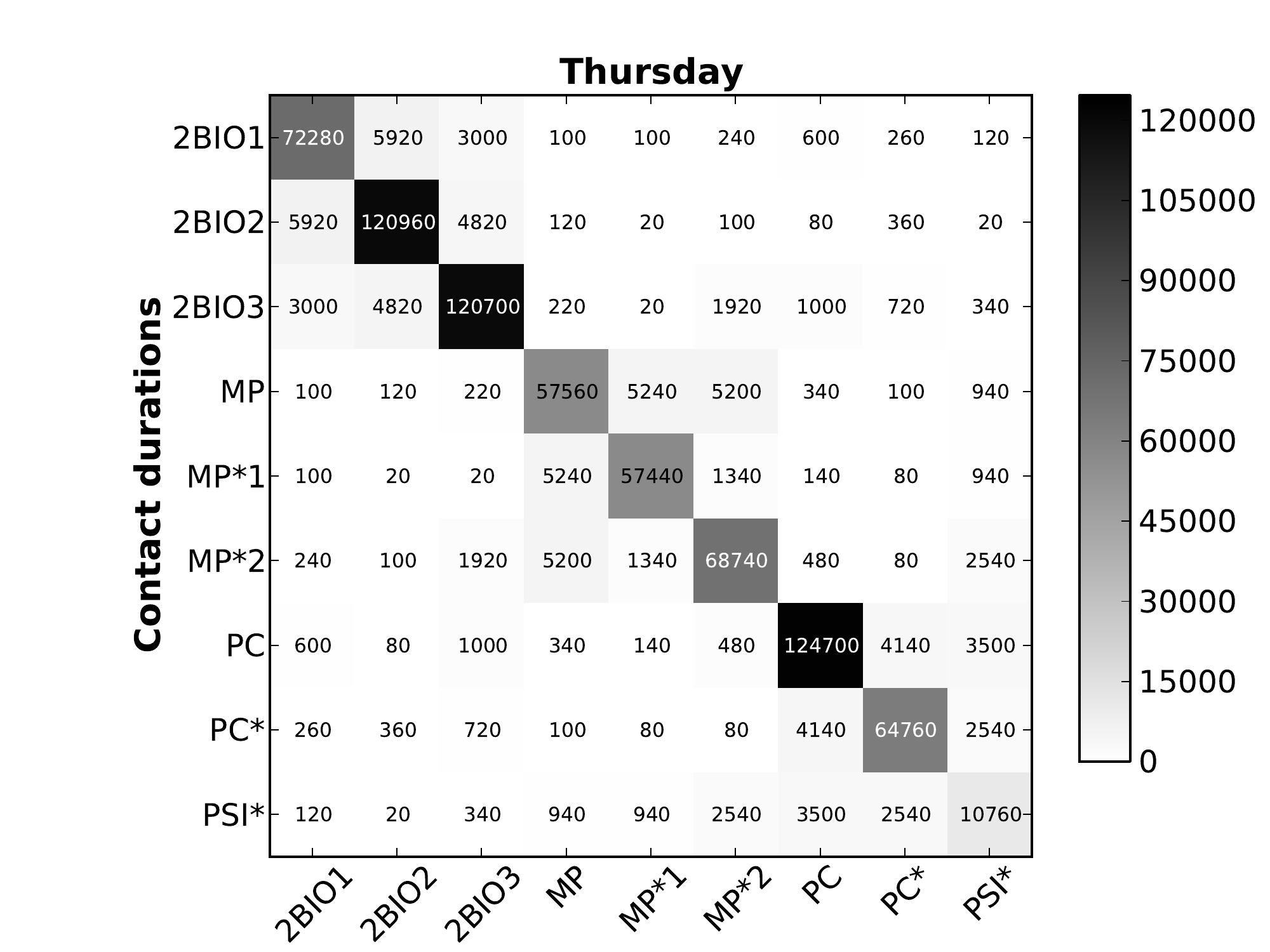}}{\includegraphics[width=0.35\textwidth]{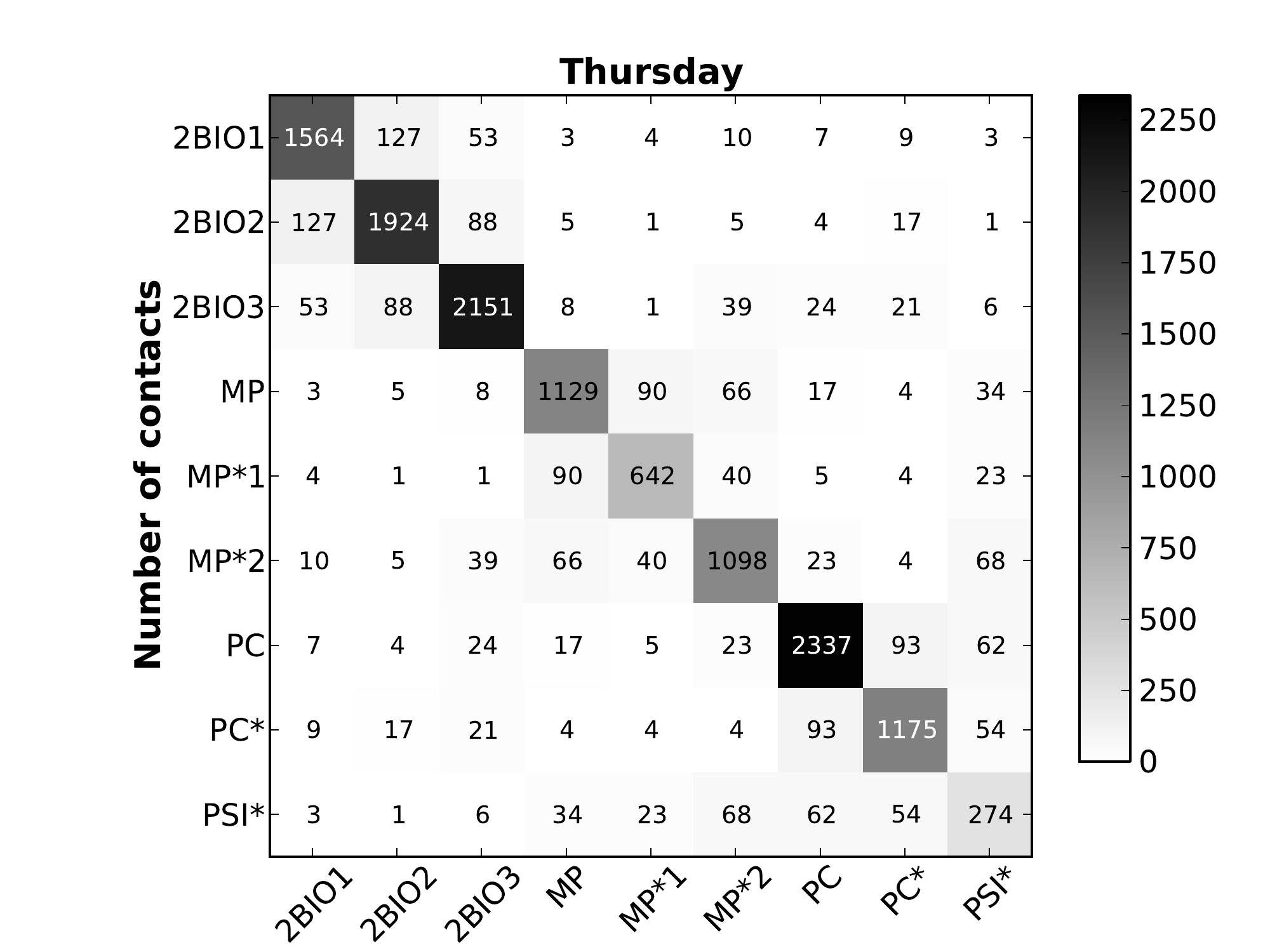}}
{\includegraphics[width=0.35\textwidth]{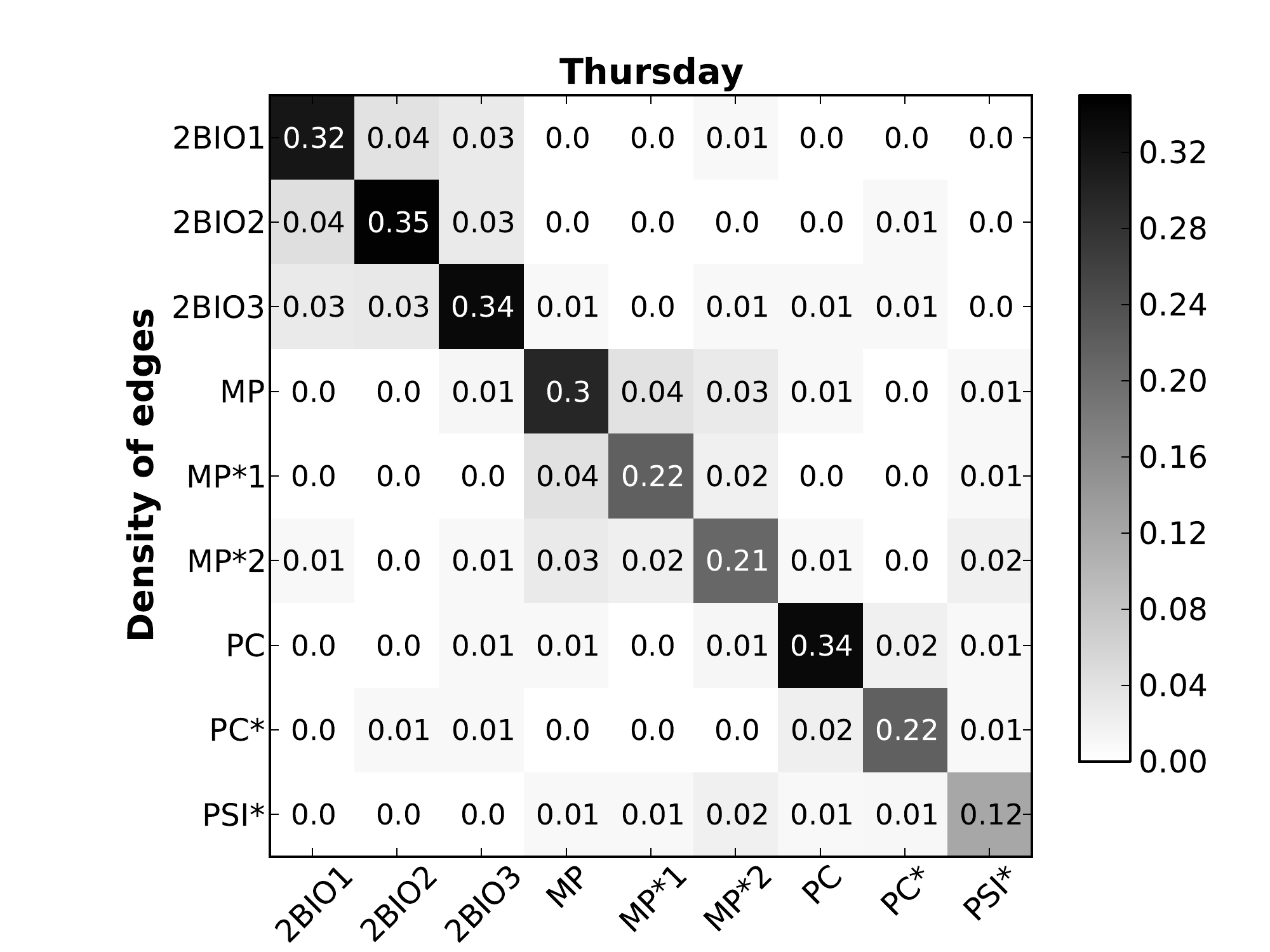}}

\hspace{-10mm}
{\includegraphics[width=0.35\textwidth]{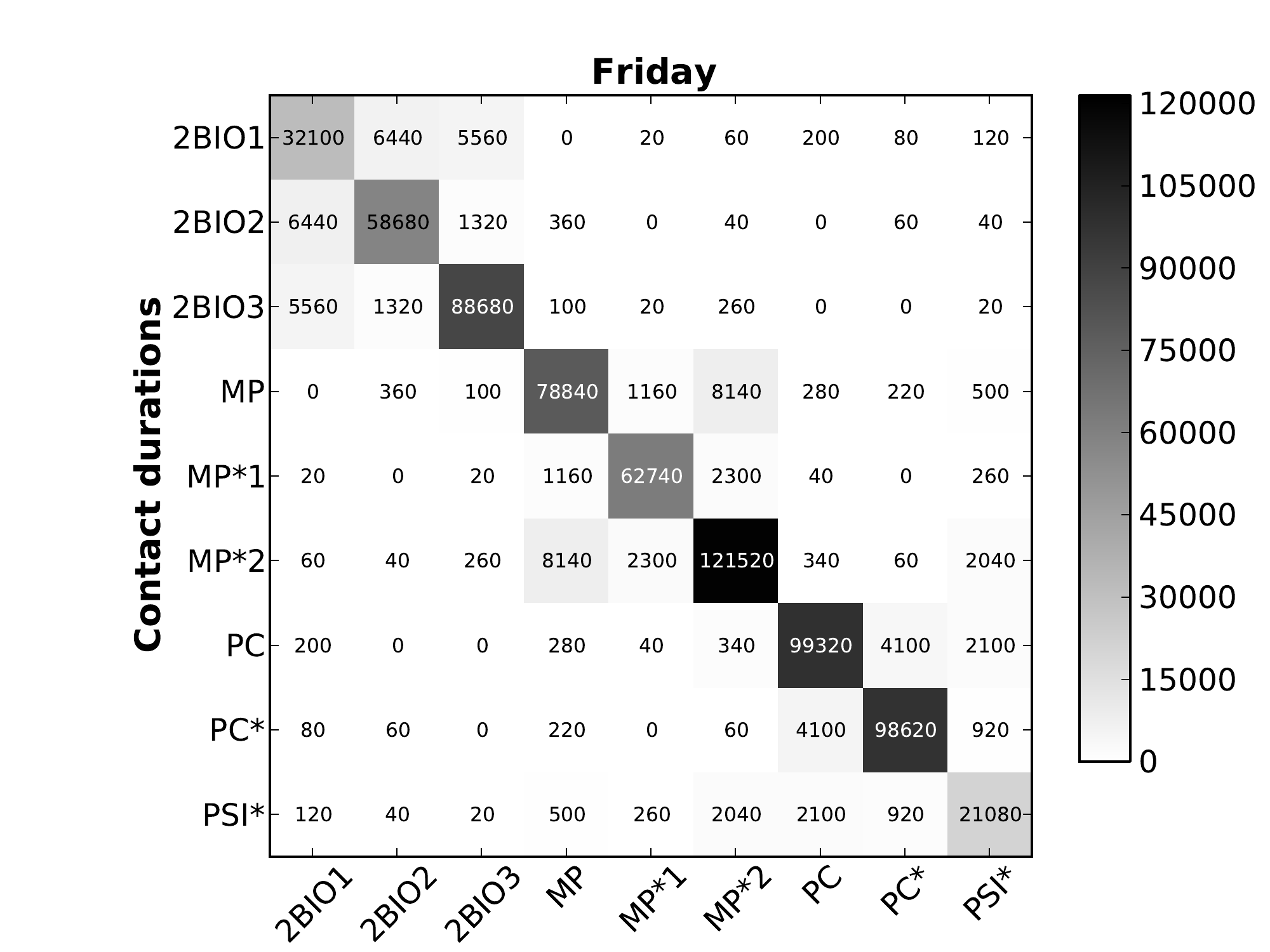}}{\includegraphics[width=0.35\textwidth]{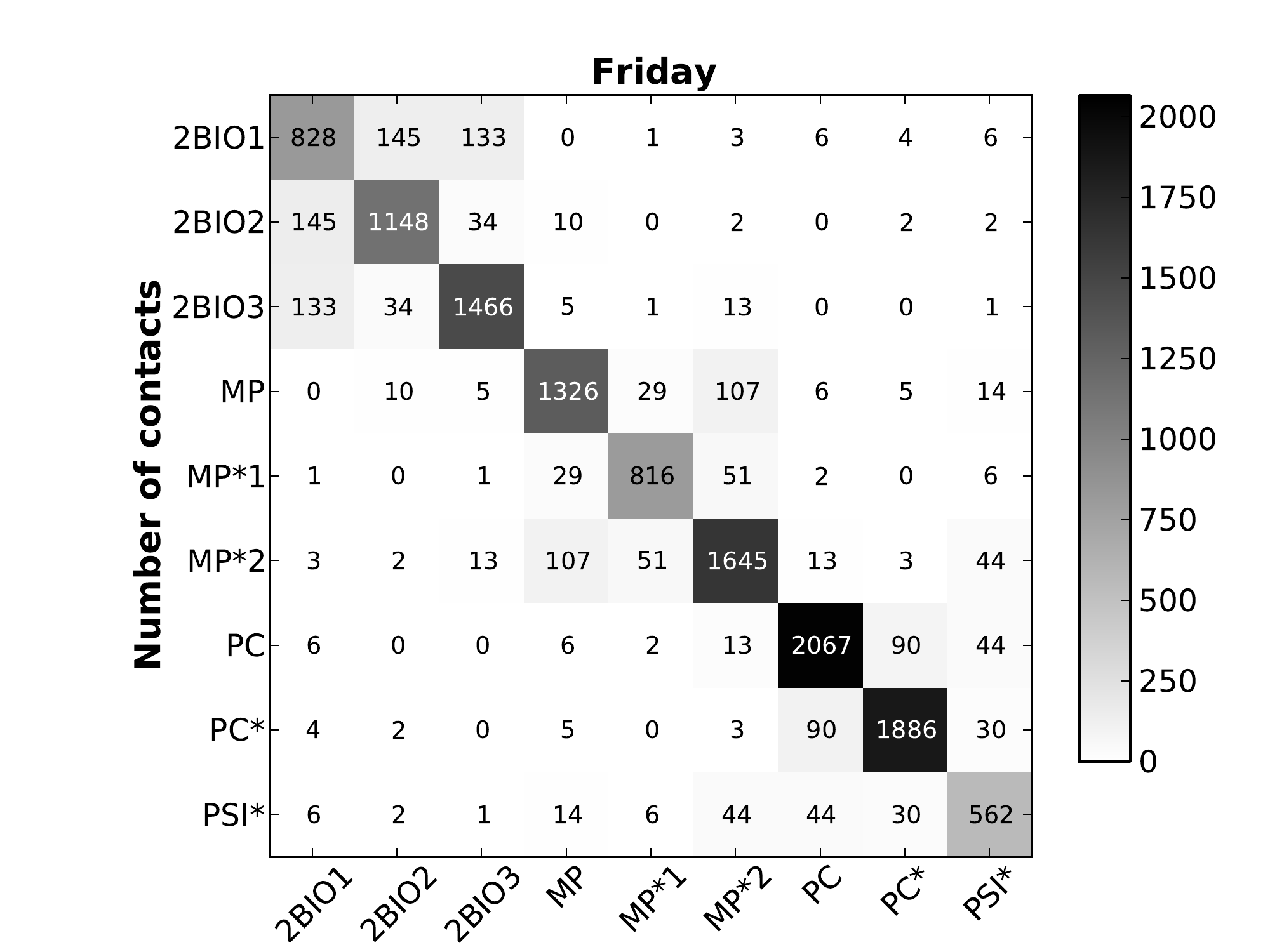}}
{\includegraphics[width=0.35\textwidth]{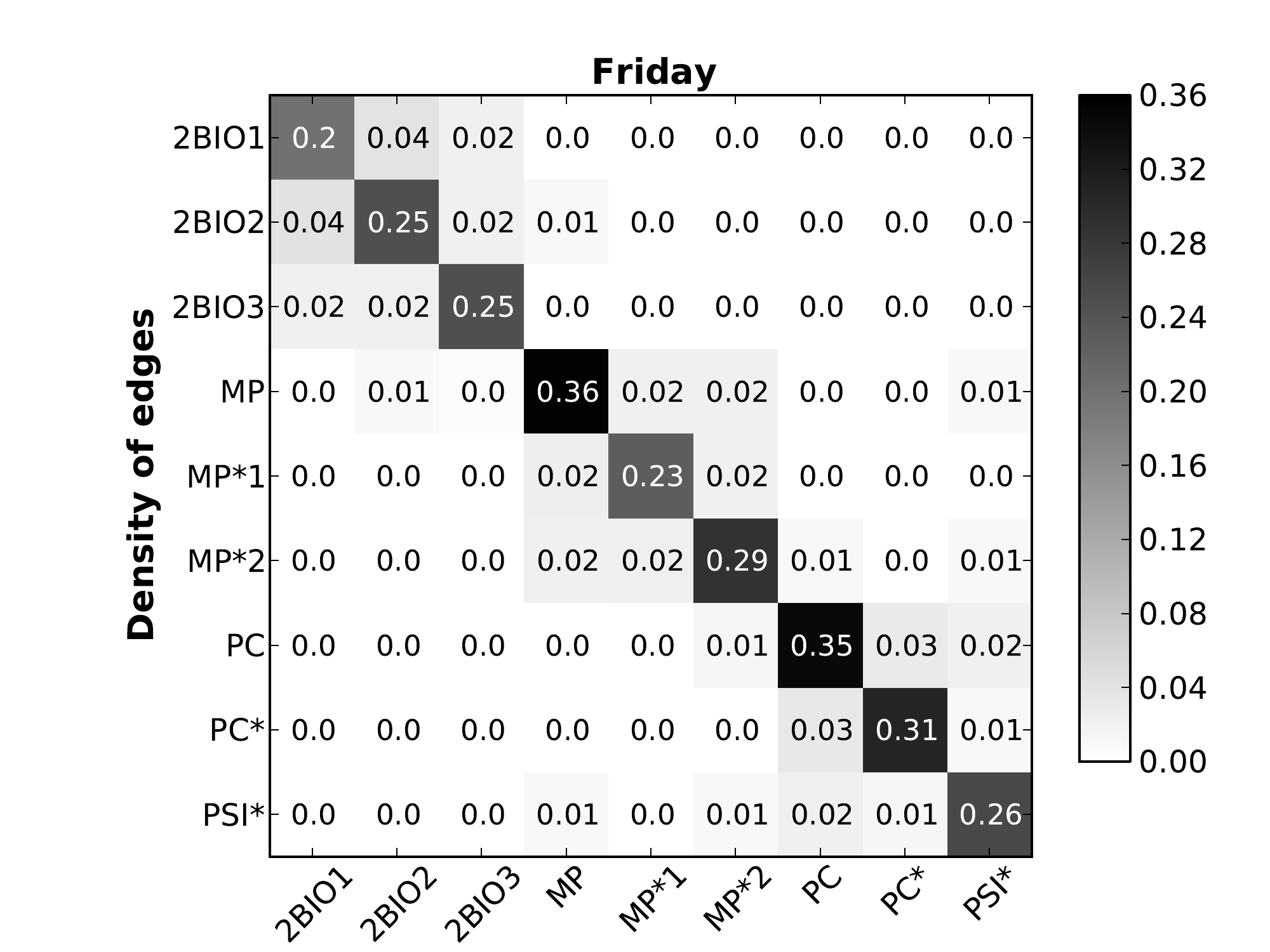}}
\caption{{\bf Contact matrices for each day of the week of study.} Contact matrices giving the cumulated durations of 
contacts (first column), the numbers of contacts (second column) and the densities of links (third column) between classes.}
\end{figure}

\clearpage
\newpage
\subsection{Comparison between the networks built from sensor data and from contact diaries}

\begin{figure}[!ht]
{\includegraphics[width=0.6\textwidth]{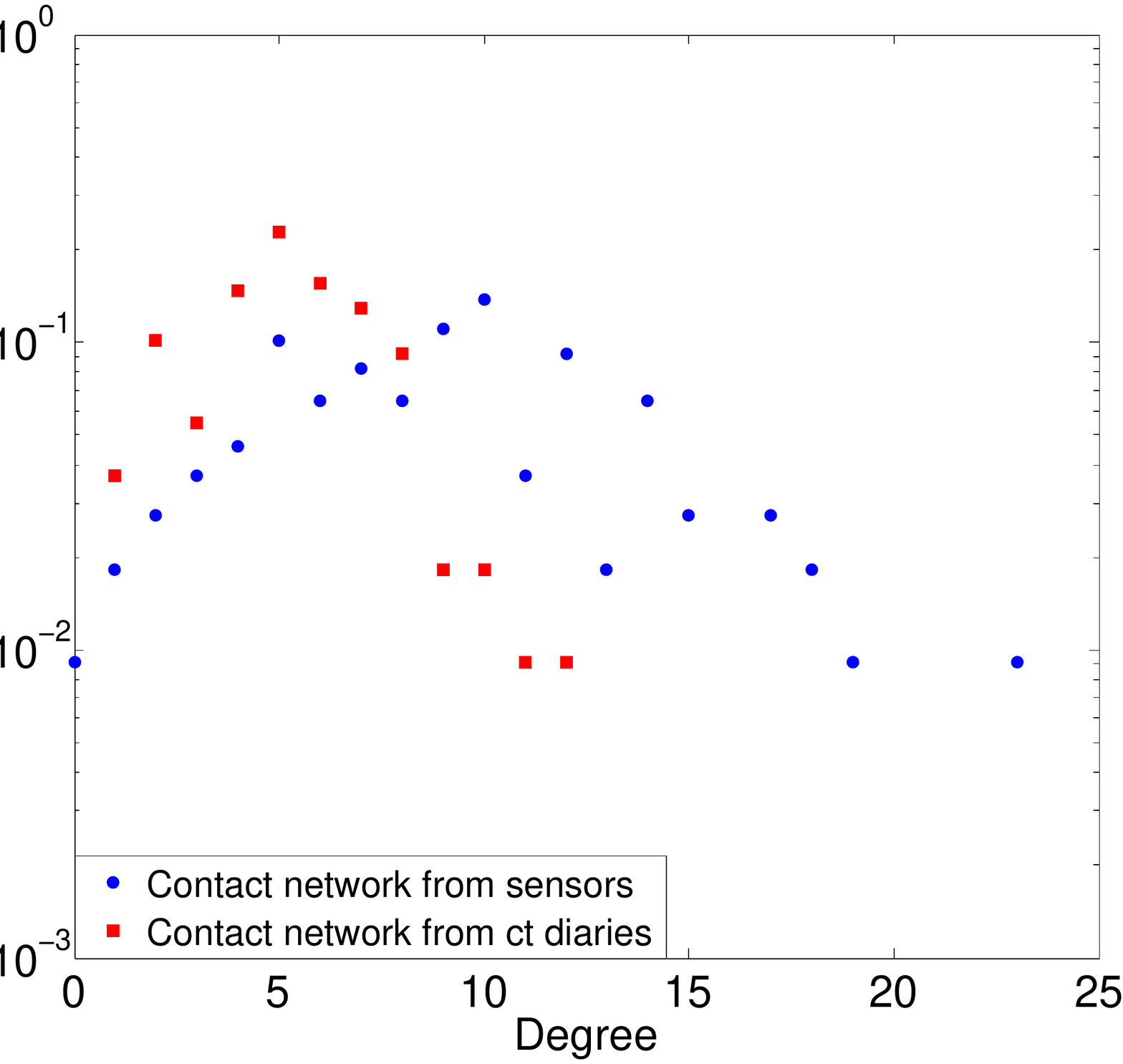}}
%\hspace{-5mm}
%\subfigure[]
%{\includegraphics[width=0.5\textwidth]{DegreeCDFmem.pdf}}
\caption{{\bf Degree distributions of the contact networks obtained by sensors and by contact diaries.} 
}
\end{figure}

\begin{figure}[!ht]
{\includegraphics[width=0.9\textwidth]{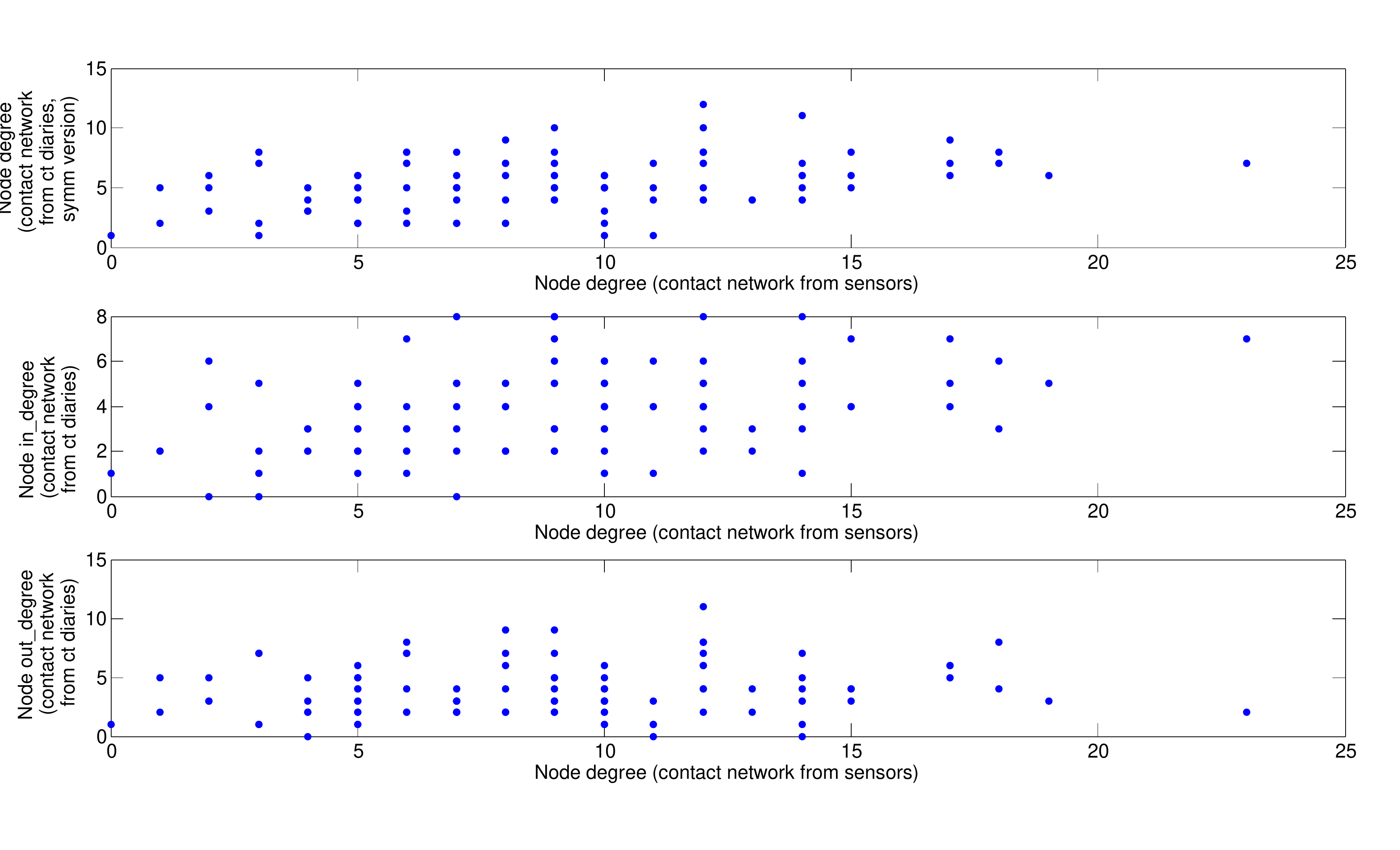}}
\caption{{\bf Comparison of the degree of individuals in the contact networks obtained by sensors and by contact diaries.} 
Scatterplot of the number of links of each node in both networks. As the network built from contact diaries is directed,
we consider the degree in its symmetrized version (top), the in-degree (middle) and the out-degree (bottom), vs. the degree
in the contact network obtained from sensor data.}
\end{figure}

%\begin{figure}[!ht]
%{\includegraphics[width=0.9\textwidth]{StrCorrMem.pdf}}
%\caption{{\bf Sensors vs. contact diaries: node strength correlations.} Correlation between node strength distribution according to the network of contacts givend by the sensors and: the nodestrength distribution according to the symmetrized version of the network of contacts given by the memory-survey (top); the node in\_strength distribution (middle) and the node out\_strength distribution (bottom) according to the network of contacts given by the memory-survey. }
%\end{figure}

\begin{table}[!ht]
\medskip
\begin{tabular}{l|l|l}
\toprule
    {\bf Threshold} &$\mathbf{k_{in}}$  &  $\mathbf{k_{out}}$   \\
\hline
\midrule
 $0s$ ($488$ links)& $0.4^{*}$ (0) & $0.14$ (0.16) \\
 $40s$ ($328$ links)& $0.44^{*}$ (0) & $0.22^{*}$ (0.02) \\
 $60s$ ($263$ links)& $0.5^{*}$ (0) & $0.2^{*}$ (0.02) \\
 $80s$ ($221$ links)& $0.49^{*} $(0) & $0.17$ (0.08) \\
 $100s$ ($203$ links)& $0.47^{*}$ (0) & $0.16$ (0.08) \\
\bottomrule
\end{tabular}
\caption{{\bf Sensors vs. contact diaries: node degree correlations.}
Correlation between the degree of a node in the contact network built from the sensor data and the in- or out-degree of the same
node in the network built from contact diaries. p-values are given in parenthesis. Each row corresponds to keeping only links
with a minimum aggregate duration of contacts (Threshold) in the contact network built from sensor data.
%The corresponding correlations for the node strength are not affected by the variation of thresholds, 
%they are always equal to: 0.21 (0.03) for the in\_strength and 0.22 (0.02) for the out\_strength case.
} \label{prob}
\end{table}

\begin{figure}[!ht]
{\includegraphics[width=0.7\textwidth]{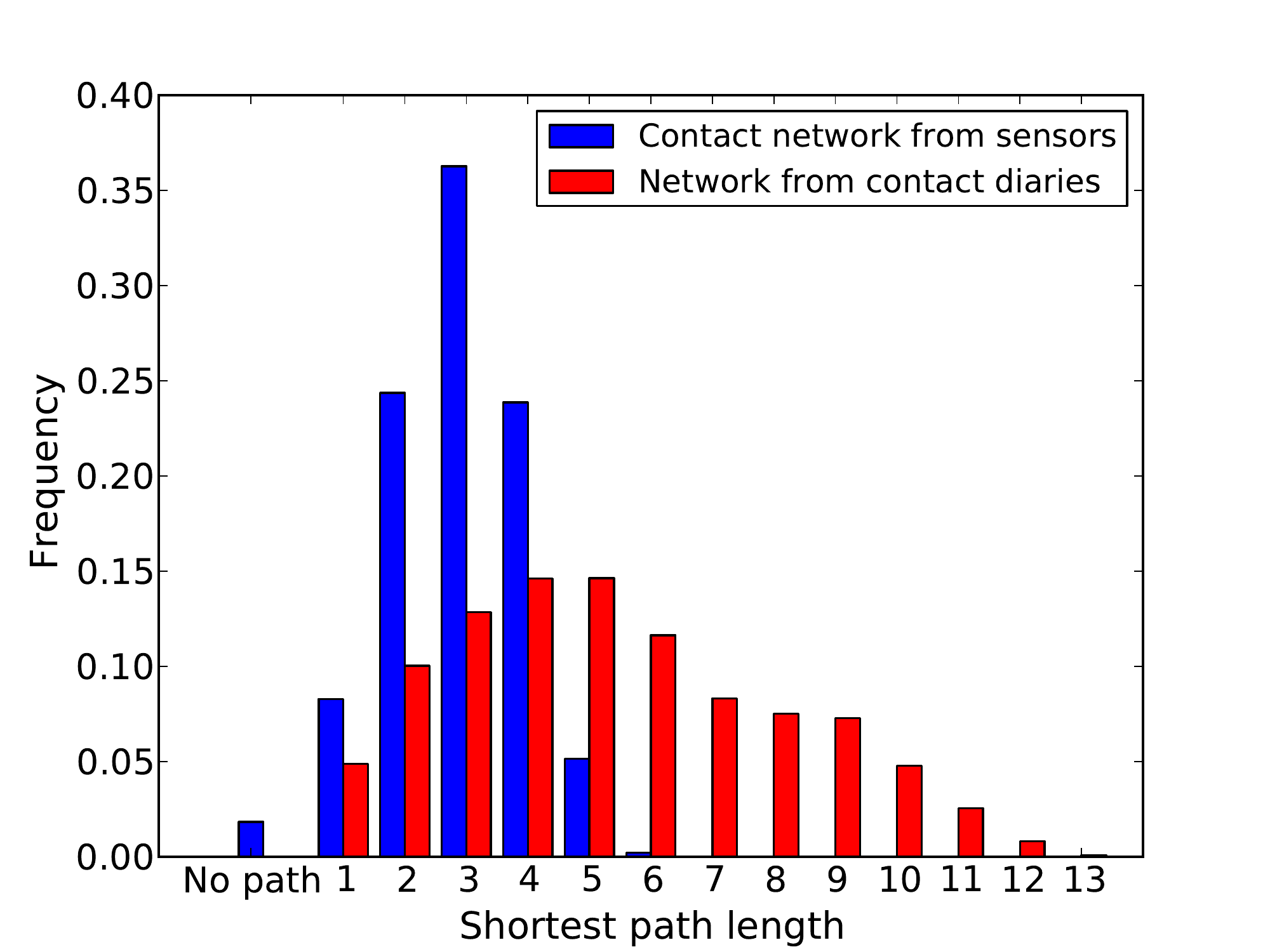}}
\caption{{\bf Sensors vs. contact diaries: shortest path length distribution} in the
networks of contacts built using the sensor (blue) and the contact diary (red) data. "No path" corresponds to isolated nodes.  }
\end{figure}

\clearpage
\newpage

\subsection{Comparison between sensor data and friendship-survey network}

\begin{figure}[!ht]
{\includegraphics[width=0.6\textwidth]{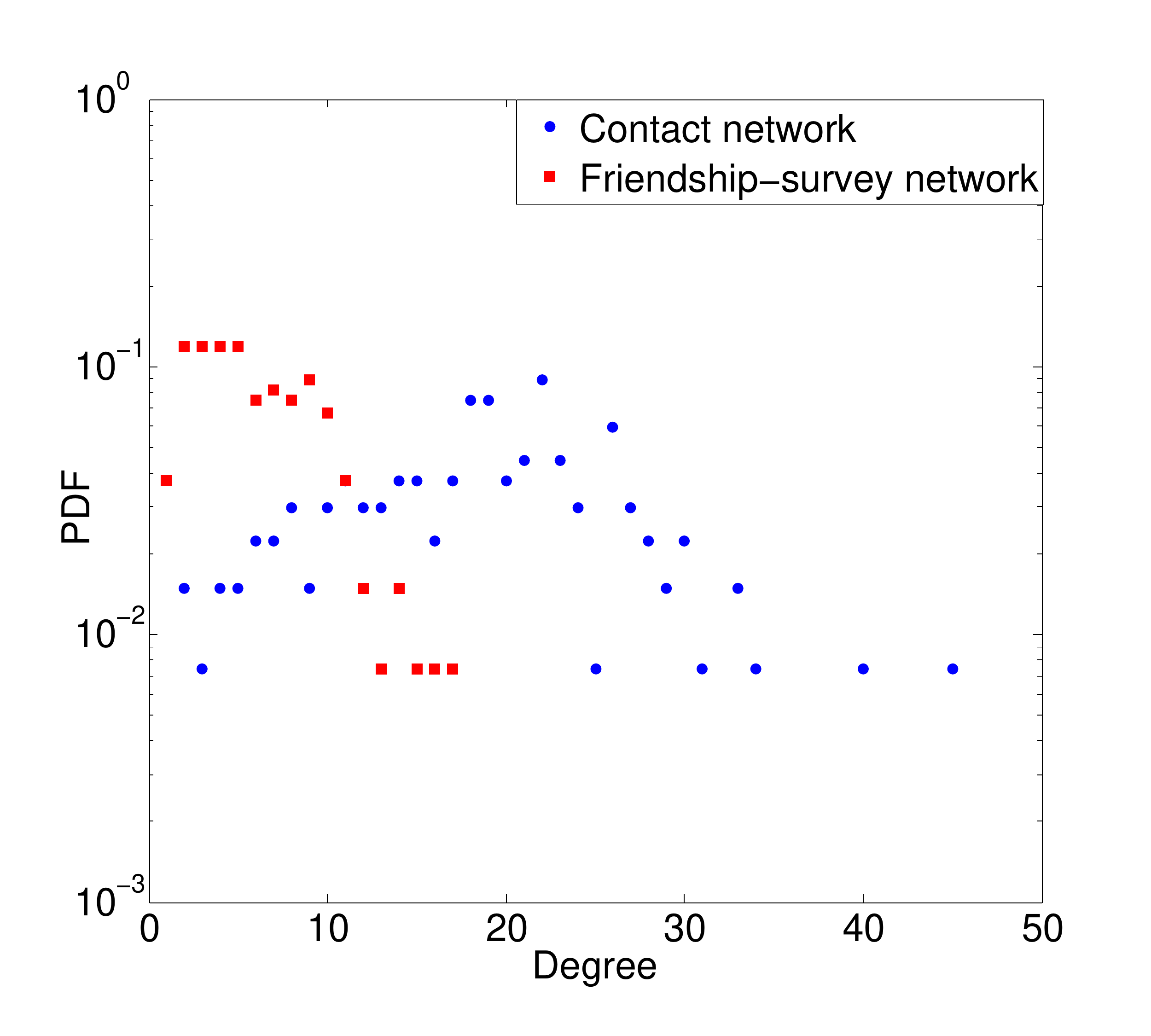}}
\caption{{\bf Sensors vs. friendship-surveys: degree distributions} in the contact network measured by the sensors and
in the network built from the friendship surveys.
}
\end{figure}

\begin{figure}[!ht]
{\includegraphics[width=1\textwidth]{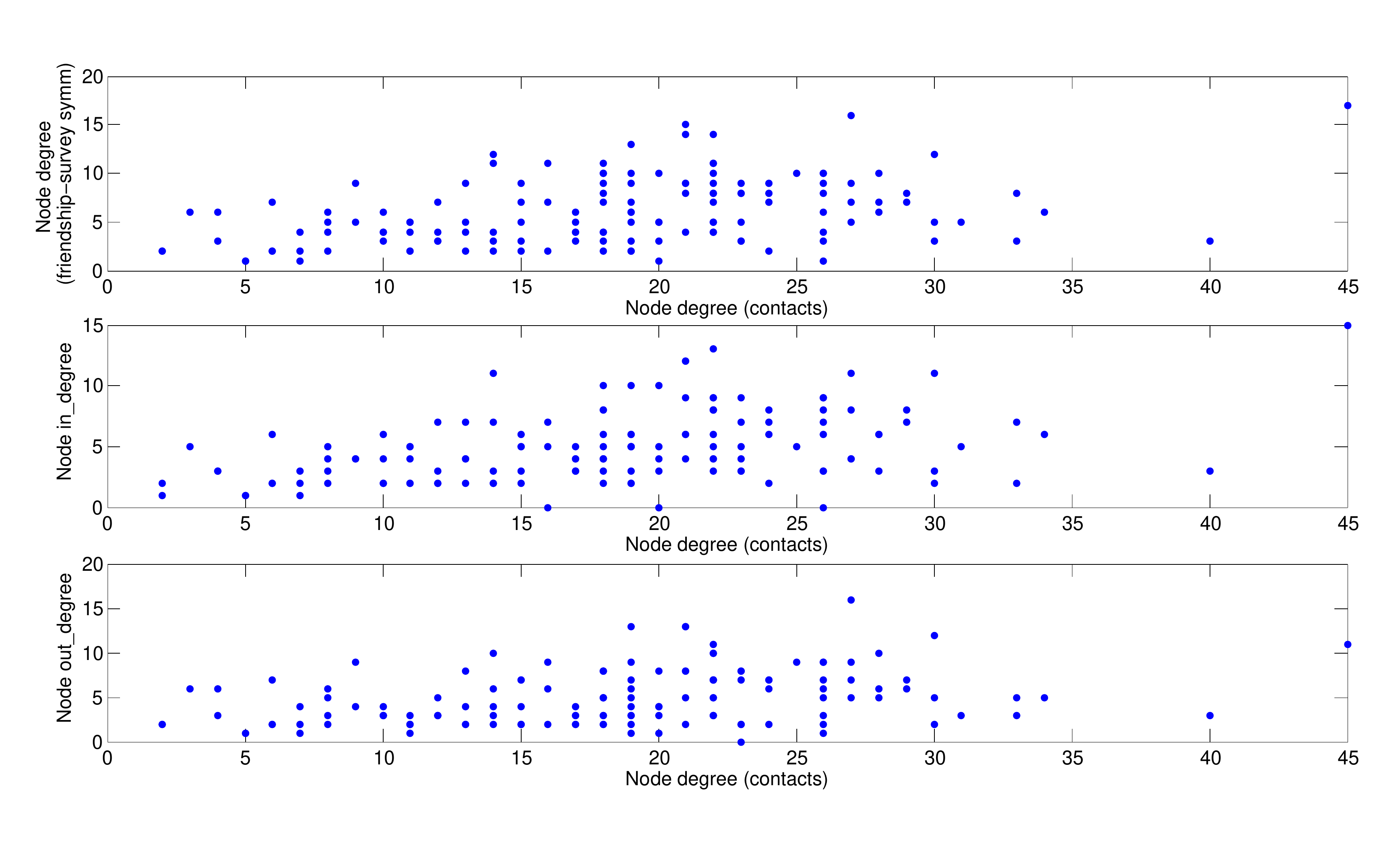}}
\caption{{\bf  Comparison of the degree of individuals in the contact networks obtained by sensors and built from the friendship surveys.}
Scatterplot of the number of links of each node in both networks. As the network of friendships is directed,
we consider the degree in its symmetrized version (top), the in-degree (middle) and the out-degree (bottom), vs. the degree
in the contact network obtained from sensor data.}
\end{figure}

\begin{table}[!ht]
\medskip

\begin{tabular}{l|l|l}
\toprule
    {\bf Threshold} &$\mathbf{k_{in}}$  &  $\mathbf{k_{out}}$   \\
\hline
\midrule
 0s ($1235$ links)& $0.42^{*}$ (0) & $0.34^{*} $(0) \\
 40s ($902$ links)& $0.51^{*}$ (0) & $0.44^{*}$ (0) \\
 60s ($765$ links)& $0.51^{*}$ (0) & $0.45^{*}$ (0) \\
 80s ($671$ links)& $0.53^{*} $(0) & $0.47^{*} (0)$ \\
 100s ($610$ links)& $0.52^{*}$ (0) & $0.47^{*} (0)$ \\
\bottomrule
\end{tabular}
\caption{{\bf Sensors vs. friendship-surveys: node degree correlations.}
Correlations between the degree of a node in the contact network built from the sensor data and the in- or out-degree of the same
node in the friendship network. p-values are given in parenthesis. Each row corresponds to keeping only links
with a minimum aggregate duration of contacts (Threshold) in the contact network built from sensor data.
}
\end{table}

\end{document}